\documentclass[sigconf, nonacm]{acmart}
\pdfoutput=1
\makeatletter
\def\@ACM@checkaffil{
    \if@ACM@instpresent\else
    \ClassWarningNoLine{\@classname}{No institution present for an affiliation}%
    \fi
    \if@ACM@citypresent\else
    \ClassWarningNoLine{\@classname}{No city present for an affiliation}%
    \fi
    \if@ACM@countrypresent\else
        \ClassWarningNoLine{\@classname}{No country present for an affiliation}%
    \fi
}
\makeatother

\usepackage{amsmath,amsfonts}
\usepackage{algorithmic}
\usepackage{tikz}
\usepackage{graphicx}
\usepackage{textcomp}
\usepackage{xcolor}
\usepackage{booktabs}
\usepackage{pifont}
\usepackage{siunitx}
\usepackage{amsmath}
\usepackage{algorithm2e}
\usepackage{multirow}
\usepackage{fancyhdr}
\usepackage{float}
\usepackage{setspace}
\usepackage{listings}
\usepackage{balance}
\usepackage{wrapfig}
\usepackage{marginnote}
\usepackage{caption}
\usepackage{boldline} %
\usepackage{colortbl} %
\usepackage{floatflt}
\usepackage{eso-pic}
\usepackage{titlesec}
\usepackage{noindentafter}
\usepackage[htt]{hyphenat}
\usepackage{lscape}
\usepackage{tablefootnote}

\definecolor{gfored}{rgb}{0.580, 0.050, 0.211}
\definecolor{ao}{rgb}{0.007, 0.520, 0.867}
\definecolor{yt}{rgb}{0.875, 0.568, 1.000}
\definecolor{moegi}{rgb}{0.357, 0.537, 0.188}
\definecolor{jl}{rgb}{1.0, 0.2, 0.8}
\definecolor{brown(web)}{rgb}{0.65, 0.16, 0.16}
\definecolor{bisque}{rgb}{1.0, 0.89, 0.77}
\definecolor{lightbisque}{rgb}{1.0, 0.95, 0.85}
\definecolor{CadetBlue}{rgb}{0.37, 0.62, 0.63}
\definecolor{whitesmoke}{rgb}{0.96, 0.96, 0.96}

\newif\ifsqueezefigs
\squeezefigsfalse

\ifsqueezefigs
\makeatletter
\g@addto@macro{\normalsize}{%
  \setlength{\abovedisplayskip}{2pt plus 1pt minus 1pt}
  \setlength{\belowdisplayskip}{2pt plus 1pt minus 1pt}
  \setlength{\abovedisplayshortskip}{0pt}
  \setlength{\belowdisplayshortskip}{0pt}
  \setlength{\intextsep}{2pt plus 1pt minus 1pt}
  \setlength{\textfloatsep}{3pt plus 1pt minus 1pt}
  \setlength{\dbltextfloatsep}{3pt plus 1pt minus 1pt}
  \setlength{\skip\footins}{4pt plus 1pt minus 1pt}}
\makeatother
\fi

\newif\ifdraft
\draftfalse

\ifdraft
    \usepackage[colorinlistoftodo s,prependcaption,textsize=small]{todonotes}
    \paperwidth=\dimexpr \paperwidth + 4cm\relax
    \oddsidemargin=\dimexpr\oddsidemargin + 2cm\relax
    \evensidemargin=\dimexpr\evensidemargin + 2cm\relax
    \marginparwidth=\dimexpr \marginparwidth + 2cm\relax

    \newcommand{\om}[1]{\textcolor{red}{#1}}
    \newcommand{\ominline}[1]{\textcolor{red}{\textbf{[@om: }#1\textbf{]}}}
    \newcommand{\ombox}[1]{\todo[size=\scriptsize, linecolor=red, bordercolor=red, backgroundcolor=white]{\textcolor{red}{\textbf{@om:} #1}}}

    \newcommand{\gf}[1]{\textcolor{blue}{#1}}

    \newcommand{\agy}[1]{\textcolor{gfored}{#1}}
    \newcommand{\agycomment}[1]{\todo[size=\scriptsize, linecolor=orange, bordercolor=orange, backgroundcolor=white]{\textcolor{gfored}{\textbf{@gy:} #1}}}
    \newcommand{\agyinline}[1]{\textcolor{gfored}{\textbf{[@agy: }#1\textbf{]}}}
    \newcommand{\agyt}[1]{{\textcolor{brown(web)}{#1}}}
    \newcommand{\agyh}[1]{{\textcolor{brown(web)}{#1}}}
    \newcommand{\agyf}[1]{{\textcolor{gfored}{#1}}}
    
    \newcommand{\atb}[1]{\textcolor{ao}{#1}}
    \newcommand{\atbcomment}[1]{\todo[size=\scriptsize, linecolor=orange, bordercolor=orange, backgroundcolor=white]{\textcolor{ao}{\textbf{@atb:} #1}}}
    
    \newcommand{\hluo}[1]{\textcolor{moegi}{#1}}
    \newcommand{\hluoinline}[1]{\textcolor{moegi}{\textbf{[@hluo: }#1\textbf{]}}}
    \newcommand{\hluobox}[1]{\todo[size=\scriptsize, linecolor=orange, bordercolor=orange, backgroundcolor=white]{\textcolor{moegi}{\textbf{@hluo:} #1}}}
    
    \newcommand{\yct}[1]{\textcolor{yt}{#1}}
    \newcommand{\yctcomment}[1]{\todo[size=\scriptsize, linecolor=orange, bordercolor=orange, backgroundcolor=white]{\textcolor{yt}{\textbf{@yct:} #1}}}

    \newcommand{\joel}[1]{\textcolor{jl}{#1}}
    \newcommand{\joelcomment}[1]{\todo[size=\scriptsize,linecolor=orange,bordercolor=orange,backgroundcolor=white]{\textcolor{jl}{\textbf{@joel:} #1}}}
    \newcommand{\param}[1]{\textcolor{red}{#1}} %

\else

    \newcommand{\om}[1]{{#1}}
    \newcommand{\ominline}[1]{}
    \newcommand{\ombox}[1]{}

    \newcommand{\gf}[1]{{#1}}
    
    \newcommand{\agy}[1]{{#1}}
    \newcommand{\agycomment}[1]{}
    \newcommand{\agyinline}[1]{}
    \newcommand{\agyt}[1]{#1}
    \newcommand{\agyh}[1]{#1}
    \newcommand{\agyf}[1]{#1}
    
    \newcommand{\atb}[1]{{#1}}
    \newcommand{\atbcomment}[1]{}
    
    \newcommand{\hluo}[1]{{#1}}
    \newcommand{\hluoinline}[1]{}
    \newcommand{\hluobox}[1]{}
    
    \newcommand{\yct}[1]{{#1}}
    \newcommand{\yctcomment}[1]{}

    \newcommand{\joel}[1]{{#1}}
    \newcommand{\joelcomment}[1]{}
    \newcommand{\param}[1]{{#1}} %

\fi

\newif\ifrebuttal
\rebuttalfalse

\ifrebuttal
\usepackage[colorinlistoftodos,prependcaption,textsize=small]{todonotes}

\definecolor{darkred}{rgb}{0.9, 0.0, 0.0}
\newcommand{\rb}[1]{\textcolor{darkred}{#1}}

\definecolor{darkblue}{rgb}{0.0, 0.0, 0.85}
\newcommand{\revision}[1]{\textcolor{darkblue}{#1}}

\newcommand{\revlabel}[1]{\marginnote{\textbf{\Large #1}}}
\newcommand{\minorrevlabel}[1]{\marginnote{\textcolor{darkred}{\textbf{\Large #1}}}}

\else

\definecolor{darkred}{rgb}{0.9, 0.0, 0.0}
\newcommand{\rb}[1]{#1}

\definecolor{darkblue}{rgb}{0.0, 0.0, 0.85}
\newcommand{\revision}[1]{#1}

\newcommand{\revlabel}[1]{}
\newcommand{\minorrevlabel}[1]{}

\fi

\newif\ifcamerareadydraft
\camerareadydraftfalse

\ifcamerareadydraft
    \usepackage[colorinlistoftodos,prependcaption,textsize=small, textwidth=1.3cm]{todonotes}
    \newcommand{\hluocr}[1]{\textcolor{blue}{#1}}
    \newcommand{\hluocrcomment}[1]{\textbf{\textcolor{moegi}{[hluo: #1]}}}

    \newcommand{\omtodo}[1]{\textbf{\textcolor{purple}{[FIXME] {#1}}}}
    
    \newcommand{\agycrcomment}[1]{\textbf{\textcolor{blue}{[@gy: #1]}}}
    \newcommand{\hluonote}[1]{\todo[size=\tiny]{\tiny{#1}}}
\else
    \newcommand{\hluocr}[1]{{#1}}
    \newcommand{\hluocrcomment}[1]{}

    \newcommand{\omtodo}[1]{}
    
    \newcommand{\agycrcomment}[1]{}
    \newcommand{\hluonote}[1]{}
\fi

\newif\ifarxiv
\arxivtrue

\newif\ifprearxiv
\prearxivfalse

\ifprearxiv
\newcommand{\prearxiv}[1]{\textcolor{blue}{#1}}
\else
\newcommand{\prearxiv}[1]{#1}
\fi

\usepackage{calc}
\newcommand*\circled[1]{\tikz[baseline=(char.base)]
{
  \node[shape=circle,fill,inner sep=0.5pt] (char) {\textcolor{white}{#1}};}
}

\newcommand*\redcircled[1]{\tikz[baseline=(char.base)]
{
  \node[shape=circle,fill=red,inner sep=0.5pt] (char) {\textcolor{white}{#1}};}
}

\newcommand*\DRAMCMD[1]{\texttt{#1}}
\newcommand*\DRAMTIMING[1]{t\textsubscript{#1}}
\newcommand*\nCHIPS{164}
\newcommand*\nMODULES{21}

\newcounter{obs}
\newcommand\observation[1]{%
   \refstepcounter{obs}
  \vspace{0.2em}
  \noindent
  \begin{tabular}{|p{0.95\linewidth}|}
       \hline
       \textbf{{Obsv. \theobs}.} {{#1}}\\
       \hline 
  \end{tabular}
  \vspace{0.2em}
}

\newcounter{tkw}
\newcommand\takeawaybox[1]{%
   \refstepcounter{tkw}
   \vspace{0.2em}

  \noindent
  \begin{tabular}{V{3}p{0.95\linewidth}V{3}}
       \hlineB{3}
       \rowcolor{whitesmoke}
       \textbf{{Takeaway \thetkw}.} {{#1}}\\
       \hlineB{3}
  \end{tabular}
    \vspace{0.2em}

}

\newcommand{\exploitingRowHammerAllCitations}[0]{\cite{rowhammer-js,  fournaris2017exploiting, poddebniak2018attacking, tatar2018throwhammer, carre2018openssl, barenghi2018software, zhang2018triggering, bhattacharya2018advanced, google-project-zero, kim2014flipping, rowhammergithub, seaborn2015exploiting, van2016drammer, gruss2016rowhammer, razavi2016flip, pessl2016drama, xiao2016one, bosman2016dedup, bhattacharya2016curious, burleson2016invited, qiao2016new, brasser2017can, jang2017sgx, aga2017good, mutlu2017rowhammer, tatar2018defeating, gruss2018another, lipp2018nethammer, van2018guardion, frigo2018grand, cojocar2019eccploit,  ji2019pinpoint, mutlu2019rowhammer, hong2019terminal, kwong2020rambleed, frigo2020trrespass, cojocar2020rowhammer, weissman2020jackhammer, zhang2020pthammer, yao2020deephammer, deridder2021smash, hassan2021utrr, jattke2022blacksmith, tol2022toward, kogler2022half, orosa2022spyhammer, zhang2022implicit, liu2022generating, cohen2022hammerscope, zheng2022trojvit, fahr2022frodo, tobah2022spechammer, rakin2022deepsteal}}

\newcommand{\exploitingRowHammerAllCitationsExceptFlipping}[0]{\cite{rowhammer-js,  fournaris2017exploiting, poddebniak2018attacking, tatar2018throwhammer, carre2018openssl, barenghi2018software, zhang2018triggering, bhattacharya2018advanced, google-project-zero, rowhammergithub, seaborn2015exploiting, van2016drammer, gruss2016rowhammer, razavi2016flip, pessl2016drama, xiao2016one, bosman2016dedup, bhattacharya2016curious, burleson2016invited, qiao2016new, brasser2017can, jang2017sgx, aga2017good, mutlu2017rowhammer, tatar2018defeating, gruss2018another, lipp2018nethammer, van2018guardion, frigo2018grand, cojocar2019eccploit,  ji2019pinpoint, mutlu2019rowhammer, hong2019terminal, kwong2020rambleed, frigo2020trrespass, cojocar2020rowhammer, weissman2020jackhammer, zhang2020pthammer, yao2020deephammer, deridder2021smash, hassan2021utrr, jattke2022blacksmith, tol2022toward, kogler2022half, orosa2022spyhammer, zhang2022implicit, liu2022generating, cohen2022hammerscope, zheng2022trojvit, fahr2022frodo, tobah2022spechammer, rakin2022deepsteal}}

\newcommand{\understandingRowHammerAllCitations}[0]{\cite{kim2014flipping, park2016statistical, park2016experiments,lim2017active, ryu2017overcoming, yun2018study, yang2019trap, walker2021ondramrowhammer, kim2020revisiting, orosa2021deeper, orosa2022spyhammer, cohen2022hammerscope, yaglikci2022understanding, khan2018analysis, agarwal2018rowhammer, li2014write, ni2018write, genssler2022reliability, mutlu2022fundamentally}}

\newcommand{\figref}[1]{Fig.~\ref{#1}}

\newcommand{\secref}[1]{§\ref{#1}}

\newcommand{\tabref}[1]{Table~\ref{#1}}

\newcommand{\shellcmd}[1]{\\\indent\indent\texttt{\footnotesize\$ #1}}

\usepackage{glossaries}
\glsdisablehyper

\newacronym{vdd}{$V_{DD}$}{supply voltage}
\newacronym{vpp}{$V_{PP}$}{wordline voltage}
\newacronym{vwl}{$V_{PP}$}{wordline voltage}
\newacronym{vgs}{$V_{GS}$}{gate-to-source voltage}
\newacronym{vth}{$V_{TH}$}{the voltage threshold that the bitline voltage should exceed for the activation to be reliably completed}
\newacronym{gnd}{$GND$}{ground}

\newacronym{ber}{$BER$}{bit error rate}
\newacronym{acmin}{$AC_{min}$}{the minimum number of total aggressor row activations to cause at least one bitflip}
\newacronym{ac}{$AC$}{activation count}
\newacronym{rblast}{$r_{Blast}$}{blast radius}
\newacronym{iqr}{$IQR$}{interquartile range}

\newacronym{trcd}{\DRAMTIMING{RCD}}{
{the minimum time between opening a row with an \DRAMCMD{ACT} command and accessing the row buffer}
}
\newacronym{trp}{\DRAMTIMING{RP}}{
{the minimum time between sending a \DRAMCMD{PRE} command and opening a row with an \DRAMCMD{ACT} command}
}
\newacronym{tras}{\DRAMTIMING{RAS}}{
{the minimum time between opening a row with an \DRAMCMD{ACT} command and closing the row with a \DRAMCMD{PRE} command}
}
\newacronym{trefi}{\DRAMTIMING{REFI}}{the \joel{default} time interval \joel{between consecutive \DRAMCMD{REF} commands}}
\newacronym{trefw}{\DRAMTIMING{REFW}}{the maximum time window between two refresh operations that target {the same} row}

\AtBeginDocument{%
  \providecommand\BibTeX{{%
    \normalfont B\kern-0.5em{\scshape i\kern-0.25em b}\kern-0.8em\TeX}}}

\author{
{Haocong Luo}\qquad
{Ataberk Olgun}\qquad
{A. Giray Ya\u{g}l{\i}k\c{c}{\i}}\qquad
{Yahya Can Tu\u{g}rul}\qquad
{Steve Rhyner}\qquad\\
{Meryem Banu Cavlak}\qquad
{Jo{\"e}l Lindegger}\qquad
{Mohammad Sadrosadati}\qquad
{Onur Mutlu}
}
\affiliation{%
  \vspace{0.7em}
  \institution{ETH Z{\"u}rich}
}

\renewcommand{\shortauthors}{Luo, et al.}

\renewcommand{\authors}{Haocong Luo, Ataberk Olgun, A. Giray Ya\u{g}l{\i}k\c{c}{\i}, Yahya Can Tu\u{g}rul, Steve Rhyner, Meryem Banu Cavlak, Jo{\"e}l Lindegger, Mohammad Sadrosadati, Onur Mutlu}

\setcopyright{none}
\begin{document}

\title{RowPress: Amplifying Read Disturbance in Modern DRAM Chips}

\begin{abstract}
    Memory isolation is critical for system reliability, security, and safety. Unfortunately, read disturbance can break memory isolation in modern DRAM chips. {For example,} RowHammer is a well-studied read-disturb phenomenon {where repeatedly opening and closing (i.e., hammering) a DRAM row \emph{many times} causes bitflips in physically nearby rows.}
    
   This paper experimentally demonstrates {and analyzes} {another widespread} {read-disturb phenomenon}, RowPress, in real DDR4 DRAM chips. {RowPress breaks memory isolation by} keeping a DRAM row open for a long period of time{, which} disturbs physically nearby rows enough to cause bitflips. 
    {We show that RowPress amplifies DRAM's vulnerability to read-disturb attacks} by significantly reducing {the number of {row activations} needed to induce a bitflip} by {one to two orders of magnitude} under realistic conditions. In extreme cases, {RowPress induces bitflips in a DRAM row when \om{an} {adjacent} row is activated \emph{\om{only \rb{once}}}.} Our detailed {characterization {of}} \nCHIPS{} real DDR4 DRAM chips {shows} that {RowPress} 1)~{affects} chips from all three major DRAM manufacturers, 2)~{gets worse as DRAM technology scales down \om{to smaller node sizes}, and} 3)~{affects a different set of DRAM cells {from RowHammer and behaves differently \om{from RowHammer} as temperature and access pattern changes. We also show that cells vulnerable to RowPress are very different from cells vulnerable to retention failures.}} 
    
    {We demonstrate in a real DDR4-based system {with RowHammer protection} that 1) a user-level program induces bitflips by leveraging RowPress while conventional RowHammer cannot do so,} and 2) a memory controller that adaptively keeps the DRAM row open for a longer period of time based on access pattern can facilitate RowPress-based attacks.
    To prevent bitflips due to RowPress, we {describe} and analyze four potential mitigation techniques, including a {new} methodology that adapts existing RowHammer mitigation {techniques} to also mitigate RowPress with low \emph{additional} performance overhead. {We evaluate this methodology and demonstrate {that} it is {effective} on a variety of workloads. We open source all our code and data to facilitate future research on RowPress.}

\end{abstract}

\maketitle

\fancypagestyle{firstpage}
{
    \fancyhead{}
    \begin{tikzpicture}[remember picture,overlay]
    \node [xshift=153mm,yshift=-10mm]
    at (current page.north west) {\href{https://www.acm.org/publications/policies/artifact-review-and-badging-current}{\includegraphics[width=1.8cm]{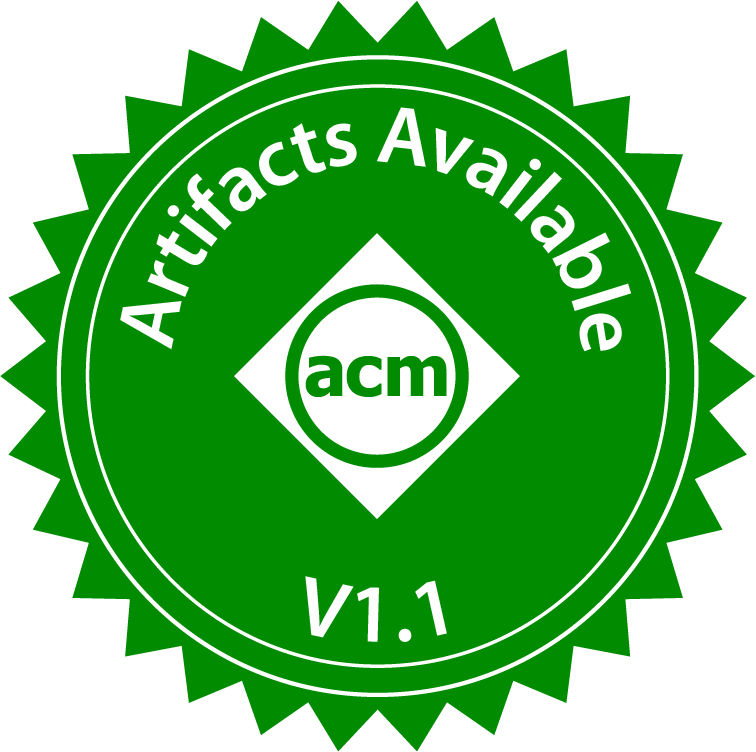}}} ;
    \node [xshift=172mm,yshift=-10mm]
    at (current page.north west) {\href{https://www.acm.org/publications/policies/artifact-review-and-badging-current}{\includegraphics[width=1.8cm]{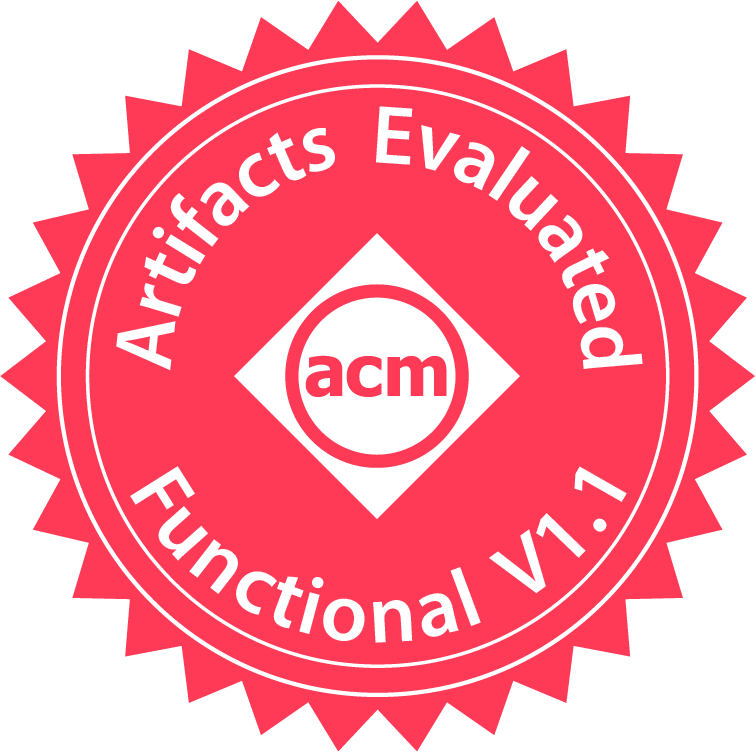}}} ;
    \node [xshift=191mm,yshift=-10mm]
    at (current page.north west) {\href{https://www.acm.org/publications/policies/artifact-review-and-badging-current}{\includegraphics[width=1.8cm]{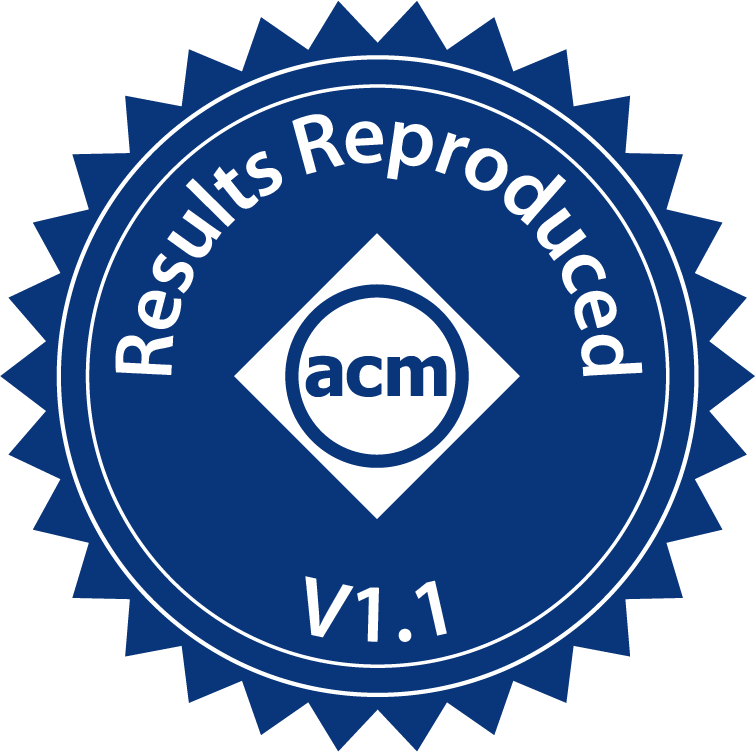}}} ;
    \end{tikzpicture}

  \renewcommand{\headrulewidth}{0pt}
  
}
\thispagestyle{firstpage}


\pagestyle{fancy}
\fancyhead{} 

\section{Introduction}
\label{sec:introduction}
{To ensure system reliability, security, and safety, it is critical to} maintain memory isolation: accessing a memory address should not cause unintended side-effects on data stored in other addresses. {Unfortunately}, with aggressive technology node scaling, {dynamic random access memory (DRAM)~{\cite{dennard1968dram}}, the prevalent {main} memory technology}, suffers from increased {{\emph{read disturbance}}: accessing (reading) a DRAM cell disturbs the operational characteristics (e.g., stored charge) of other physically close DRAM cells.} 

\emph{RowHammer} {is an example} {read-disturb phenomenon {where}} repeatedly {opening and closing (i.e., hammering) a DRAM row ({called} aggressor row) \emph{many times} ({e.g.,} tens of thousands times)} can cause bitflips in physically nearby rows ({called} victim rows)~{\cite{kim2014flipping, kim2020revisiting}}.

RowHammer {is} a critical security vulnerability as attackers can {induce and exploit the bitflips to take over a system or leak private or security-critical data}~\exploitingRowHammerAllCitations{}. {Prior works~{\cite{kim2014flipping, kim2020revisiting}} experimentally demonstrate that RowHammer significantly worsens {as} DRAM {manufacturing technology scales to smaller nodes. For example,} \gls{acmin} {has reduced} {by $14\times$} in less than a decade~\cite{kim2020revisiting}. {T}o ensure reliable, secure, and safe operation in {modern and future} DRAM-based systems, it is critical to develop a {rigorous} understanding of {read disturbance} {effects like RowHammer}.}

In this paper, we experimentally demonstrate {another widespread} {read-disturb phenomenon},
\emph{RowPress}, in real DDR4 DRAM chips. We show that keeping a DRAM row {(i.e., aggressor row)} open for a {long period of time} (i.e., a large aggressor row on time, \DRAMTIMING{AggON}) disturbs physically nearby DRAM rows.\footnote{The industry is aware that keeping a DRAM row open for a long period of time can cause read disturbance: Micron mentions ``RAS Clobber'' in two earlier patents~\cite{ito2017Apparatus, Wolff2018wordline}, while Samsung calls this ``Passing Gate Effect'' in a very recent work {placed} on arXiv while our paper has been under review~\cite{hong2023dsac}. We name this phenomenon “RowPress”, which we believe is an intuitive name that immediately shows the difference compared to RowHammer in a figurative way: we ``press'' (i.e., {keep} open for a long period of time) instead of ``hammer'' (i.e., repeatedly open and close) the row.} {Doing so} induces bitflips {in the victim row} \emph{without} {requiring {(tens of)} thousands of activations to the} aggressor row. {We characterize RowPress {in} \nCHIPS{} off-the-shelf DDR4 DRAM chips from all three major manufacturers, and find that RowPress significantly amplifies DRAM’s vulnerability to read-disturb attacks {(i.e., {greatly} reduces \glsdesc{acmin}, \gls{acmin})}.}

\begin{figure}[h]
    \centering
    \includegraphics[width=\linewidth]{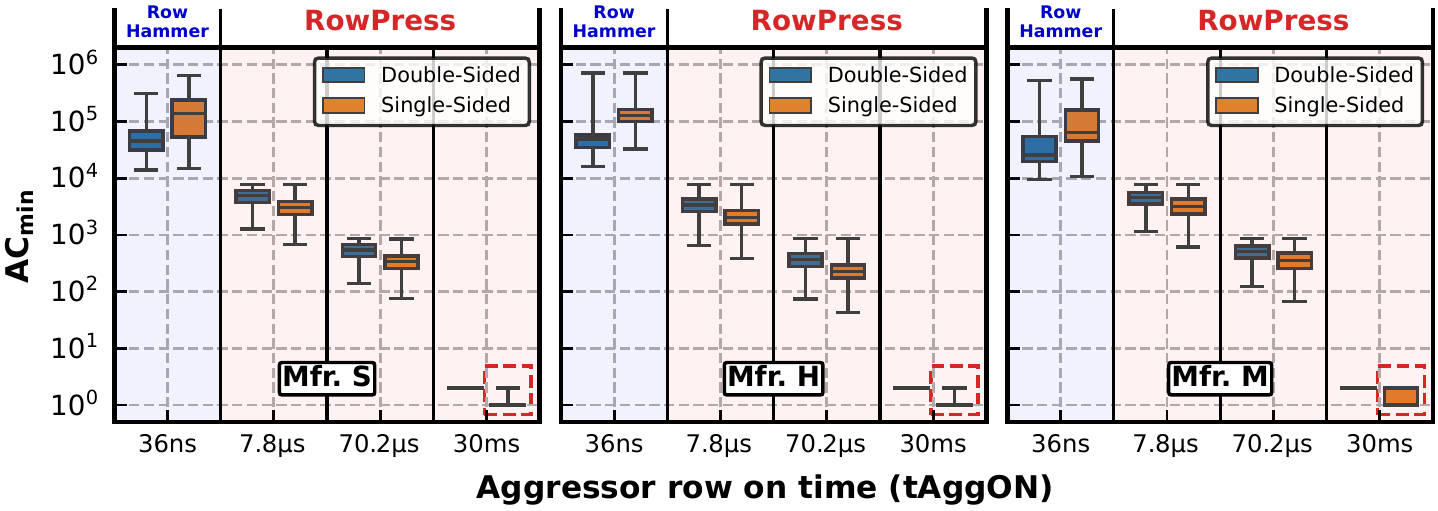}
    \caption{{$AC_{min}$ {distributions} of conventional RowHammer (RH) and three representative cases of RowPress (RP) at $80^{\circ}C$ \om{across 164 DDR4 chips {from manufacturers S, H, and M}}.}}
    \label{fig:hcf_intro}
\end{figure}

{To illustrate this,} {\figref{fig:hcf_intro} shows {the distribution of \gls{acmin} (y-axis) {we measure {in \nCHIPS{}} DRAM chips across all three major DRAM manufacturers} when the aggressor row stays open as much as \DRAMTIMING{AggON} {(x-axis)}} {between consecutive activations} at \SI{80}{\celsius} {with one (single-sided) and two (double-sided) aggressor row(s)}} {in a {box-and-whiskers} plot.\footnote{\label{fn:boxplot}{{The box is lower-bounded by the first quartile (i.e., the median of the first half of the ordered set of data points) and upper-bounded by the third quartile (i.e., the median of the second half of the ordered set of data points).
The \gls{iqr} is the distance between the first and third quartiles (i.e., box size).
Whiskers show the minimum and maximum values.}}} We study the single- and double-sided RowPress access patterns in detail in \om{\secref{sec:sen_acc_pattern}}.}

{The two leftmost boxes in each {plot} shows the distribution of \gls{acmin} {for the conventional single-sided ({orange}) and double-sided (blue) RowHammer pattern, where} the aggressor row is open for the minimum amount of time {($\DRAMTIMING{AggON} = \DRAMTIMING{RAS} = 36ns$)}\footnote{\atb{Manufacturer-recommended minimum row open time ($t_{RAS}$) ranges from \SI{32}{\nano\second} to \SI{35}{\nano\second} in DDR4~\cite{jedec2017ddr4}. We use a \SI{36}{\nano\second} minimum \DRAMTIMING{AggON} 1) to cover the whole range of $t_{RAS}$ values and 2) due to the {limited DRAM command bus frequency of our testing infrastructure (i.e., we can only send a DRAM command at every \SI{1.5}{\nano\second})}~\cite{olgun2022drambender}.}} allowed by the DRAM specification~\cite{jedec2017ddr4}\om{, as done in conventional RowHammer attacks~\exploitingRowHammerAllCitations{}}. {We observe that as \DRAMTIMING{AggON} increases, compared to the most effective RowHammer pattern, the most effective RowPress pattern reduces \gls{acmin}}}
{1)~by \rb{$17.6\times$} on average (up to \rb{$40.7\times$}) when \DRAMTIMING{AggON} {is as large as} the refresh interval (\SI{7.8}{\micro\second})\footnote{{{\emph{Refresh interval}} is the time interval between two consecutive refresh commands that a DRAM row can be kept open~\cite{jedec2017ddr4, jedec2012ddr3}}.},
{2)~by \rb{$159.4\times$} on average (up to \rb{$363.8\times$}) when \DRAMTIMING{AggON} is \SI{70.2}{\micro\second}, the maximum \agyf{allowed} \DRAMTIMING{AggON}~\cite{jedec2017ddr4},} and {3)~down to {\emph{only one}} activation for {an extreme} \DRAMTIMING{AggON} of \SI{30}{\milli\second} {(highlighted by dashed red {boxes})}.} }

{Our {detailed} {characterization results and sensitivity stud\om{ies}} suggest that RowPress has a different underlying error mechanism compared to {{the RowHammer phenomenon} in DRAM}~{\cite{kim2014flipping, kim2020revisiting, mutlu2017rowhammer, yang2019trap, mutlu2019rowhammer, walker2021ondramrowhammer, park2016experiments, park2016statistical, yaglikci2022understanding, mutlu2022fundamentally}}.} We experimentally demonstrate that 1)~{only less than \rb{ \param{0.013}\SI{}{\percent}} of the DRAM cells that {exhibit} RowPress bitflips also {exhibit} RowHammer bitflips \om{(\secref{sec:relationship})}}, and 2) RowPress behaves \om{very} differently \om{from} RowHammer \om{with} temperature {(\secref{sec:sen_temperature})} and access pattern {(\secref{sec:sen_acc_pattern})} changes. {We also show detailed results {demonstrating} that cells vulnerable to RowPress are very different from cells vulnerable to retention failures (only less than $0.34\%$ overlap).}

{{We demonstrate that a user-level program can induce RowPress \gf{bitflips} in a real DDR4-based system {that already employs RowHammer protection}.} {The program accesses \emph{multiple different} columns of the aggressor DRAM row so that the memory controller keeps the aggressor row open for a longer period of time to serve these accesses.} {As a result, the program exercises RowPress} and {induces bitflips, while conventional RowHammer {cannot},} in the presence of in-DRAM RowHammer mitigation mechanisms~(\secref{sec:real}).} We believe this program can be the basis of a proof-of-concept RowPress attack.

Our characterization results suggest that DRAM-based systems need to take RowPress into account to {maintain the fundamental security{/safety/reliability} property of memory isolation}. Based on our findings, we discuss and evaluate the implications of RowPress {on} existing {{read-disturb} mitigation} mechanisms that consider \om{\emph{only}}  RowHammer. {We propose a methodology to adapt \agyf{RowHammer} mitigation techniques} to {also} mitigate RowPress with low \emph{additional} performance overhead by both 1) limiting the \emph{maximum row-open time}, and 2) configuring the RowHammer defense to account for the {RowPress-induced} reduction in \gls{acmin}. {We {experimentally} demonstrate that by applying our proposed methodology to {two major techniques (}PARA~\cite{kim2014flipping} and Graphene~\cite{park2020graphene}), we can mitigate both RowHammer and RowPress with {an average (maximum)} \emph{additional} slowdown of only {$3.6\%$ ($13.1\%$)} and {$-0.63\%$ ($4.6\%$)}, respectively.}

We make the following contributions in this paper:
\begin{itemize}
    \item {To our knowledge, this is the first work to} \om{experimentally} {demonstrate} {the RowPress phenomenon} and {its widespread existence in {real} DDR4 DRAM chips from all three major manufacturers.}
    
    \item {We provide an extensive characterization of RowPress on \nCHIPS{} {real} DRAM chips. {Our results show that RowPress 1)} significantly amplifies DRAM’s vulnerability to read-disturb attacks, {2)} gets worse as DRAM technology scales down,} {and 3) is \om{very} different from RowHammer {and retention failures} in {terms of} the DRAM cells it affects and in the way it behaves as temperature and access pattern changes.}

    \item {We demonstrate that a simple user-level program induces RowPress \gf{bitflips} on a real DDR4-based system, {while} a state-of-the-art RowHammer program cannot.}
    
    \item We {describe, analyze, and evaluate} four potential ways to mitigate {read-disturb attacks exploiting RowPress.} {We introduce} a methodology to adapt existing RowHammer mitigation {techniques} to also mitigate RowPress with low \emph{additional} performance overhead.

    \item We open-source~{\cite{rowpress-artifact-github}} all our infrastructure, test programs, and raw data {to enable 1) reproduction and replication of our results, and 2) further research on RowPress.}

\end{itemize}

\section{Background \& Motivation}
\label{sec:background}
We provide a high-level introduction \om{to DRAM organization}~({\secref{sec:dram_org}}), \om{major} DRAM operations~({\secref{sec:dram_op}}), DRAM timing parameters involved in this work~({\secref{sec:dram_timing}}), and \om{r}ead-\om{d}isturb  \om{m}echanisms in DRAM~(\secref{sec:read_disturb_background}).

\subsection{DRAM Organization}
\label{sec:dram_org}

\begin{figure}[h]
    \centering
    \includegraphics[width=0.95\linewidth]{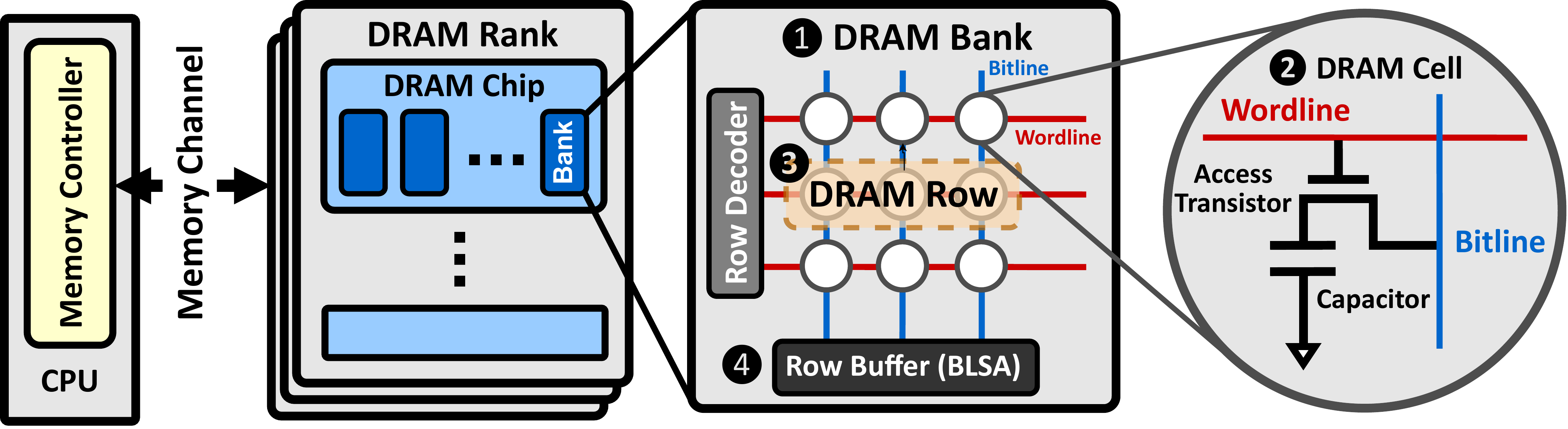}
    \caption{Hierarchical organization of modern DRAM.}
    \label{fig:dram_org}
\end{figure}

\figref{fig:dram_org} shows the hierarchical organization of modern DRAM-based main memory. The CPU's \emph{memory controller} communicates with a \emph{DRAM module} over a \emph{memory channel}. A module contains one or multiple \emph{DRAM ranks} that share the memory channel. A rank is made up of multiple \emph{DRAM chips} that are operated in a lock-step manner (i.e.,~all chips receive and process the same command at the same time). Each DRAM chip contains multiple \emph{DRAM banks}~\circled{1} that can be accessed independently.

Inside a DRAM bank, \emph{DRAM cells} are organized into a two-dimensional array, addressed by rows and columns. A DRAM cell~\circled{2}consists of 1) a capacitor, which stores one bit of information in the form of {electrical} charge level, and 2) an access transistor\om{,} which connects the capacitor to a bitline, controlled by a wordline.
{When the row decoder {(including wordline drivers)} drives a wordline high, the access transistors of all DRAM cells in the row~\circled{3} are enabled, electrically connecting each cell in the row to {its corresponding} bitline.}
DRAM cells {in the same column share} a bitline, which is used to read from and write to the cell{s} {via} the row buffer~\circled{4} ({which contains} bitline sense amplifiers, BLSA).

\subsection{\om{Major} DRAM Operations}
\label{sec:dram_op}

\noindent\textbf{DRAM Access.}
Accessing DRAM consists of three steps. First, the memory controller issues an \DRAMCMD{ACT} (activate) command together with a row address to the bank. The row decoder drives the wordline of that row to open the row {(i.e., enables the access transistors)}. {Data} is then transferred from the DRAM cells {in the row} to the row buffer through the bitlines. Second, once the data is in the row buffer, {the memory controller can send \DRAMCMD{RD}/\DRAMCMD{WR} commands to read/write data {from/to} the opened row.} Third, the memory controller sends a \DRAMCMD{PRE} (precharge) command to close the opened row before accessing another row in the same bank.

\noindent\textbf{DRAM Refresh.}
{DRAM cells lose charge over time, risking \emph{retention failure} \om{induced} \gf{bitflips} if the\om{ir} charge is not restored in time. \om{To avoid this,} the memory controller periodically restores each DRAM {row's charge levels} by sending \DRAMCMD{REF} (refresh) commands.}
Before issuing a \DRAMCMD{REF} command, the memory controller must send a \DRAMCMD{PRE} command to close any {open} row to prepare the bank for refresh.

\subsection{Key DRAM Timing Parameters}
\label{sec:dram_timing}
\begin{figure}[h]
    \centering
    \includegraphics[width=1.0\linewidth]{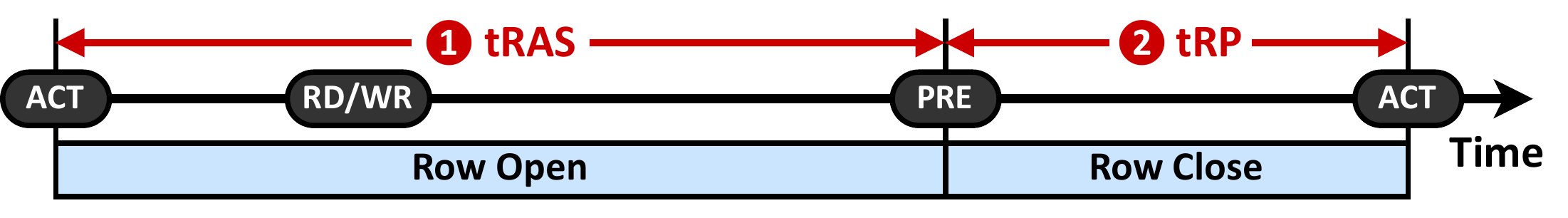}
    \caption{Timeline of \gf{k}ey DRAM \gf{a}ccess \gf{o}perations.}
    \label{fig:dram_op}
\end{figure}

To guarantee correct operation, the memory controller must {time DRAM commands according to certain \emph{timing parameters}}~\cite{jedec2012ddr3,jedec2015lpddr4,jedec2017ddr4,jedec2020ddr5}.
{\figref{fig:dram_op} shows a timeline of the key DRAM access operations.} We describe {four} key timing parameters involved in this work: 
1)~\acrshort{tras}, 2)~\acrshort{trp}, 3)~\acrshort{trefi}, and 4)~\acrshort{trefw}.

\glsdisp{tras}{\acrshort{tras} is \glsdesc{tras} (\redcircled{1}in \figref{fig:dram_op}).}
\glsdisp{trp}{\acrshort{trp} is \glsdesc{trp} (\redcircled{2}in \figref{fig:dram_op}).}
\glsdisp{trefi}{\acrshort{trefi} is \glsdesc{trefi}}.
\glsdisp{trefw}{\acrshort{trefw} is \glsdesc{trefw}}.

A majority of DRAM timing parameters define lower bounds for the time intervals between {pairs of} DRAM commands. {For example,} \DRAMTIMING{RAS} is the \emph{minimum} amount of time that the memory controller has to wait before issuing a \DRAMCMD{PRE} command to close {an open(ed)} DRAM row. 
{The memory controller may keep the DRAM row open \emph{longer}} than \DRAMTIMING{RAS} to serve more \DRAMCMD{RD}/\DRAMCMD{WR} commands \om{(in anticipation of future requests to the same row{~\cite{mutlu2008parbs, rixner00, zuravleff1997controller, mutlu2007stall}})}{, depending on the memory controller's implementation and the workload's access pattern}.
{In general}, {if the memory controller does \emph{not} postpone \DRAMCMD{REF} commands,} a DRAM row can be open for a  {duration} of \DRAMTIMING{REFI} before it has to be closed to serve a \DRAMCMD{REF} command. {Otherwise, a DRAM row can be open for up to $9\times$ \DRAMTIMING{REFI} because} the JEDEC DDR4 standard~\cite{jedec2017ddr4} {allows postponing} up to {eight} \DRAMCMD{REF} command\gf{s}. Under normal operating conditions {(i.e., within the temperature range of {$0^{\circ}C$ to $85^{\circ}C$)}}, \DRAMTIMING{REFI} is \gf{\SI{7.8}{\micro\second}} for commodity DDR4 chips.

\subsection{\hluo{Motivation}}
\label{sec:read_disturb_background}
There are three \om{major causes of} bitflips in DRAM \om{cell arrays}: 1) \om{soft errors caused} by charged and/or energetic particle \hluo{strike}\joel{s} \hluo{\cite{baumann2005radiation, may1979alpha, Lantz1996Soft, Gorman1994cosmic}}, 2) data retention failures \hluo{due to the volatile and leaky nature of DRAM cells{~\cite{liu2012raidr, liu2013experimental, patel2017reaper, khan2014efficacy, khan2016parbor}}}, and 3) read disturbance (e.g., RowHammer~\understandingRowHammerAllCitations{}) caused by {undesirable} {interactions} between circuit components{. Both retention failures and RowHammer get worse as DRAM technology scales down {to smaller node sizes}.}

Read disturbance \hluo{has significant implications} for system reliability, security, and safety because it is a widespread issue and can be exploited to break memory isolation ~\exploitingRowHammerAllCitations{}. Therefore, it is important to \hluo{identify and understand read disturbance mechanisms in DRAM.} {\textbf{Our goal} is to 1) rigorously and comprehensively characterize and investigate the read disturbance caused by increased {aggressor row on time (\DRAMTIMING{AggON})}, and 2) understand its implications for secure, reliable, and safe operation of DRAM-based systems.}
\section{Methodology}
\label{sec:methodology}

{We describe} \hluo{our} DRAM testing infrastructure \yct{and} the \hluo{real} DDR4 DRAM chips tested. \agy{We explain the methodology of {our characterization experiments} in their respective sections {(under~\secref{sec:characterization})}.}

\subsection{DRAM Testing Infrastructure}
\label{subsec:methodology_infra}

\begin{figure}[h]
\centering
\includegraphics[width=1.0\linewidth]{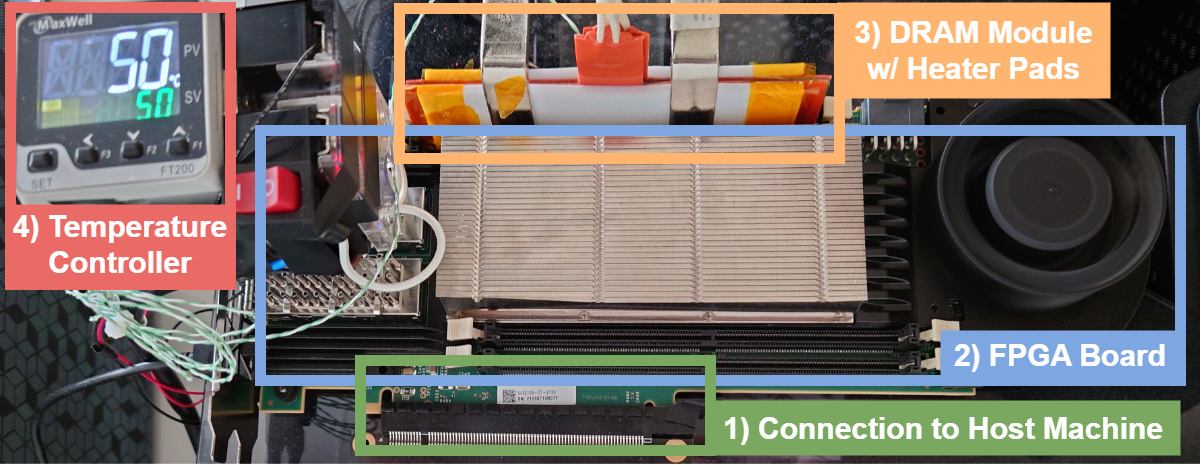}
\caption{Our {DDR4 DRAM} testing infrastructure.}
\label{fig:hostmachine}
\end{figure}

\agy{We test {commodity} DDR4 DRAM chips using an FPGA-based DRAM testing infrastructure that consists of four main components \gf{(as Fig.~\ref{fig:hostmachine} illustrates)}: 1)~a host machine that generates the test \hluo{program} and \hluo{collects experiment results}, 2)~an FPGA development board (Xilinx Alveo U200~\cite{alveou200}), programmed with DRAM Bender~\cite{olgun2022drambender, drambendergithub} {(based on SoftMC~\cite{hassan2017softmc, softmcgithub})}, \hluo{to execute our test programs}, 3)~a thermocouple temperature sensor and a pair of heater pads pressed against the DRAM chips {to maintain a {target temperature level}}, and 4)~a PID temperature controller (MaxWell FT200~\cite{maxwellFT200}) that controls the heaters and keeps the temperature at the desired level.}

\noindent\textbf{Disabling Interference Sources.}
\agy{To observe RowPress' effects {at the} circuit level,} we disable potential sources of interference following a methodology {similar to} prior works~\cite{kim2020revisiting, orosa2021deeper, yaglikci2022understanding, hassan2021utrr}.
First, \agy{we disable periodic refresh} during the execution of our test programs to \agy{1)~}keep the timings of {our test programs} precise and 2)~disable any existing on-die RowHammer defense mechanisms (e.g., TRR)~\agy{\cite{frigo2020trrespass, hassan2021utrr}} {so as to} observe the \agy{DRAM chip's} {fundamental read disturbance} behavior at the circuit level.
Second, we \agy{bound our test programs' execution time \hluo{strictly} within a refresh window}
(i.e., 64ms \DRAMTIMING{REFW}) of the tested DRAM chips \agy{to {prevent} data retention failures {from interfering} with read-disturb failures.} 
\joel{Third, we ensure that the tested DRAM modules and chips have neither rank-level nor on-die ECC.} \hluo{Doing so ensures that} \atb{we directly observe \hluo{and analyze} all {circuit-level} bitflips \hluo{without interference} {from architecture-level correction and mitigation mechanisms}.}

\subsection{Commodity DDR4 DRAM Chips Tested}
\label{subsec:methodology_dramchips}
\agy{Table~\ref{tab:dram_chip_list} shows} \hluo{the} \nCHIPS{} \agy{(\nMODULES{})} \hluo{real} DDR4 DRAM chips \agy{(modules) that we test} from \agy{all three major DRAM manufacturers.} 
To \hluo{{demonstrate} that RowPress is intrinsic to the DRAM technology \atb{and is a widespread phenomenon {across manufacturers},}} 
we test a variety of DRAM chips spanning different die densities and die revisions from each DRAM chip manufacturer\gf{.}\footnote{The technology node that a DRAM chip is \gf{manufactured} with is usually not publicly available. We assume that two DRAM chips from the same manufacturer have the same technology node only if they \hluo{share} both the same die density and die revision code. A die revision code of X indicates that there is no public information available about the die revision (e.g., the original DRAM chip manufacturer's markings have been removed by the DRAM module vendor and the DRAM stepping field in the SPD is \texttt{0x00}). {More details on the tested chips and a summary of their RowPress and RowHammer characteristics are in Appendix B.}}%

\begin{scriptsize}
\begin{table}[h!]
  \centering
  \footnotesize
  \captionsetup{justification=centering, singlelinecheck=false, labelsep=colon}
  \caption{Tested DDR4 DRAM Chips.}
    \begin{tabular}{lcccccc}
        \toprule
            {{\bf Mfr.}} & \textbf{\#DIMMs} & {{\bf  \#Chips}}  & {{\bf Density}} & {{\bf Die Rev.}}& {{\bf Org.}}& {{\bf Date}}\\
        \midrule
\multirow{4}{4em}{Mfr. S (Samsung)} & 2 & 8  & 8Gb  & B   & x8  & 20-53  \\                           
  & 1 & 8  & 8Gb  & C   & x8  & N/A   \\                         
   & 3 & 8  & 8Gb  & D   & x8  & 21-10  \\                           
       & 2 & 8  & 4Gb  & F   & x8  & N/A   \\                         
        \midrule
\multirow{4}{4em}{Mfr. H (\mbox{SK Hynix})}     & 1 & 8  & 4Gb  & A   & x8  & 19-46  \\     
     & 1 & 8  & 4Gb  & X & x8  & N/A   \\               
     & 2 & 8  & 16Gb & A   & x8  & 20-51  \\                           
     & 2 & 8  & 16Gb & C   & x8  & 21-36  \\                           
     \midrule                  
\multirow{5}{4em}{Mfr. M (Micron)}     & 1 & 16 & 8Gb  & B   & x4  & N/A   \\       
    & 2 & 4  & 16Gb & B   & x16 & 21-26  \\                            
    & 1 & 16 & 16Gb & E   & x4  & 20-14  \\     
    & 2 & 4  & 16Gb & E   & x16 & 20-46  \\ 
    & 1 & 4  & 16Gb & F   & x16 & 21-50  \\                            
        \bottomrule
    \end{tabular}
    \label{tab:dram_chip_list}
\end{table}
\end{scriptsize}

\agy{To account for in-DRAM row address mapping~\cite{kim2014flipping, smith1981laser, horiguchi1997redundancy, keeth2001dram, itoh2013vlsi, liu2013experimental,seshadri2015gather, khan2016parbor, khan2017detecting, lee2017design, tatar2018defeating, barenghi2018software, cojocar2020rowhammer,  patel2020beer}, we reverse-engineer the physical row address layout, following the {methodology of} prior works~{\cite{kim2020revisiting, orosa2021deeper, yaglikci2022understanding, hassan2021utrr}}.}

\section{\om{Major} RowPress Characterization}
\label{sec:characterization}
We characterize RowPress by \om{analyzing} 1) how DRAM’s vulnerability to read disturbance changes as \DRAMTIMING{AggON} increases, and 2) \om{properties} of RowPress bitflips \om{that distinguish them from RowHammer and retention {failure} bitflips}. \om{We evaluate the {sensitivity of RowPress biflips} to {temperature, access pattern, and aggressor row \emph{off} time (i.e., \DRAMTIMING{AggOFF})} in \secref{sec:sensitivity}.} {Appendix~\secref{sec:extended_characterization_results} provides further results and plots.}

\subsection{Experiment Methodology}
\label{sec:characterization_methodology}

\noindent\textbf{Metric.}
To characterize how RowPress amplifies DRAM's vulnerability to read disturbance, we examine how \glsfirst{acmin} changes as \DRAMTIMING{AggON} increases. A lower \gls{acmin} means more \agyf{vulnerability} to \hluo{read disturbance}.

\begin{figure}[h]
    \centering
    \vspace{0.2em}
    \includegraphics[width=0.95\linewidth]{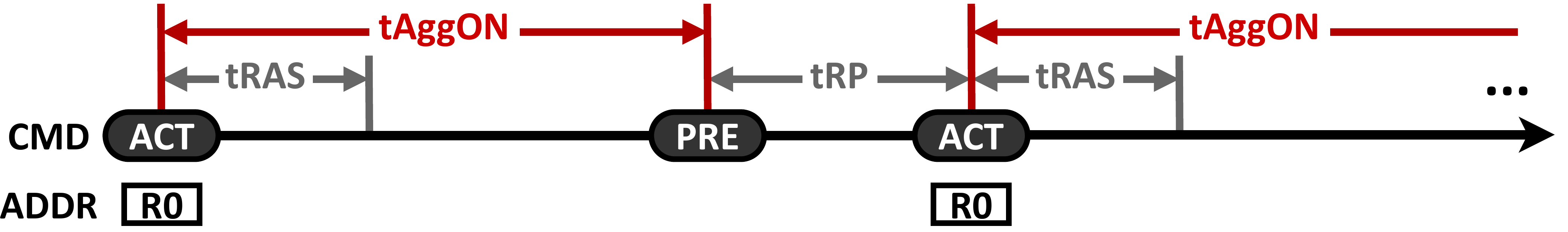}
    \caption{{Single-sided} RowPress access pattern used to characterize how \gls{acmin} changes as \DRAMTIMING{AggON} increases.}
    \label{fig:characterization_single_pattern}
\end{figure}

\noindent\textbf{Access Pattern.}
\figref{fig:characterization_single_pattern} illustrates \agy{our \hluo{RowPress} access pattern targeting a {single} aggressor row \hluo{(single-sided)} to induce bitflips. We 1)~activate (ACT) the aggressor row (R0), 2)~keep the aggressor row on for a \hluo{certain amount of} time (\DRAMTIMING{AggON}), and 3)~close the row \hluo{with} a precharge (PRE) command. To respect the timing constraints, we wait until precharge latency \DRAMTIMING{RP} is satisfied before repeating the same access pattern.} We sweep \DRAMTIMING{AggON} from the minimum possible value of \gf{\SI{36}{\nano\second}} ({i.e., the} \agy{nominal} \DRAMTIMING{RAS} value) up to \gf{\SI{30}{\milli\second}}. {Note that for \DRAMTIMING{AggON} $=$ \SI{36}{\nano\second}, our single-sided RowPress pattern is identical to a single-sided RowHammer access pattern.\footnote{\agy{\joel{The} RowHammer access pattern activates {an} aggressor row as frequently as possible, and thus closes the row {(i.e., precharges the bank)} as soon as it can, which is \SI{36}{\nano\second} ($=$ \DRAMTIMING{RAS}) after the row is opened.}}} We test {3072} rows (the first, the middle, and the last {1024} rows) in bank 1 \agy{for} each DRAM module.

\noindent\textbf{Algorithm.} For every \DRAMTIMING{AggON} value we {evaluate}, we \om{find} the \gls{acmin} \agy{for each tested row} using a {modified version of the bisection-method} algorithm used by prior works~\cite{orosa2021deeper, yaglikci2022understanding}. {Instead of a fixed \gls{acmin} accuracy (e.g., 100 in~\cite{yaglikci2022understanding} and 512 in ~\cite{orosa2021deeper}), we enable an accuracy of 1\%, rounded up to the next integer (i.e., we terminate the search for \gls{acmin} when the difference between the current and previous {measurements} of \gls{acmin} is no larger than 1\% of the previous {measurements})}. {We report that we could not induce any bitflip if the test program's execution time exceeds 60ms ({which is} strictly smaller than the refresh window of \gf{\SI{64}{\milli\second}} in DDR4~\cite{jedec2017ddr4}).} For every tested row, we repeat the \gls{acmin} search five times and report the minimum \gls{acmin} value {we observe}. 

\noindent\textbf{Data Pattern.} We use a \atb{checkerboard} data pattern~\cite{vandegoor2002address} \joel{where} we fill the aggressor row with \texttt{0xAA} \atb{and victim rows with \texttt{0x55}.} \atb{We \joel{consider} three adjacent rows \joel{on each side} of the aggressor row as \joel{victim} rows. We use this data pattern for all {our} characterization and sensitivity studies. {{We study the data pattern sensitivity of RowPress bitflips in detail in \secref{sec:sen_data_pattern}.}}}

\noindent\textbf{Temperature.} We maintain the DRAM chip temperature at a normal operating condition of $50^{\circ}C$. {We study the temperature sensitivity of RowPress bitflips in \secref{sec:sen_temperature}.}

\subsection{Vulnerability to Read Disturbance}
\label{sec:vulnerability_to_readdisturbance}

\begin{figure}[h]
    \centering
    \includegraphics[width=1.0\linewidth]{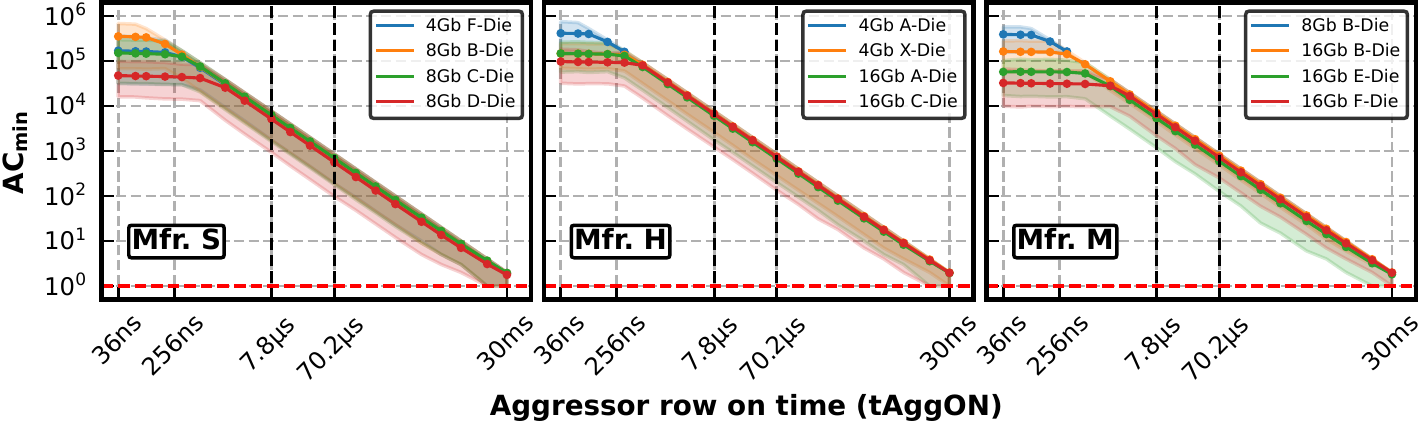}
    \caption{\gls{acmin} as \DRAMTIMING{AggON} \atb{increases}\om{;} single-sided RowPress at $50^{\circ}C$.}
    \label{fig:acmin_characterization}
\end{figure}

\figref{fig:acmin_characterization} shows the \gls{acmin} {distribution} \agy{(y-axis)} {of different die revisions for all three major DRAM manufacturers} as we sweep \DRAMTIMING{AggON} \agy{(x-axis)} from \SI{36}{\nano\second} to \SI{30}{\milli\second} in log-log scale. {For each manufacturer ({i.e., each plot}), we group the data based on the die revision \atb{(different colors)} and aggregate the \gls{acmin} values from all the rows we test \atb{in all chips with the same die revision}}. Each data point shows the mean \gls{acmin} value and the error band shows the {minimum and maximum of \gls{acmin} values} \hluo{across all tested rows}. We highlight the \DRAMTIMING{AggON} values of \SI{7.8}{\micro\second} (\DRAMTIMING{REFI}) and \SI{70.2}{\micro\second} ($9\times$\DRAMTIMING{REFI}) on the x-axis\joel{,} as they are the two potential upper bounds of \DRAMTIMING{AggON}, as \om{dictated by} the JEDEC DDR4 standard~\cite{jedec2017ddr4}.\footnote{{Whether \SI{7.8}{\micro\second} or \SI{70.2}{\micro\second} is the upper bound {for \DRAMTIMING{AggON}} depends on the memory controller's implementation. If the memory controller does \emph{not} allow any refresh commands to be postponed, the upper bound is \SI{7.8}{\micro\second}. Otherwise, because the JEDEC DDR4 standard{~\cite{jedec2017ddr4}} allows \emph{up to} eight refresh commands to be postponed {(Section 4.26 in~\cite{jedec2017ddr4})}, the upper bound can be as high as \SI{70.2}{\micro\second}.}} {We mark \gls{acmin} $=$ 1 on the y-axis.} We make three {major} observations from \joel{\figref{fig:acmin_characterization}}.

\observation{RowPress significantly reduces \gls{acmin} as \DRAMTIMING{AggON} increases.\label{obsv:RP_acminreduction}}

 For example, for almost all {(10 \agy{of} 12) die revisions from all three DRAM manufacturers,\footnote{The only exceptions are \agy{Mfr. \rb{H}'s} 4Gb A-Dies and \agy{Mfr. \rb{M}'s} 8Gb B-Dies, \agy{none of which exhibit} any bitflips when \DRAMTIMING{AggON} is larger than \SI{336}{\nano\second} with the single-sided RowPress pattern at $50^{\circ}C$.}} we observe that \agy{\gls{acmin} reduces by $21\times$} \hluo{on average} when \DRAMTIMING{AggON} increases from \gf{\SI{36}{\nano\second}} to \gf{\SI{7.8}{\micro\second}}. For modules with 8Gb B-Dies from Mfr. \rb{S}, the reduction in mean \gls{acmin} can reach up to \rb{$59\times$}. If \DRAMTIMING{AggON} increases from \gf{\SI{36}{\nano\second}} to \gf{\SI{70.2}{\micro\second}}, the reduction in mean \gls{acmin} is \rb{$190\times$}, and the maximum reduction reaches \rb{$537\times$}, as observed in modules with 8Gb B-Dies from Mfr. \rb{S}.\ombox{Double check all these numbers}

\observation{In extreme cases, RowPress \om{causes} bitflips with only \rb{one} aggressor row activation (i.e., \gls{acmin}$=1$).\label{obsv:RP_extreme}}

We observe that for almost all die revisions from all three manufacturers, \agy{1)~}we can \emph{always} induce bitflips as we continue to increase \DRAMTIMING{AggON} {until \SI{30}{\milli\second}}, and {2) for $13.1\%$} of the {tested} rows that experience bitflips, only a single activation of an aggressor row (i.e., \gls{acmin}$=1$), is needed to induce bitflips \agy{when \DRAMTIMING{AggON} is \SI{30}{\milli\second} at $50^{\circ}C$}. We conclude that, \om{unlike RowHammer,} RowPress does not {have to} rely on repeatedly accessing the aggressor row \om{\emph{many}} times to induce bitflips.

\observation{RowPress is a common DRAM vulnerability across all three major DRAM manufacturers.\label{obsv:RP_common}}

We observe that the \gls{acmin} trends across {almost all} die revisions from all three major DRAM manufacturers follow a consistent pattern. {First, \gls{acmin} decreases slowly as \DRAMTIMING{AggON} starts to increase.} {For example, when \DRAMTIMING{AggON} increases by $5.17\times$ from \gf{\SI{36}{\nano\second}} to \gf{\SI{186}{\nano\second}},} {\gls{acmin} reduces on average by only $1.17\times, 1.04\times$, and $1.08\times$ for Mfr. S, H, and M, respectively.} 
Second, as \DRAMTIMING{AggON} continues to increase {(e.g., beyond \SI{7.8}{\micro\second})}, \gls{acmin} decreases drastically {for all three manufacturers}, following an \agy{approximately} straight line in log-log scale. We find that the \gls{acmin} trend lines {when \DRAMTIMING{AggON} $\geq$ \gf{\SI{7.8}{\micro\second}}} for all three manufacturers have very similar slopes: \rb{$-1.020$, $-1.013$, and $-1.013$} for Mfr. S, H, and M, respectively. %
\agy{Given the similarity in \gls{acmin} reduction {with} increasing \DRAMTIMING{AggON} across all tested die revisions from all three major manufacturers \om{spanning \nCHIPS{} chips},} we conclude that RowPress is \agy{an} intrinsic \agy{read-disturb phenomenon} to the DRAM technology. 

\begin{figure}[h]
    \centering
    \includegraphics[width=1.0\linewidth]{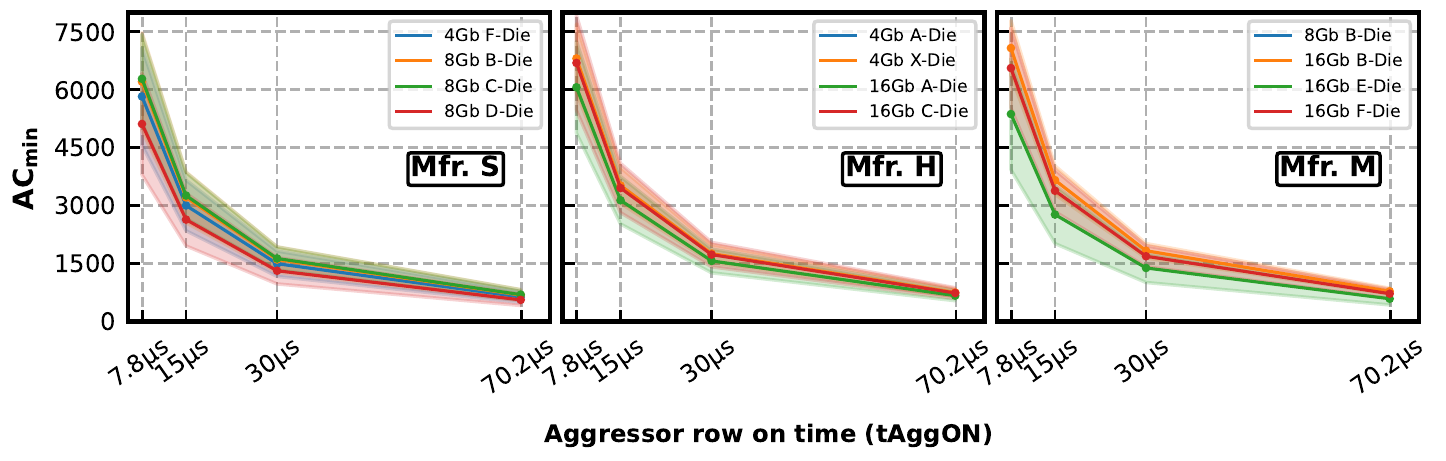}
    \caption{\gls{acmin} for \DRAMTIMING{AggON} between \SI{7.8}{\micro\second} and \SI{70.2}{\micro\second} in linear-linear scale\om{;} single-sided RowPress at $50^{\circ}C$.}
    \label{fig:acmin_50C_zoomed}
\end{figure}

{Note that a slope close to $-1$ in log-log scale does \emph{not} mean that \gls{acmin} reduces linearly as \DRAMTIMING{AggON} reduces. {\figref{fig:acmin_50C_zoomed} shows a portion of the \gls{acmin} distribution from~\figref{fig:acmin_characterization} with a smaller range of \DRAMTIMING{AggON} values (from \SI{7.8}{\micro\second} to \SI{70.2}{\micro\second}) in \emph{linear-linear} scale.}}

{We observe that as \DRAMTIMING{AggON} increases, the reduction rate of \gls{acmin} decreases. The average \gls{acmin} reduction for Mfr. S, H, and M when \DRAMTIMING{AggON} increases from \SI{7.8}{\micro\second} to \SI{15}{\micro\second} are \SI{-0.37}{\micro\second}$^{-1}$, \SI{-0.41}{\micro\second}$^{-1}$, and \SI{-0.39}{\micro\second}$^{-1}$, respectively, but only \SI{-0.021}{\micro\second}$^{-1}$, \SI{-0.023}{\micro\second}$^{-1}$, and \SI{-0.021}{\micro\second}$^{-1}$, respectively, when \DRAMTIMING{AggON} increases from \SI{30}{\micro\second} to \SI{70.2}{\micro\second}. \hluocr{We conclude that \gls{acmin} does \emph{not} reduce linearly as \DRAMTIMING{AggON} increases.}}

\begin{figure}[h]
    \centering
    \includegraphics[width=1.0\linewidth]{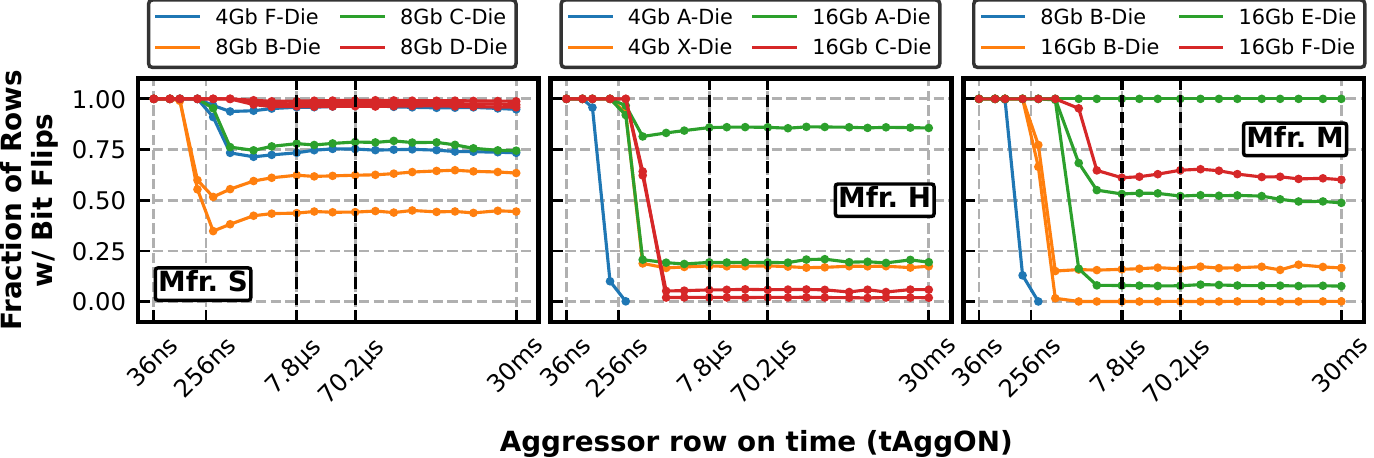}
    \caption{The fraction of rows that experience at least one bitflip\om{;} single-sided RowPress at $50^{\circ}C$.}
    \label{fig:acmin_coverage_50C}
\end{figure}

\figref{fig:acmin_coverage_50C} shows the fraction of the tested rows that \om{have} at least one RowPress bitflip \agy(y-axis) as we sweep \DRAMTIMING{AggON} \agy{(x-axis)}. \agyf{Each {plot} corresponds to a different manufacturer.}
\agy{Each curve represents a different DRAM module and \om{is} colored by its die revision.}

\observation{RowPress worsens as DRAM technology node scales down.\label{obsv:RP_gets_worse}}

In general, the more advanced the technology node\agy{\footnote{\agy{For a given manufacturer and die density, the later in the alphabetical order the die revision code is, the more likely the chip has a more advanced technology node.}}} (as indicated by the die revision), the more rows are vulnerable to RowPress. For example, for the three 8Gb Dies from Mfr. \rb{S}, as \DRAMTIMING{AggON} increases, almost 100\% of the tested rows of the D-Dies experience RowPress bitflips, {which drops to below 80\% for the C-Dies  and \om{below} 60\% for the B-Dies.}

\takeawaybox{RowPress \agy{1)}~is a common read-disturb phenomenon {in DRAM chips} that \agyf{exacerbates} \hluo{DRAM's vulnerability to read disturbance} \agy{and 2)~}get\agy{s} worse as DRAM technology scales down {to smaller node sizes}.}

\begin{figure}[h]
    \centering
    \includegraphics[width=1.0\linewidth]{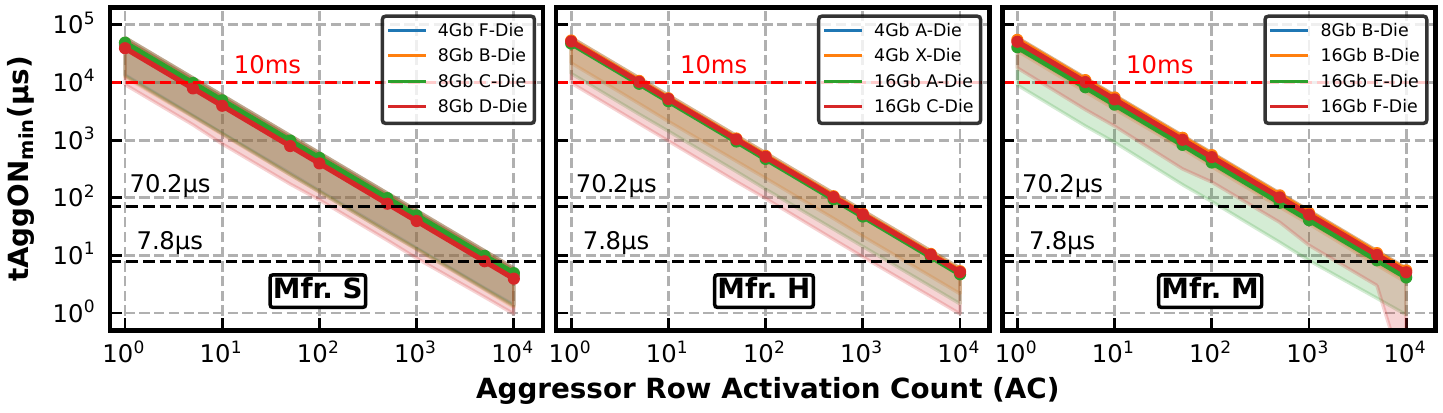}
    \caption{{\DRAMTIMING{AggON}$_{\scriptstyle{\text{min}}}$} as aggressor row activation count (AC) increases; single-sided RowPress at $50^\circ C$.}
    \label{fig:mintaggon_50C}
\end{figure}

{To further understand the relationship between \DRAMTIMING{AggON} and aggressor row activation count ($AC$) {of RowPress}, we examine the {\emph{minimum \DRAMTIMING{AggON}}} ({\DRAMTIMING{AggON}$_{\scriptstyle{\text{min}}}$}) to induce at least one bitflip for a given activation count using {the single-sided RowPress pattern}. \figref{fig:mintaggon_50C} shows how {\DRAMTIMING{AggON}$_{\scriptstyle{\text{min}}}$} changes as we sweep activation count from 1 to 10K. {The error band shows the minimum and maximum \DRAMTIMING{AggON}$_{\scriptstyle{\text{min}}}$ values.} We highlight the two potential upper-bound \DRAMTIMING{AggON} values of \SI{7.8}{\micro\second} (\DRAMTIMING{REFI}) and \SI{70.2}{\micro\second} ($9\times$\DRAMTIMING{REFI}) on the y-axis.}

\observation{{{\DRAMTIMING{AggON}$_{\scriptstyle{\text{min}}}$} significantly decreases as $AC$ increases.\label{obsv:aggon_decreases_linearly}}
}

{As $AC$ increases from 1 to 10000, the average {\DRAMTIMING{AggON}$_{\scriptstyle{\text{min}}}$} decreases from \SI{43.3}{\milli\second} to \SI{4.3}{\micro\second}, from \SI{48.3}{\milli\second} to \SI{4.8}{\micro\second}, and from \SI{44.5}{\milli\second} to \SI{4.5}{\micro\second} for Mfr. S, H, and M, respectively.\footnote{We observe no bitflips in modules with Mfr. H 4Gb A-Die and Mfr. M 8Gb B-Die in this experiment.} The decreasing {\DRAMTIMING{AggON}$_{\scriptstyle{\text{min}}}$} trend lines are very similar across all three manufacturers. Their slopes are -1.000, -0.999, and -1.000 for Mfr. S, H, and M, respectively, in~\figref{fig:mintaggon_50C}.}\footnote{{Note that for Mfr. M's 16Gb F-Die (colored red), when $AC=10^4$, we observe a minimum \DRAMTIMING{AggON}$_{\scriptstyle{\text{min}}}$ of only \SI{66}{\nano\second} (cropped in \figref{fig:mintaggon_50C}).}}

\observation{In extreme cases, RowPress can induce bitflips for \DRAMTIMING{AggON} {values} less than \SI{10}{\milli\second} with only a single aggressor row activation (i.e., $AC=1$).\label{obsv:aggon_less_than_tenms}}

We observe that, for the Mfr. S 8Gb D-Dies, the Mfr. H 16Gb C-Dies, and the Mfr. M 16Gb E-Dies, there are one, two, and two rows out of the 3072 rows we test {experience} bitflips with $AC=1$ {at a} {\DRAMTIMING{AggON}$_{\scriptstyle{\text{min}}}$} {value} less than \SI{10}{\milli\second} {(highlighted with dashed red lines)}. The minimum {\DRAMTIMING{AggON}$_{\scriptstyle{\text{min}}}$} observed for these three dies are \SI{9.2}{\milli\second}, \SI{9.8}{\milli\second}, and \SI{9.0}{\milli\second}, respectively.

\subsection{Distinguishing Characteristics of RowPress}
\label{sec:relationship}
\begin{figure}[h]
    \centering
    \includegraphics[width=1.0\linewidth]{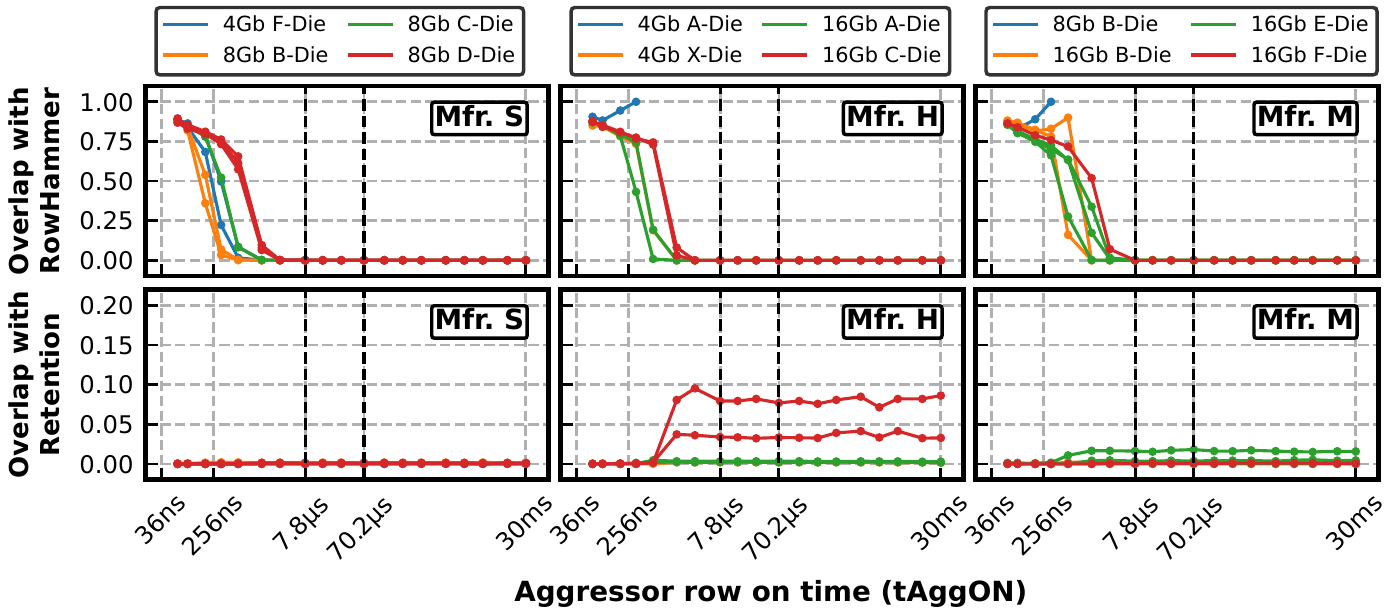}
    \caption{{Overlap ratio of RowPress-vulnerable cells @ \gls{acmin} with RowHammer-vulnerable cells @ \gls{acmin} (first row of plots) and retention failures (second row of plots).}}
    \label{fig:overlap}
\end{figure}

\noindent \textbf{\om{Cells Vulnerable to RowPress vs\gf{.}} RowHammer and Retention Failure.}
We compare the set of DRAM cells that experience bitflips from our search for \gls{acmin} as we sweep \DRAMTIMING{AggON} {beyond \SI{36}{\nano\second}} (i.e., for each \DRAMTIMING{AggON}, the set of cells that experience bitflips with the minimum number of aggressor row activations {that causes bitflips for that \DRAMTIMING{AggON})} with two other {sets} of cells: 1)~the set of cells that experience RowHammer bitflips (i.e., when \DRAMTIMING{AggON} equals \DRAMTIMING{RAS} \gf{\SI{36}{\nano\second}}%
), and 2)~the set of cells that exhibit bitflips in a \agy{data} retention failure test.\footnote{We initialize the DRAM rows with the same checkerboard data pattern \agyf{as} in~\secref{sec:vulnerability_to_readdisturbance}, {and disable auto-refresh for four seconds \agyf{at \SI{80}{\celsius} to induce retention-failure bitflips}},
\agy{similar to prior work~\cite{patel2017reaper}}.} 
\agy{\figref{fig:overlap} shows how increasing \DRAMTIMING{AggON} (x-axis) changes} \yct{the fraction of RowPress-vulnerable cells (y-axis) that also experience RowHammer (retention) failure in the first (second) row of \agy{subplots}. \agy{Similar to \figref{fig:acmin_coverage_50C}, each curve represents a different DRAM module, color-coded based on its die revision.}}

\observation{\om{An overwhelming} majority of the DRAM cells vulnerable to RowPress are {\emph{not}} vulnerable to RowHammer or data retention failures.\label{obsv:overlap}}

 \om{{{For \DRAMTIMING{AggON} $\geq$ \SI{7.8}{\micro\second}}, on average, only less than $0.013\%$} of DRAM cells vulnerable to RowPress overlap with those vulnerable to RowHammer, and less than \rb{$0.34\%$} overlap with retention failures. Therefore, an overwhelming majority of RowPress bitflips are different {from} those caused by RowHammer and retention failures.\footnote{Prior works~\cite{kim2014flipping, kim2020revisiting} already show that RowHammer bitflips have little overlap with retention failure bitflips.} These results suggest that
\agyf{different failure mechanisms lead to RowPress and RowHammer bitflips.}
}

\begin{figure}[h]
    \centering
    \includegraphics[width=1.0\linewidth]{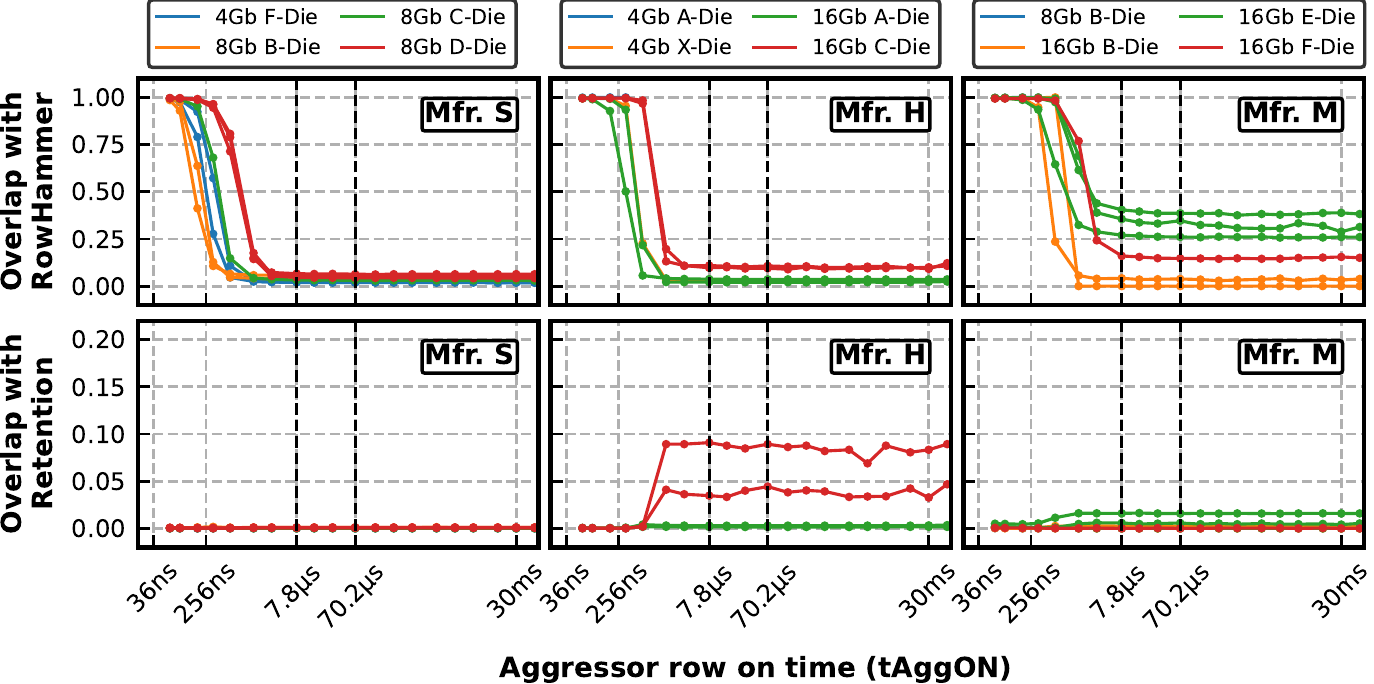}
    \caption{{Overlap ratio of RowPress-vulnerable cells @ $AC_{max}$ with RowHammer-vulnerable cells @ $AC_{max}$ (first row of plots) and retention failures (second row of plots).}}
    \label{fig:overlap_acmax}
\end{figure}

{
\figref{fig:overlap_acmax} shows the overlap ratio of the set of cells that experience bitflips when we activate the aggressor row as many times as possible (i.e., at $AC_{max}$) for each \DRAMTIMING{AggON} value with the RowHammer-vulnerable cells (also at $AC_{max}$, first row {of plots}) and retention failures (second row {of plots}). Similar to \figref{fig:overlap}, we observe {that} the overlap between RowPress-vulnerable cells and RowHammer vulnerable cells significantly decreases as \DRAMTIMING{AggON} increases.}
 
\begin{figure}[h]
    \centering
    \includegraphics[width=1.0\linewidth]{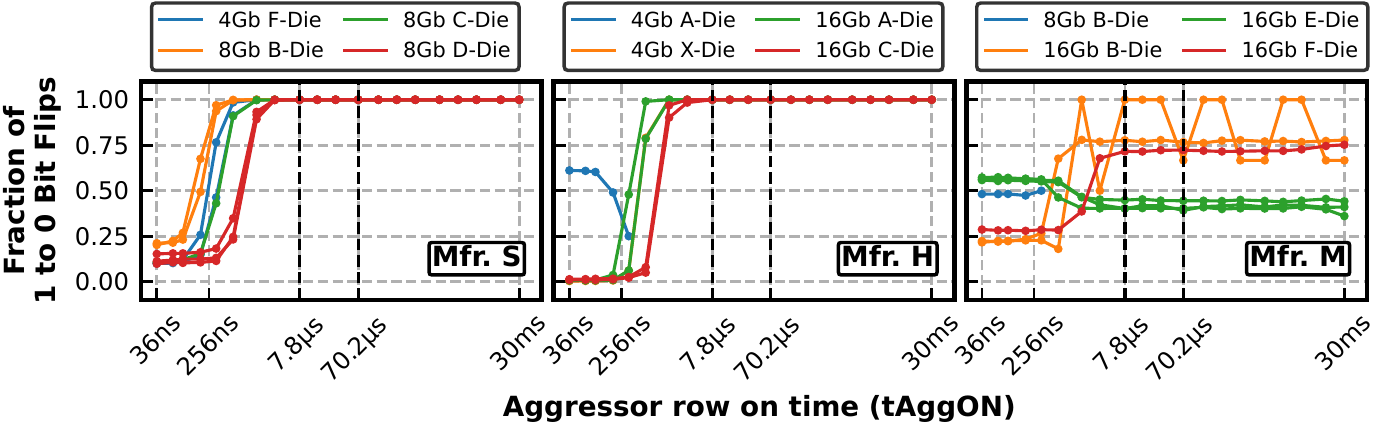}
    \caption{Fraction of 1 to 0 bitflips.}
    \vspace{-2em}
    \label{fig:bitflip_direction}
\end{figure}

\noindent \textbf{Bitflip Direction.}
\figref{fig:bitflip_direction} \agy{shows} the fraction of \agy{1 to 0} bitflips \agy{across} all the bitflips \om{we observe} \agy{(y-axis)} as we sweep \DRAMTIMING{AggON} \agy{(x-axis)}. \agy{Similar to \figref{fig:acmin_coverage_50C}, each curve represents a different DRAM module, color-coded based on its die revision.}

\observation{\hluo{RowPress and RowHammer bitflips have opposite directions}.\label{obsv:direction}}

\hluo{With the checkerboard data pattern we test, the dominant bitflip direction for RowHammer (i.e., when \DRAMTIMING{AggON} is \gf{\SI{36}{\nano\second}})  %
is 0 to 1.} As \DRAMTIMING{AggON} increases (i.e., for RowPress), for almost all die revisions from Mfr. \rb{S} and \rb{H} (except for \agy{Mfr. \rb{H}'s} 4Gb A-Die \agy{chips} that do not show any bitflip), the dominant bitflip direction changes to 1 to 0. For example, the fraction of 1 to 0 bitflips \agy{reaches $100\%$} for \DRAMTIMING{AggON} $\geq$ \gf{\SI{7.8}{\micro\second}}.  %
\agy{Similarly, the fraction of 1 to 0 bitflips in Mfr. \rb{M}'s 16Gb {B-Die} and F-Die chips reaches $75\%$ in this region of \DRAMTIMING{AggON}.}\footnote{{In a concurrent work~\cite{hong2023dsac}, {DRAM engineers} from Samsung claim that {the bitflips caused by RowHammer and the passing gate effect (caused by increased \DRAMTIMING{AggON})} have opposite directionality because RowHammer \emph{injects} electrons into the victim cell while {the passing gate effect} \emph{attracts} electrons from the victim cell. We call for more detailed device-level modeling and analysis on this {topic}.}}
As an exception, \agy{Mfr. \rb{M}'s} 16Gb E-Die \agy{chip}s show an opposite trend: the fraction of 1 to 0 bitflips decreases as \DRAMTIMING{AggON} increases. The reason for this \agy{opposite behavior} could be \agy{a different \hluo{layout} of true- and anti-cells}
compared to {that in other chips}.\footnote{{A fully charged {(discharged)} DRAM cell does not necessarily imply that the stored value is 1 {(0)}. A cell is called {true (anti) cell} if a fully charged state represents a value of 1 {(0)}~\cite{liu2013experimental}}\gf{.}}

\takeawaybox{RowPress has a different failure mechanism {from} RowHammer and data retention failures in DRAM. There is almost no overlap between RowPress, RowHammer, and data retention bitflips, and the directionality of RowHammer and RowPress bitflips show opposite trends.}

\section{RowPress Sensitivity Study}
\label{sec:sensitivity}

{We examine the sensitivity of RowPress bitflips to} 1)~temperature, 2)~access pattern, and 3)~aggressor row off time (\DRAMTIMING{AggOFF}). {We study the repeatability of RowPress bitflips in Appendix~\ref{sec:repeatability}.} 

\subsection{Temperature}
\label{sec:sen_temperature}
\begin{figure}[h]
    \centering
    \includegraphics[width=1.0\linewidth]{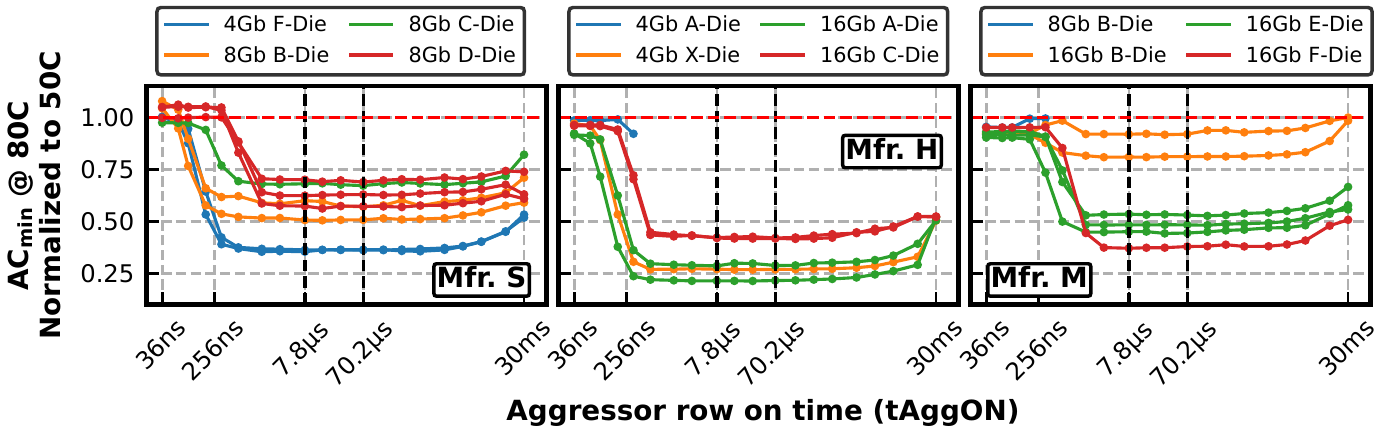}
    \caption{\gls{acmin} at $80^{\circ}C$ normalized to $50^{\circ}C$; single-sided RowPress.}
    \label{fig:temperature_agg}
\end{figure}

\noindent\textbf{Methodology.}
To investigate how {RowPress bitflips} change as DRAM chip temperature changes, we repeat the \gls{acmin} experiments (as described in~\ref{sec:characterization_methodology}) {except we} increase the temperature from $50^{\circ}C$ to $80^{\circ}C$. \figref{fig:temperature_agg} shows the mean \gls{acmin} values we observe at $80^{\circ}C$ normalized to $50^{\circ}C$ as we sweep \DRAMTIMING{AggON} at $80^{\circ}C$ in {linear (y-axis) - log (x-axis)} scale.

\observation{{As temperature increases, RowPress reduces \gls{acmin} more.}}

We observe that for all die revisions vulnerable to RowPress, \gls{acmin} consistently reduces for the same \DRAMTIMING{AggON} value as temperature increases from $50^{\circ}C$ to $80^{\circ}C$. {For example, when \DRAMTIMING{AggON} is \gf{\SI{7.8}{\micro\second}}, the average \gls{acmin} at $80^{\circ}C$ is only 0.55$\times$, 0.32$\times$, and 0.59$\times$ of that at $50^{\circ}C$, for Mfr. S, H, and M, respectively.}
Across all manufacturers, \gls{acmin} reduces by $48\times$ on average (up to \rb{$122\times$}, observed in 8Gb B-Dies from Mfr. \rb{S}) when \DRAMTIMING{AggON} {increases from \SI{36}{\nano\second} to \SI{7.8}{\micro\second}} at $80^{\circ}C$. When \DRAMTIMING{AggON} increases {from \SI{36}{\nano\second}} to \gf{\SI{70.2}{\micro\second}}, \gls{acmin} reduces by \rb{$438\times$} on average  (up to \rb{$1106\times$}) at $80^{\circ}C$. 
{In contrast}, at $50^{\circ}C$, the reduction in \gls{acmin} is only $21\times$ on average (up to \rb{$59\times$}) when \DRAMTIMING{AggON} {increases from \SI{36}{\nano\second} to \SI{7.8}{\micro\second}} and \rb{$190\times$} (up to \rb{$537\times$}) {when \DRAMTIMING{AggON} increases from \SI{36}{\nano\second} to} \gf{\SI{70.2}{\micro\second}}. \rb{For a \DRAMTIMING{AggON} of \SI{30}{\milli\second}, $82.8\%$ of the rows with bitflips experience {an} \gls{acmin} of \emph{only one} {(not shown in~\figref{fig:temperature_agg})} at $80^{\circ}C$ (only $13.1\%$ at $50^{\circ}C$).} 
{
We provide more results involving \gls{acmin} at $65^{\circ}C$ in Appendix~\ref{sec:65C}.
}

\figref{fig:coverage_80C} shows the fraction of rows that have at least one RowPress bitflip as we sweep \DRAMTIMING{AggON} at $80^{\circ}C$.

\begin{figure}[h]
    \centering
    \includegraphics[width=1.0\linewidth]{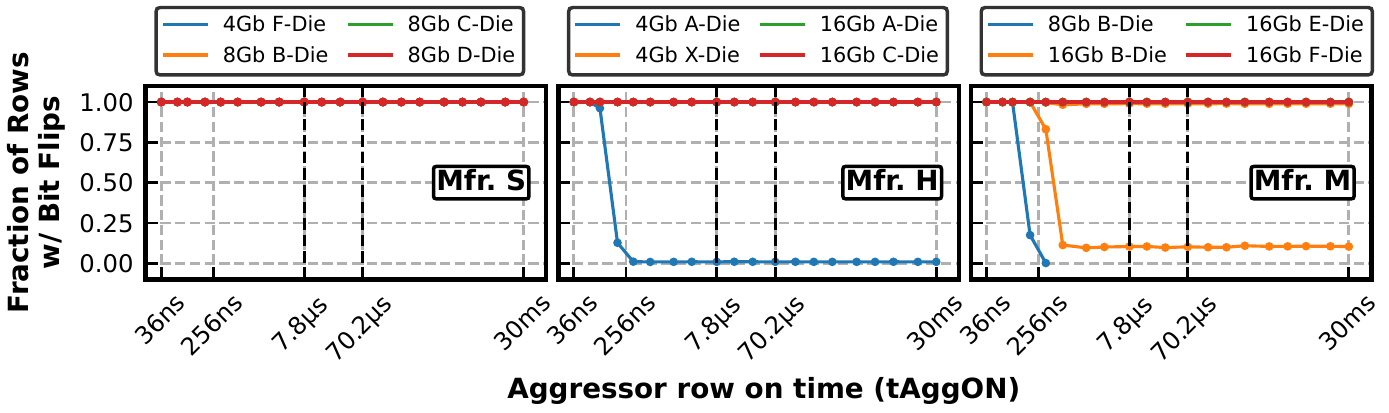}
    \caption{{Fraction} of rows that experience at least one bitflip at $80^{\circ}C$; single-sided RowPress.}
    \label{fig:coverage_80C}
\end{figure}

\observation{{Fraction} of rows that have at least one RowPress bitflip significantly increases as temperature increases.}

We observe that almost all die revisions from all three manufacturers that are vulnerable to RowPress have their {fractions of rows with at least one bitflip} {increase} to almost $100\%$ at $80^{\circ}C$. Note that, for 4Gb A-Die from Mfr. \rb{H} where we observe {\emph{no bitflips at all}} for \DRAMTIMING{AggON} $>$ \gf{\SI{336}{\nano\second}} %
at $50^{\circ}C$, we {are able to} observe bitflips {in a small fraction of rows (on average, 0.86\% of all tested rows)} with larger \DRAMTIMING{AggON} values up to \gf{\SI{30}{\milli\second}} %
at $80^{\circ}C$.

\begin{figure}[h]
    \centering
    \includegraphics[width=1.0\linewidth]{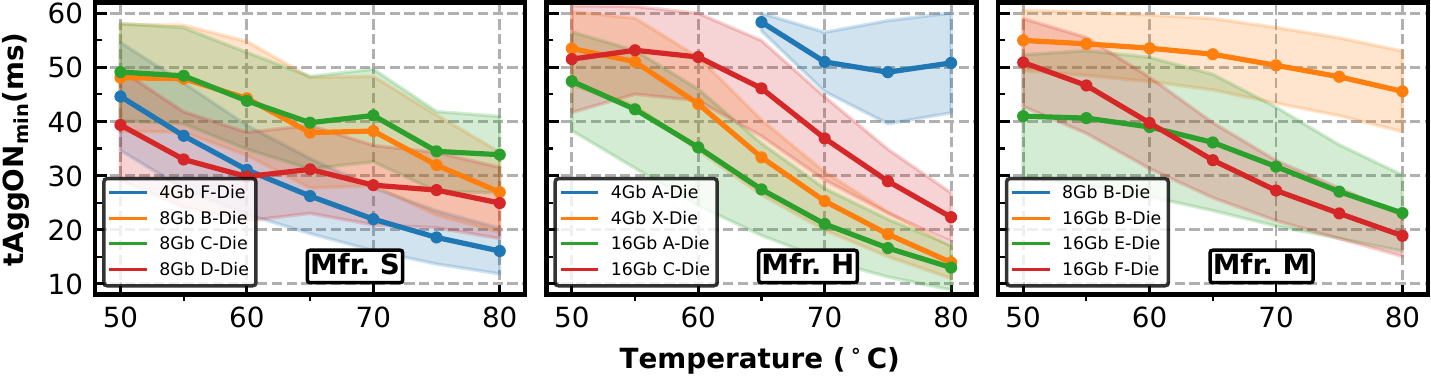}
    \caption{{\DRAMTIMING{AggON}$_{min}$ when $AC=1$ as we sweep temperature from $50^{\circ}C$ to $80^{\circ}C$ with $5^{\circ}C$ steps; single-sided RowPress.}}
    \label{fig:taggonmin_temp_sweep}
\end{figure}

To study the effect of increasing temperature on \DRAMTIMING{AggON}$_{\scriptsize{min}}$ (i.e., the minimum \DRAMTIMING{AggON} to induce at least one bitflip) when $AC=1$, we sweep temperature from $50^\circ C$ to $80^\circ C$ with a step size of $5^\circ C$ and show the results in~\figref{fig:taggonmin_temp_sweep}.\footnote{We do not sweep the temperature with the fine-grained step size $5^\circ C$ for the other experiments because of the prohibitively long experiment {times}.} {The error band shows the standard deviation of \DRAMTIMING{AggON}$_{\scriptsize{min}}$.}

\observation{As temperature increases, {\DRAMTIMING{AggON}$_{\scriptstyle{\text{min}}}$} significantly decreases.\label{obsv:aggonmin_reduces_with_temperature}}

{We observe that} \DRAMTIMING{AggON}$_{min}$ significantly decreases as we gradually increase temperature from $50^\circ C$ to $80^\circ C$. For Mfr. S, H, and M, the average (minimum) \DRAMTIMING{AggON}$_{min}$ reduces by $1.78\times$ ($1.90\times$), $2.84\times$ ($3.24\times$), and $1.64\times$ ($1.95\times$), respectively, {going from $50^\circ C$ to $80^\circ C$}. For example, for 16Gb A-Dies from Mfr. H, across all tested rows, the average (minimum) \DRAMTIMING{AggON}$_{min}$ is \gf{\SI{47.4}{\milli\second}} (\gf{\SI{14.3}{\milli\second}}) at $50^\circ C$, and reduces to only \gf{\SI{13.0}{\milli\second}} (\gf{\SI{3.0}{\milli\second}}) at $80^\circ C$. Note that for Mfr. H's 4Gb A-Die, {where} we could not induce any bitflip even when $AC = 10000$ at $50^\circ C$ (\figref{fig:mintaggon_50C}), we {are able to} induce RowPress bitflips {when} $AC=1$ {at temperatures} $\geq$ $65^\circ C$.

\takeawaybox{RowPress gets significantly worse as temperature increases. This behavior is very different from how RowHammer bitflips change with temperature~\cite{kim2014flipping, orosa2021deeper}.}
\vspace{-1em}

\subsection{Access Pattern}
\label{sec:sen_acc_pattern}

\begin{figure}[h]
    \centering
    \vspace{0.3em}
    \includegraphics[width=0.95\linewidth]{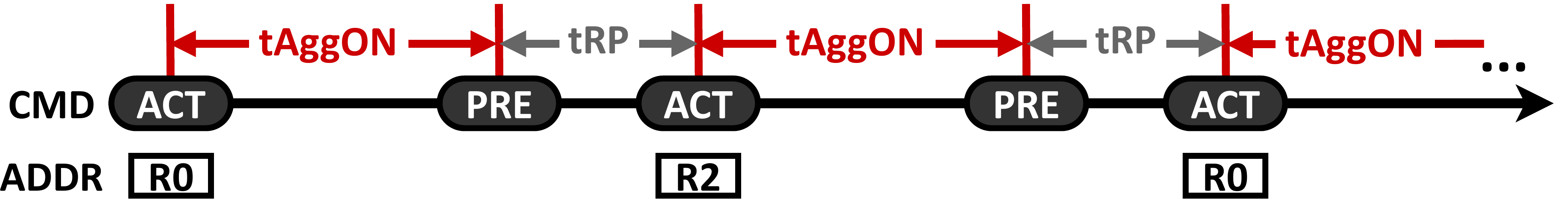}
    \caption{{Double-sided} RowPress {access} pattern.}
    \label{fig:double-pattern}
\end{figure}

\noindent\textbf{Methodology.}
To investigate how {the bitflips induced by RowPress} change as access pattern changes, we repeat the \gls{acmin} experiments  (described in~\secref{sec:characterization_methodology}) {except we} use a {\em double-sided} RowPress pattern involving two aggressor rows, as shown in \figref{fig:double-pattern}. In the double-sided RowPress pattern, we replace the row address of every other aggressor row activation {in the single-sided access pattern (shown in \figref{fig:characterization_single_pattern})} from R0 to R2. We treat the row R1 between R0 and R2 and three adjacent rows before R0 {(i.e., R-1, R-2, R-3)} and after R2 (i.e., R3, R4, R5) as the victim rows. We conduct the test at both $50^{\circ}C$ and $80^{\circ}C$.

\begin{figure}[h]
    \centering
    \vspace{0.3em}
    \includegraphics[width=1.0\linewidth]{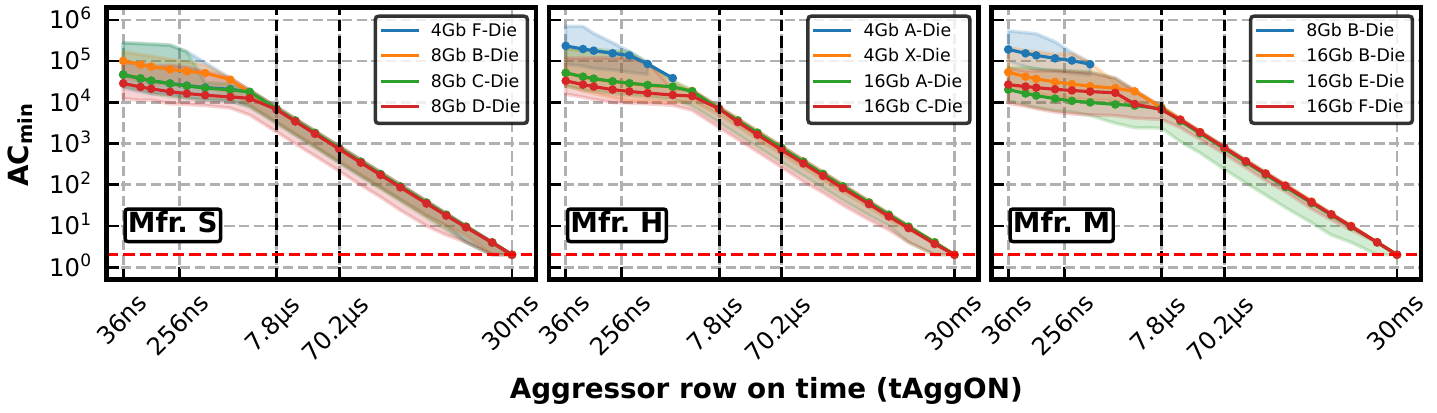}
    \caption{\revision{\gls{acmin} of double-sided RowPress; $50^{\circ}C$.\revlabel{S5}}}
    \label{fig:double_acmin_50C}
\end{figure}

We show how \gls{acmin} changes with the double-sided RowPress pattern at $50^{\circ}C$ as we sweep \DRAMTIMING{AggON} in \figref{fig:double_acmin_50C}. {The error band shows the minimum and maximum \gls{acmin} values.}

\observation{As \DRAMTIMING{AggON} increases, double-sided RowPress exhibits a similar decreasing \gls{acmin} trend as single-sided.}

As \DRAMTIMING{AggON} increases, \gls{acmin} significantly decreases with the double-sided RowPress pattern. The slopes of the overlapping \gls{acmin} trend lines in~\figref{fig:double_acmin_50C} for \DRAMTIMING{AggON} $\geq$ \gf{\SI{7.8}{\micro\second}} %
of Mfr\gf{.} S, H, M are \rb{$-1.015$, $-1.010$, and $-1.011$}, respectively. %
{Compared to the single-sided RowPress pattern, the decrease in \gls{acmin} is much larger with the double-sided RowPress pattern.} For example, on average, when \DRAMTIMING{AggON} increases from \gf{\SI{36}{\nano\second}} to \gf{\SI{186}{\nano\second}}, \gls{acmin} reduces by $1.62\times$, $1.56\times$, and $1.64\times$ for Mfr. S, H, and M, respectively, with the double-sided pattern, compared to {only} $1.17\times$, $1.04\times$, and $1.08\times$ of the single-sided pattern.

\begin{figure}[h]
    \centering
    \includegraphics[width=1.0\linewidth]{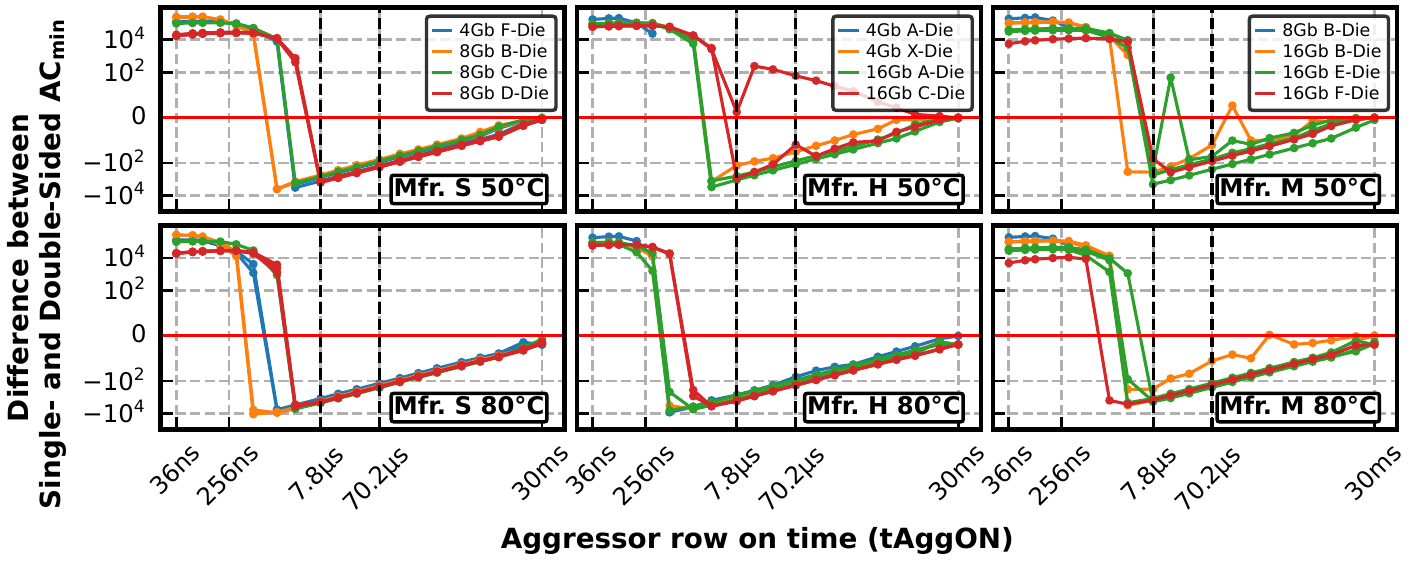}
    \caption{Single-sided \gls{acmin} minus double-sided \gls{acmin} at $50^{\circ}C$ (first row) and $80^{\circ}C$ (second row).}
    \label{fig:temperature_singledouble_diff}
\end{figure}

To comprehensively investigate how the access pattern and {the} temperature of the DRAM chip affect \gls{acmin}, we plot the difference between single- and double-sided \gls{acmin} (i.e., \gls{acmin}$(single)$ - \gls{acmin}$(double)$) at $50^{\circ}C$ (first row) and $80^{\circ}C$ (second row) in \figref{fig:temperature_singledouble_diff}. A data point \revision{below\revlabel{S5}} $0$ means that the single-sided RowPress pattern needs fewer aggressor row activations in total to induce a bitflip compared to double-sided. 

\observation{Single-sided RowPress becomes more effective {at inducing bitflips} as \DRAMTIMING{AggON} increases beyond a certain value compared to \gf{d}ouble-sided RowPress.}

We observe that, as \DRAMTIMING{AggON} increases, double-sided RowPress {is} initially more effective compared to single-sided at $50^{\circ}C$ (e.g., the single-sided pattern requires at least $10^4$ more aggressor row activations to cause bitflips for almost all die revisions when \DRAMTIMING{AggON} $<$ \gf{\SI{1536}{\nano\second}}).
However, as \DRAMTIMING{AggON} continues to increase {beyond \SI{1536}{\nano\second}}, single-sided RowPress becomes more effective compared to double-sided for some die revisions. For example, for \DRAMTIMING{AggON} $=$ \gf{\SI{1536}{\nano\second}}, single-sided RowPress requires \rb{4210} less aggressor row activations on average to induce bitflips compared to double-sided for the 8Gb B-Dies from Mfr. \rb{S} at $50^{\circ}C$. As temperature increases from $50^{\circ}C$ to $80^{\circ}C$, we observe that: 1) single-sided RowPress becomes even more effective{, for} example, for the 8Gb B-Dies from Mfr. \rb{S}, {the single-sided RowPress pattern} needs \rb{8699} less aggressor row activations on average for \gls{acmin} $=$ \gf{\SI{1536}{\nano\second}} compared to the double-sided RowPress pattern, and 2) for almost all die revisions from all manufacturers, single-sided \gls{acmin} is \emph{consistently} smaller than double-sided for \DRAMTIMING{AggON} values larger than \gf{\SI{7.8}{\micro\second}.} {
We provide more results involving \gls{acmin} at $65^{\circ}C$ in Appendix~\ref{sec:65C}.
}

Note that this behavior is very different from RowHammer\joel{,} where double-sided RowHammer is strictly {more effective at inducing bitflips} than single-sided~\cite{kim2014flipping}. \figref{fig:hcf_intro} summarizes the {\gls{acmin}} results we observe for single-sided and double-sided patterns for RowHammer and RowPress at $80^{\circ}C$.  

\takeawaybox{RowPress behaves very differently from RowHammer as we change the access pattern from single-sided to double-sided. {As \DRAMTIMING{AggON} increases beyond a certain value, {RowPress} needs {fewer} aggressor row activations to induce bitflips with the single-sided pattern compared to {the double-sided pattern}.}}

\subsection{Data Pattern}
\label{sec:sen_data_pattern}

{
\noindent\textbf{Methodology.}
To investigate how RowPress bitflips are affected by the data pattern of the victim and aggressor rows (i.e., what is the most effective data pattern to induce RowPress bitflips?), we repeat the \gls{acmin} experiments with more data patterns, summarized in Table~\ref{tab:datapattern}. We {denote} the inverse of a data pattern with the suffix ``I''. Due to the large search space of all \DRAMTIMING{AggON} values, we test a {set} of representative \DRAMTIMING{AggON} values: \gf{\SI{36}{\nano\second}} (=\DRAMTIMING{RAS}), \gf{\SI{66}{\nano\second}}, \gf{\SI{636}{\nano\second}}, \gf{\SI{7.8}{\mu\second}} (=\DRAMTIMING{REFI}), $9\times$\gf{\SI{7.8}{\mu\second}}, \gf{\SI{300}{\mu\second}}, and \gf{\SI{6}{\milli\second}}.
}

\begin{table}[h]
\centering
\small
\caption{{Tested data patterns}}
\label{tab:datapattern}
\begin{tabular}{@{}cccc@{}}
\toprule
\multirow{2}{*}{\textbf{Row Type}} & \multicolumn{3}{c}{\textbf{Data Pattern}} \\
                                   & \textbf{C}hecker\textbf{B}oard (\textbf{I})  & \textbf{R}ow\textbf{S}tripe (\textbf{I})  & \textbf{C}ol\textbf{S}tripe (\textbf{I}) \\ \midrule 
Aggressor                          & \texttt{0xAA} (\texttt{0x55})       & \texttt{0xFF} (\texttt{0x00})    & \texttt{0x55} (\texttt{0xAA})      \\
Victim                             & \texttt{0x55} (\texttt{0xAA})       & \texttt{0x00} (\texttt{0xFF})    & \texttt{0x55} (\texttt{0xAA})  \\ \bottomrule
\end{tabular}
\end{table}

{
\noindent\textbf{Metric.}
To quantify the effectiveness of different data patterns for a die revision, we normalize their average \gls{acmin} (across all rows we test) to the average \gls{acmin} value of the CheckerBoard (CB) pattern. A value lower (higher) than 1.00 means the data pattern is more (less) effective than the CB pattern {at} inducing bitflips. 
}

\begin{figure}[h]
    \centering
    \includegraphics[width=1.0\linewidth]{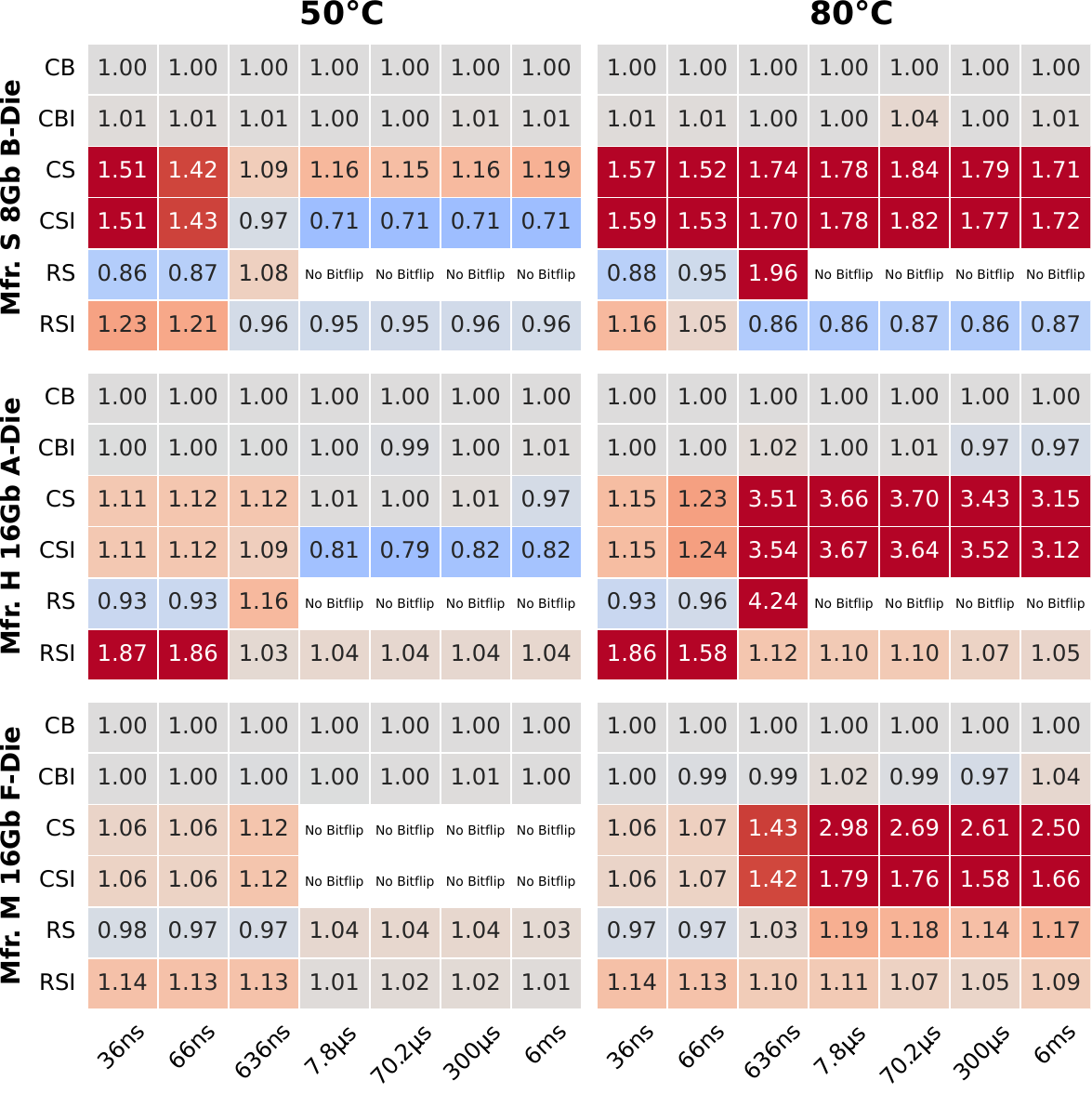}
    \caption{\gls{acmin} of different data patterns normalized to the CB data pattern at different \DRAMTIMING{AggON} values from three representative die revisions from the three manufacturers; Single-sided access pattern; $50^\circ C$ (left column) and $80^\circ C$ (right column);  A value lower (higher) than 1.00 means the data pattern is more (less) effective than the CB pattern {at} inducing bitflips, {colored as blue (red).}}
    \label{fig:dp_example_single}
\end{figure}

{
\figref{fig:dp_example_single} shows the normalized \gls{acmin} values of different data patterns (y-axis) at different \DRAMTIMING{AggON} values (x-axis) from three representative die revisions from the three manufacturers\footnote{We find {that the remaining} die revisions {behave similarly to} one of the three representative die revisions.} using a single-sided access pattern at $50^\circ C$ (left column) and $80^\circ C$ (right column). A red (blue) cell means at {a given} $x$ \DRAMTIMING{AggON}, the $y$ data pattern is less (more) effective {at} inducing bitflips compared to the baseline CheckerBoard pattern. Certain data patterns could not induce any bitflip at certain \DRAMTIMING{AggON} values, even with the maximum {possible} activation count {(within 60ms, which is strictly smaller than the 64ms refresh window)}. We mark these cases as ``No Bitflip'' (white cell) in the figure. We make the following two observations.
}

\observation{CheckerBoard pattern is in general the most effective RowPress data pattern {among the ones tested}.}

{
We observe that, in most cases, the CheckerBoard pattern is the most effective {at} inducing RowPress bitflips {among the tested data patterns} for the following two reasons. First, we can \emph{always} induce bitflips with the CheckerBoard pattern as we increase \DRAMTIMING{AggON}. In comparison, although the RowStripe pattern in Mfr. S 8Gb B-Die and Mfr. H 16Gb A-Die is more effective with low \DRAMTIMING{AggON} values (i.e., up to $13\%$ smaller \gls{acmin} when \DRAMTIMING{AggON} is \gf{\SI{66}{\nano\second}}), it cannot induce \emph{any} bitflip for \DRAMTIMING{AggON} larger than \gf{\SI{636}{\nano\second}}. Second, compared to the other data patterns, the CheckerBoard pattern is less affected by the increase in temperature. For example, although the ColumnStripeI pattern is the most effective for large \DRAMTIMING{AggON} values ($\geq$ \gf{\SI{7.8}{\micro\second}}) for Mfr. S 8Gb B-Die and Mfr. H 16Gb A-Die at $50^\circ C$ (up to $29\%$ smaller \gls{acmin}), it becomes the least effective (up to $267\%$ increase in \gls{acmin}) at $80^\circ C$. 
}

\observation{The most effective RowHammer data pattern is not necessarily the most effective RowPress pattern.}

{
For all three representative die revisions shown in \figref{fig:dp_example_single}, RowStripe is the most effective data pattern to induce Rowhammer bitflips (i.e., \DRAMTIMING{AggON} = \gf{\SI{36}{\nano\second}}). However, as we increase \DRAMTIMING{AggON}, it becomes significantly less effective compared to the other patterns. For Mfr. S 8Gb B-Die and Mfr. H 16Gb A-Die, the RowStripe pattern {\emph{cannot}} induce any bitflip for \DRAMTIMING{AggON} $>$ \gf{\SI{636}{\nano\second}}, even at $80^\circ C$.
}

\begin{figure}[h]
    \centering
    \includegraphics[width=1.0\linewidth]{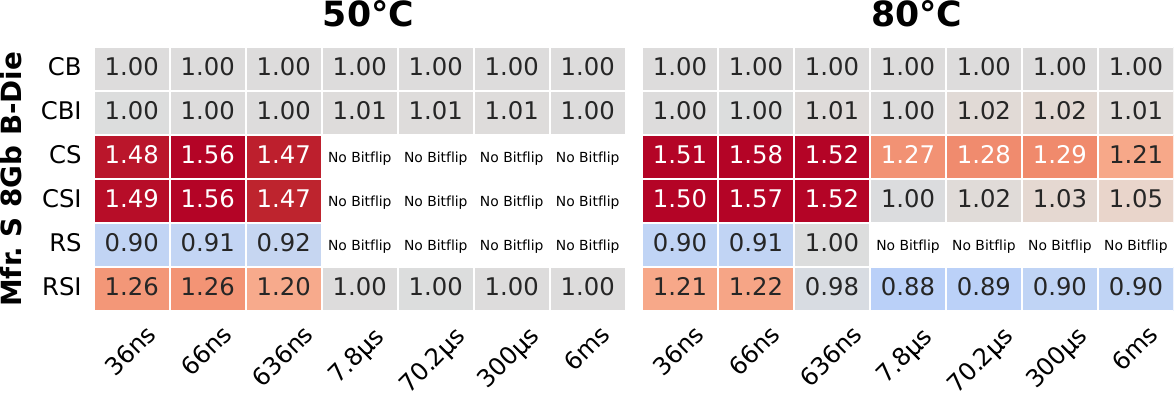}
    \caption{Normalized \gls{acmin} of different data patterns of Mfr. S 8Gb B-Die; Double-sided access pattern; $50^\circ C$ (left column) and $80^\circ C$ (right column).}
    \label{fig:dp_example_double}
\end{figure}

\figref{fig:dp_example_double} shows the normalized \gls{acmin} values of different data patterns from Mfr. S 8Gb B-die using a double-sided access pattern at $50^\circ C$ and $80^\circ C$. We observe that the effectiveness of ColumnStripe {and ColumnStripeI} patterns increases as \DRAMTIMING{AggON} increases in the double-sided access pattern, in contrast to the decreasing effectiveness as we show in \figref{fig:dp_example_single}. {Note that this is the only case where we observe any major difference comparing single-sided to double-sided. The other die revisions behave similarly for the double-sided access pattern compared to single-sided.}

{We believe the data pattern dependence of RowPress and RowHammer require more and deeper study to fully understand and model the effect on the two read disturbance phenomena.}

\subsection{tAggON vs tAggOFF}
\label{sec:sen_ftber}

\revision{Prior works on {\emph{device-level}} mechanisms of RowHammer~\cite{park2014active, yang2019trap} show that increasing \DRAMTIMING{AggON} has little impact on DRAM {read disturbance}, while doing the opposite, increasing \DRAMTIMING{AggOFF} (i.e., the aggressor row off time), worsens {read disturbance}. This seems to contradict our results in~\secref{sec:vulnerability_to_readdisturbance} and~\secref{sec:sen_acc_pattern}. However, the methodology of those {prior} works~\cite{park2014active, yang2019trap} is limited because they only {test} 1) a very small range of \DRAMTIMING{AggON} and \DRAMTIMING{AggOFF} values (up to \SI{50}{\nano\second} in~\cite{yang2019trap} and \SI{72.5}{\nano\second} in~\cite{park2014active}), and 2) a single-sided access pattern.}

\begin{figure}[h]
    \centering
    \vspace{0.4em}
    \includegraphics[width=0.95\linewidth]{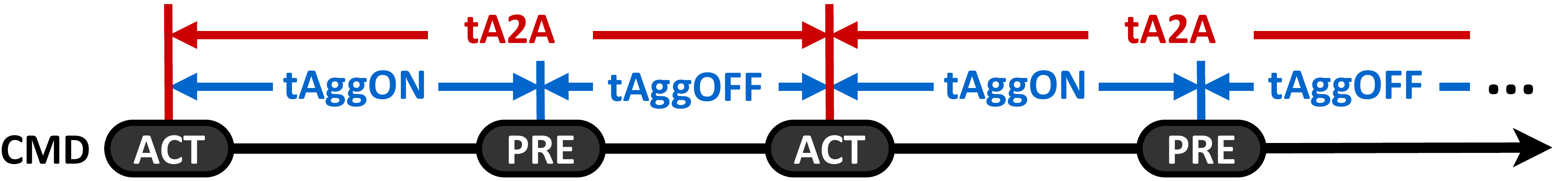}
    \caption{The RowPress-ONOFF pattern.}
    \label{fig:ftber-pattern}
\end{figure}

\noindent\textbf{Access Pattern.}
\revision{To compare RowPress to the read-disturb mechanisms discussed in prior works~\cite{park2014active, yang2019trap}}, we design the RowPress-ONOFF access pattern shown in \figref{fig:ftber-pattern}\revision{, based on the pattern proposed in~\cite{park2014active}}. In this pattern, we can adjust \DRAMTIMING{AggON} and \DRAMTIMING{AggOFF} by changing: 1)  when we issue the \DRAMCMD{PRE} command to close the aggressor row, and 2) when we issue the \DRAMCMD{ACT} command to open the aggressor row. We denote the time interval between two consecutive \DRAMCMD{ACT} commands as \DRAMTIMING{A2A}. Notice that since \DRAMTIMING{A2A} = \DRAMTIMING{AggON} + \DRAMTIMING{AggOFF}, the {minimum} possible value of \DRAMTIMING{A2A} is min(\DRAMTIMING{AggON}) + min(\DRAMTIMING{AggOFF}) = \DRAMTIMING{RAS} + \DRAMTIMING{RP} = \DRAMTIMING{RC}.

\noindent\textbf{Methodology.}
We fix the activation frequency {of a row} by fixing \DRAMTIMING{A2A}. We increase \DRAMTIMING{A2A} beyond \DRAMTIMING{RC} by {$\Delta \DRAMTIMING{A2A} = \{240$, $600$, $1200$, $2400$, $6000$$\}$ ns}. For each \DRAMTIMING{A2A} {value}, we sweep the {fraction} of $\Delta \DRAMTIMING{A2A}$ that contributes to \DRAMTIMING{AggON} {from $0\%$ to $100\%$ (with a step size of $25\%$)}. For example, 25\% means \DRAMTIMING{AggON} = 25\%  $\Delta \DRAMTIMING{A2A}$ + \DRAMTIMING{RAS}, and \DRAMTIMING{AggOFF} = 75\% $\Delta \DRAMTIMING{A2A}$ + \DRAMTIMING{RP}. For all configurations, we activate the aggressor row(s) as many times as possible to {induce the most number of bitflips} without exceeding the experiment time limit of \gf{\SI{60}{\milli\second}}. {We conduct the experiments at $50^{\circ}C$ and $80^{\circ}C$.}

\noindent\textbf{Metric.}
We measure the {\gls{ber}, i.e., the fraction of DRAM cells in a DRAM row that experience bitflips.} We repeat the experiment five times and report the highest \gls{ber} to {evaluate} the worst-case scenario.

\begin{figure}[h]
    \centering
    \vspace{0.4em}
    \includegraphics[width=1.0\linewidth]{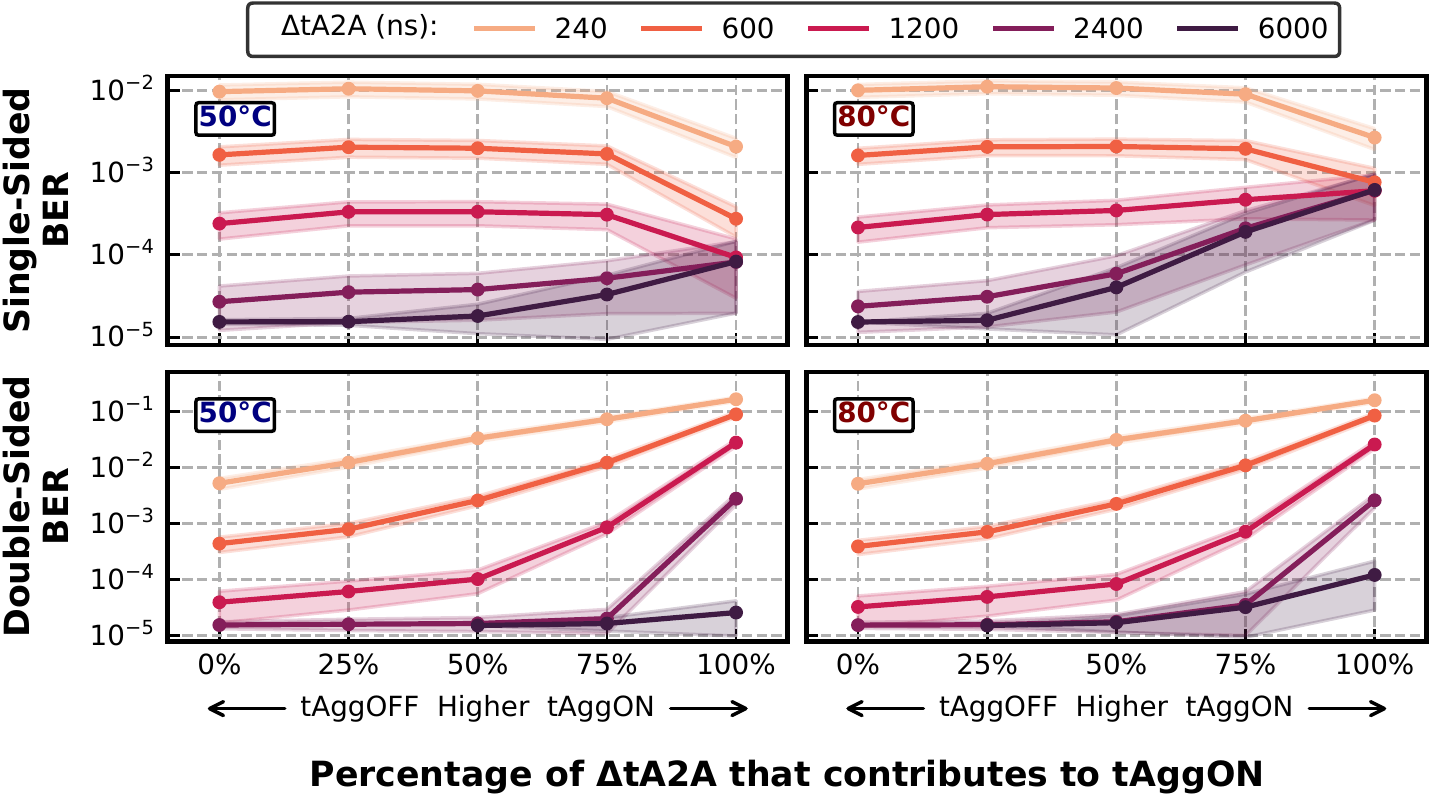}
    \caption{BER of the representative Mfr. S 8Gb D-Die; single- (top row) and double-sided (bottom row) RowPress-ONOFF pattern at $50^{\circ}C$ (left column) and $80^{\circ}C$ (right column).}
    \label{fig:ftber}
\end{figure}

\figref{fig:ftber} shows the \gls{ber} (y-axis) for both single-{sided} (top row) and double-sided (bottom row) RowPress-ONOFF pattern \hluo{for a representative\footnote{We observe a similar trend for {almost all} other die revisions. We {show only} one representative {die revision} to {illustrate} the results more clearly. We show all other die revisions in {Appendix~\secref{sec:appendix-allonoff}.}} die revision (8Gb D-Die from Mfr. S)}. We sweep $\Delta \DRAMTIMING{A2A}$ (different {lines} in each {plot}) and the percentage of $\Delta \DRAMTIMING{A2A}$ that contributes to \DRAMTIMING{AggON} (x-axis) at $50^{\circ}C$ ({left} column) and $80^{\circ}C$ ({right} column). {The error band shows the standard
deviation of \gls{ber}.} We make the following three observations.

\observation{For {the single-sided access} pattern, increasing \DRAMTIMING{AggON} ({i.e.,} decreasing \DRAMTIMING{AggOFF}) {with small (large) $\Delta \DRAMTIMING{A2A}$ values} mitigates (exacerbates) read disturbance.}

For $\Delta \DRAMTIMING{A2A}$ {values} {$\leq$} \gf{\SI{1200}{\nano\second}} {(i.e., the upper three lines in the top two plots)}, we observe that BER decreases as we increase \DRAMTIMING{AggON} ({and} thus decrease \DRAMTIMING{AggOFF}) with the single-sided pattern. This agrees with prior device-level works~\cite{park2016experiments, yang2019trap}\footnote{Injected charge (from diffused channel electrons~\cite{park2016experiments} and charge traps~\cite{yang2019trap}) needs sufficient amount of time to be recombined at the victim cell and fully exhausted \emph{after} the row is closed (i.e., longer \DRAMTIMING{AggOFF})} that test a small range of \DRAMTIMING{AggON}/\DRAMTIMING{AggOFF} values (up to \SI{50}{\nano\second} in~\cite{yang2019trap} and \SI{72.5}{\nano\second} in~\cite{park2014active},
respectively). As $\Delta \DRAMTIMING{A2A}$ takes larger values (e.g., \gf{\SI{2400}{\nano\second}} %
and \gf{\SI{6000}{\nano\second}}), %
we observe an \emph{opposite trend} {to what we observe with smaller \DRAMTIMING{A2A} values}: BER increases as we increase \DRAMTIMING{AggON} ({and} thus decrease \DRAMTIMING{AggOFF}). {This is {neither observed nor} explained by prior device-level works~\cite{park2016experiments, yang2019trap}. 
}

\observation{{For {the single-sided access} pattern, increasing temperature exacerbates read disturbance for large $\Delta$\DRAMTIMING{A2A} and \DRAMTIMING{AggON} values.}}

{For the single-sided pattern, we observe that as temperature increases from $50^{\circ}C$ to $80^{\circ}C$, BER significantly increases (remains almost unchanged) for large (small) $\Delta$\DRAMTIMING{A2A} and \DRAMTIMING{AggON} values. For example, the average BER increases by 7.5$\times$ (only $1.04\times$) from $50^{\circ}C$ to $80^{\circ}C$ when $\Delta$\DRAMTIMING{A2A} $=$ \SI{6000}{\nano\second} (\SI{240}{\nano\second}) and $100\%$ of $\Delta$\DRAMTIMING{A2A} contributes to \DRAMTIMING{AggON}. {At the inflection point of} $\Delta$\DRAMTIMING{A2A} $=$ \SI{1200}{\nano\second}, {when $50\%$ {to} $100\%$ of $\Delta$\DRAMTIMING{A2A} contributes to \DRAMTIMING{AggON}, BER \emph{decreases} at $80^{\circ}C$, in contrast to \emph{increasing} at $50^{\circ}C$.} This observation is {\emph{not}} fully explained by prior device-level works~\cite{park2016experiments, yang2019trap} because they do \emph{not} change $\Delta$\DRAMTIMING{A2A}, \DRAMTIMING{AggON}, and \DRAMTIMING{AggOFF} when investigating {the effect of temperature on read disturbance.}}  

\observation{For {the} double-sided pattern, read disturbance consistently worsens as \DRAMTIMING{AggON} increases and \DRAMTIMING{AggOFF} decreases.}

For {\emph{all}} $\Delta \DRAMTIMING{A2A}$ values we test {with the double-sided access pattern}, we observe that BER consistently increases as \DRAMTIMING{AggON} increases ({i.e.,} as \DRAMTIMING{AggOFF} decreases){, unlike} the single-sided case where we observe \emph{opposite} trends for small and large $\Delta \DRAMTIMING{A2A}$ values. Such a difference {in the bit error rate behavior of single-sided and double-sided access patterns} is \emph{not} covered by prior device-level works~\cite{park2016experiments, yang2019trap}. {Our observations indicate that} access pattern plays an important role in \agyf{RowPress's} device-level failure mechanisms {and further device-level investigation is necessary to develop a better understanding of RowPress}.

\takeawaybox{\yct{RowPress is a read-disturb phenomenon that existing device-level studies {do not} fully explain. {We call for more device-level research to provide {fundamental} lower-level understanding of the RowPress phenomenon.}}}

\vspace{-1em}

\section{Real System Demonstration of RowPress}
\label{sec:real}

We {experimentally} demonstrate {that a simple user-level C++ program can induce RowPress bitflips} on a real DDR4-based system {despite the existence of periodic {auto-refresh} and {in-DRAM {target row refresh (TRR)} mechanisms employed by the manufacturer}}.

\vspace{-.5em}

\subsection{{Experimental Setup}}
\label{sec:real_setup}

\noindent\textbf{System Configuration.}
We use an Ubuntu 18.04 system (Linux kernel 5.4.0-131-generic~\cite{linux-kernel-540-131}) with an Intel i5-10400 {(Comet Lake)} processor~{\cite{intel-comet-lake}} and a {16GB} {dual rank DDR4} DRAM module~\cite{samsung-real-datasheet} {from Mfr. {S} {(Samsung)}}. This DRAM module has target row refresh (TRR)~{\cite{hassan2021utrr,frigo2020trrespass}}, a widely adopted in-DRAM {RowHammer} mitigation mechanism employed by DRAM manufacturers.

\noindent\textbf{Memory Address Mapping.}
We reverse engineer the {processor's address mapping from physical memory addresses to DRAM rank, bank, row, and column} {addresses} using DRAMA~\cite{pessl2016drama}, similar to prior works (e.g.,~\cite{frigo2020trrespass,jattke2022blacksmith,deridder2021smash}). We allocate {a} {1GB} page using hugepage support~\cite{hugepage-linux} to directly manipulate the least significant 30 physical address {bits} {that contain all of} the DRAM rank {and} bank address bits {and part of the row address bits}. %
{We carefully generate pointers to aggressor and victim rows within {the {1GB} page} to precisely place them in physically adjacent DRAM rows.}\footnote{Although we leverage {a {1GB} hugepage} for this {real-system} demonstration {of RowPress}, {hugepages are not necessary for allocating physically adjacent DRAM rows {and inducing bitflips}, as} prior works~\cite{zhang2020pthammer,lipp2018nethammer,kogler2022half,kwong2020rambleed} on system-level RowHammer attacks experimentally demonstrate. {One can extend our {real-system demonstration} program to 
{avoid using} hugepages.}} 

\vspace{-.5em}
\subsection{RowPress on Real Systems}
\label{sec:real_rowpress}

\noindent{\textbf{Challenges.}} {We face two challenges in inducing RowPress bitflips in {a} real system.} {First, TRR can detect aggressor rows in a RowPress access pattern and prevent us from inducing bitflips by refreshing the victim rows. However, TRR mechanisms typically keep track of \emph{only} a few aggressor rows~\cite{hassan2021utrr,frigo2020trrespass} and these mechanisms can be bypassed by certain access patterns that access many other dummy aggressor rows (called dummy rows~\cite{hassan2021utrr,frigo2020trrespass}) besides the real aggressor rows. Such access patterns aim to trick a TRR mechanism into detecting \emph{only} the dummy rows and allow the real aggressor rows to remain undetected.}

{Second, the memory controller needs to keep the aggressor row on for a long duration (i.e., large \DRAMTIMING{AggON}) such that we can perform RowPress. Ensuring that a DRAM row remains open for a large \DRAMTIMING{AggON} value is not straightforward because we do not have fine-grain\gf{ed} control over the timing parameters used and the command sequences scheduled by the memory controller in {a} real system ({in contrast} to our real chip characterization setup {in~\secref{subsec:methodology_infra}}). However, carefully-designed access patterns can {make} the memory controller {keep} the DRAM row open for a long duration. For example, if a DRAM row is open, the memory controller can serve memory requests that target different cache blocks in the row at high data transfer rates~\cite{jedec2017ddr4}. Therefore, if an access pattern issues memory requests to different cache blocks in the {\emph{same}} DRAM row, we hypothesize that the memory controller will keep the DRAM row open to serve subsequent memory requests in the access pattern (we verify this hypothesis in \secref{sec:real_verify}).}

\noindent{\textbf{Test Program.}} {Algorithm~\ref{alg:rowpress-program} shows {the key part} of our} test program. {We mark the input parameters of the program in red.} {To overcome the first challenge, the program is} {based on an} access pattern described in~\cite{hassan2021utrr}, {which can} induce {RowHammer} bitflips in the presence of TRR.
{This access pattern uses 16 dummy rows that are activated shortly after the aggressor rows\footnote{Dummy rows are placed at least 100 rows away from the victim row~\cite{hassan2021utrr} to ensure that {activating them does} not cause bitflips on the victim row.} to {prevent} the in-DRAM {TRR} mechanism {from detecting} the aggressor row activations~\cite{hassan2021utrr, frigo2020trrespass, jattke2022blacksmith, deridder2021smash}.} {To overcome the second challenge and use large \DRAMTIMING{AggON} values, we access multiple {(i.e., \texttt{NUM\_READS})} cache blocks in {each} aggressor row.} {In every iteration}, the access pattern 1) activates the two aggressor rows {adjacent to a victim row multiple {(i.e., \texttt{NUM\_AGGR\_ACTS} {in line 7})} times (i.e., performs double-sided RowPress {with varying \DRAMTIMING{AggON}})}, and 2) activates each of the 16 dummy rows four times {(line 17)}~\cite{hassan2021utrr}. 

\renewcommand{\lstlistingname}{Algorithm} %
\begin{lstlisting}[caption={RowPress test program.},captionpos=b,label={alg:rowpress-program}, basicstyle=\scriptsize]
   // find two neighboring aggressor rows based on physical address mapping
   AGGRESSOR1, AGGRESSOR2 = find_aggressor_rows(|\textcolor{red}{{VICTIM}}|);
   // initialize the aggressor and the victim rows
   initialize(|\textcolor{red}{{VICTIM}}|, 0x55555555);
   initialize(AGGRESSOR1, AGGRESSOR2, 0xAAAAAAAA);
   // Synchronize with refresh
   for (iter = 0 ; iter < |\textcolor{red}{{NUM\_ITER}}| ; iter++):
     for (i = 0 ; i < |\textcolor{red}{{NUM\_AGGR\_ACTS}}| ; i++):
       // access multiple cache blocks in each aggressor row
       // to keep the aggressor row open longer
       for (j = 0 ; j < |\textcolor{red}{{NUM\_READS}}| ; j++): *AGGRESSOR1[j];
       for (j = 0 ; j < |\textcolor{red}{{NUM\_READS}}| ; j++): *AGGRESSOR2[j];
       // flush the cache blocks of each aggressor row
       for (j = 0 ; j < |\textcolor{red}{{NUM\_READS}}| ; j++):
         clflushopt (AGGRESSOR1[j]); 
         clflushopt (AGGRESSOR2[j]);
       mfence ();
     activate_dummy_rows();
   record_bitflips[|\textcolor{red}{{VICTIM}}|] = check_bitflips(|\textcolor{red}{{VICTIM}}|);
\end{lstlisting}

{The test program first initializes} the victim and the aggressor rows using the same checkerboard data pattern we {evaluated} in our {DRAM chip} characterization studies (lines \atb{4}--\atb{5}). We use this data pattern as it is reported~\cite{kim2020revisiting} to have the highest average read disturbance error coverage across DDR4 chips from three manufacturers. 
Second, the test program executes one or multiple (depending on the \texttt{NUM\_READS} parameter)
memory load instructions targeting {different cache blocks of} each aggressor row {(lines {10, 11})}. {Executing multiple memory load instructions to different cache blocks keeps {an aggressor row} open for a long time, whereas switching between different aggressor rows} open{s} and close{s} the two {aggressor} rows as {they} are in the same bank {(\secref{sec:background})}. Third, the program executes one or multiple \texttt{clflushopt} instructions to flush the cache blocks of {each} aggressor row to DRAM (lines {13}--15). {{Doing so} ensures that subsequent memory accesses (i.e., using load instructions) to the aggressor rows will {access DRAM instead of processor caches.}} Fourth, the program executes an \texttt{mfence} instruction (line 16) to ensure that the data is fully flushed before any subsequent memory load instruction is executed~\cite{kim2014flipping}. {Fifth, the program {accesses the 16 dummy rows, four times each,} to bypass TRR (line 17).} For every victim row, we execute this access pattern {for 800K iterations (i.e., \texttt{NUM\_ITER=800K} in line 6)} {to gather statistically significant results} and record the bitflips in the victim row (line 18).

\noindent{\textbf{Methodology.}}
{{We run our program using \texttt{NUM\_AGGR\_ACTS\\=\{1,2,3,4\}}, and \texttt{NUM\_READS=\{{1,2,4,16,32,48,64,80,128}\}}}\footnote{A DRAM row in the module we test has 128 cache blocks.} on 1500 arbitrarily selected victim rows. To reduce experiment time, we do not test \texttt{NUM\_READS$>$48(80)} for \texttt{NUM\_AGGR\_ACTS=4(3)} because the access pattern would not fit in a \DRAMTIMING{REFI} window. We synchronize our access pattern with the refresh commands, {similarly} to prior works~\cite{deridder2021smash, jattke2022blacksmith}, to increase the chance of bypassing TRR.}

\begin{figure}[h]
    \centering
    \includegraphics[width=1.0\linewidth]{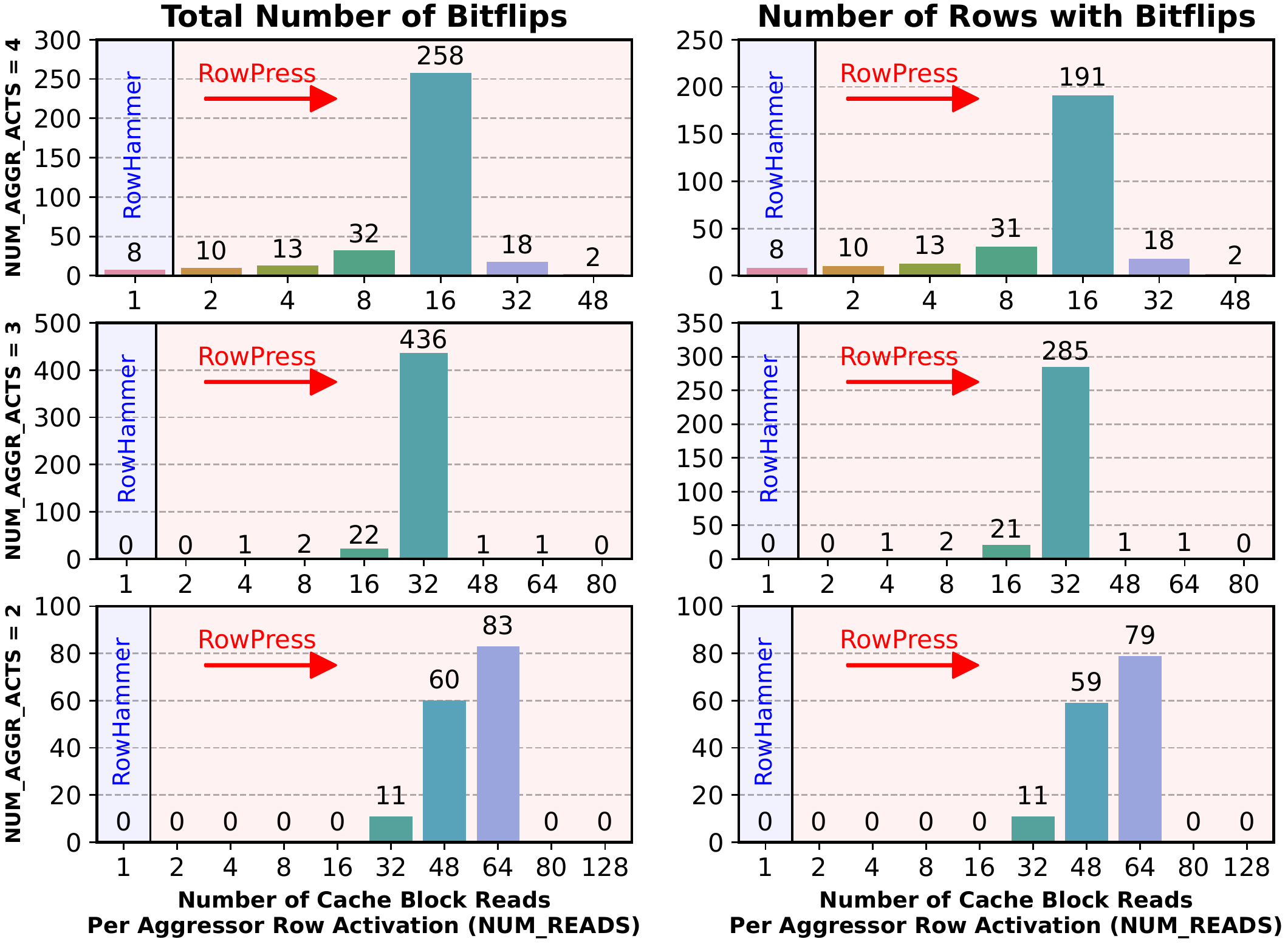}

    \caption{Number of RowHammer vs. RowPress bitflips (left) and number of rows with bitflips (right) we observe after running our test program {with four (top), three (middle), and two (bottom) activations per aggressor row per iteration}.}
    \label{fig:real_rowpress_bitflips}
    \vspace{-1em}
\end{figure}

\noindent{\textbf{Results.}} 
{{\figref{fig:real_rowpress_bitflips} shows the total number of bitflips (left) and the number of rows with bitflips (right) for different {number of cache blocks read per aggressor row activation} ({\texttt{NUM\_READS};} x-axis) when {we activate each aggressor row four (top {plots}), three (middle {plots}), and two (bottom {plots}) times per iteration}. We do not plot \texttt{NUM\_AGGR\_ACTS=1} because we do not observe any bitflips for all \texttt{NUM\_READS} we test.} {The leftmost bar in {each graph} shows the number of \atb{\emph{conventional RowHammer-induced}} bitflips, \atb{where we read \emph{only a single} cache block {per aggressor row activation}, as done in prior works that induce RowHammer bitflips (e.g., via proof-of-concept programs~\cite{kim2014flipping} and RowHammer attacks~\exploitingRowHammerAllCitationsExceptFlipping{}), \atb{such that the aggressor row is kept open for a short time}}. {Remaining} bars {in each graph}  show results for RowPress-induced bitflips (with \joel{an} increasing number of cache block reads {from left to right}, such that the \atb{aggressor} row is kept open for an increasing amount of time).}}

\observation{{Our test program leveraging RowPress induces bitflips when RowHammer cannot.}}

\observation{{Our test program leveraging RowPress induces many more bitflips compared to RowHammer, {at} the same aggressor row activation count.}}

{Our test program leveraging RowPress {induces} a significant number of bitflips in many DRAM rows while RowHammer \emph{cannot} induce \emph{any} bitflip when \texttt{NUM\_AGGR\_ACTS=\{2,3\}} (i.e., the program activates each aggressor row two/three times per iteration). {The program} induces up to 83 bitflips in 79 rows when \texttt{NUM\_AGGR\_ACTS=2} and \texttt{NUM\_READS=64} (i.e., the program reads 64 cache blocks per aggressor row activation), and up to 436 bitflips in 285 rows when \texttt{NUM\_AGGR\_ACTS=3} and \texttt{NUM\_READS=32}.} 

{When \texttt{NUM\_AGGR\_ACTS=4}, our test program leveraging RowPress induces significantly more bitflips compared to RowHammer. For example, the program induces up to 258 bitflips in 191 rows when \texttt{NUM\_READS=16}. In comparison, RowHammer induces only 8 bitflips in 8 rows with the same aggressor row activation count. }

\takeawaybox{{Leveraging RowPress, a user-level program 1) {induces} bitflips when RowHammer cannot{, and 2) induces many more} bitflips compared to RowHammer, at the same aggressor row activation count.}}

\observation{{In a real system, our test program does not always induce more bitflips as the number of cache blocks read per aggressor row activation increases.}}

{We observe that the number of bitflips and DRAM rows with bitflips first increases significantly as we increase \texttt{NUM\_READS}, but then decreases significantly after \texttt{NUM\_READS} reaches a certain point. For example, when \texttt{NUM\_AGGR\_ACTS=4}, the number of bitflips (rows with bitflips) keeps increasing from 8 (8) to 258 (191) as \texttt{NUM\_READS} increases from 1 to 16, but then decreases to 18 (18) when \texttt{NUM\_READS} is 32, and only 2 (2) when \texttt{NUM\_READS} is 48.}

{We attribute the increase in the number of bitflips and rows with bitflips when \texttt{NUM\_READS} increases to two reasons. First, the increase {in} \texttt{NUM\_READS} {causes} the memory controller keep the DRAM row open for a longer period of time, which leads to {an} increase in \DRAMTIMING{AggON}. Second, the increase of \texttt{NUM\_READS} reduces the activation frequency of the real aggressor rows compared to the dummy rows, which reduces the {probability} of real aggressor rows being detected by the TRR mechanism. We hypothesize that the {reasons} for the decrease in the number of bitflips and rows with bitflips after \texttt{NUM\_READS} increases beyond a certain value {are} that {1) the access pattern becomes too long, making it difficult to synchronize with the refresh commands, {and} 2) the activation frequency of the aggressor rows becomes too low to induce a large number of bitflips.}

{We conclude that, with a user-level program on \om{a} real DDR4-based {Intel} 
system with TRR protection, {1) RowPress induces bitflips when RowHammer cannot,} 2) RowPress induces many more bitflips than RowHammer{, and} 3) {increasing \DRAMTIMING{AggON} {up to a certain value} increases} RowPress{-induced} bitflips and number of rows with such bitflips. {Thus,} read-disturb-based attacks on real systems (e.g.,~\cite{frigo2020trrespass,jattke2022blacksmith}) can leverage RowPress to be more effective.}

{We investigate a variant of our RowPress test program that induces even more bitflips in more rows in Appendix ~\secref{sec:realvariant}.}

\subsection{Verifying {\DRAMTIMING{AggON} Increase}}
\label{sec:real_verify}
We assumed in our real system experiment in {the previous section} that {accessing different cache blocks in a DRAM row can keep the row open for a long time.} We now {briefly} describe how we verify that this is indeed the case. We develop a simple program that 1) flush{es} all cache blocks of {a} {tested} DRAM row from the processor's caches using \texttt{clflushopt} instructions\footnote{{We disable all hardware prefetchers of the processor by modifying model-specific register values~\cite{intel-swd} before running the verification program. Doing so, together with the \texttt{clflushopt} instructions that flushes all cache blocks in the tested DRAM row in the program, makes sure subsequent accesses to the remaining cache blocks (i.e., after accessing the first cache block) of the row are served from DRAM.}}, 2) accesses a different row in the same bank {as the tested row} to ensure that the memory controller sends a precharge command to close the open row, and 3) record{s} how many processor cycles it takes to access each cache block in the {tested} DRAM row. We run this program 100K times to collect statistically significant results.

{\atb{\figref{fig:real_system_access} shows the frequency histogram of latency values \atb{(observed using Intel time stamp counter~\cite{intel-swd})} for 1)~accessing the first cache block (green bars) and 2)~accessing the subsequent (i.e., the {remaining} 127) cache blocks (blue bars). We mark the median latency values for these two types of accesses with dashed red lines.}}

\begin{figure}[h]
    \centering
    \includegraphics[width=1.0\linewidth]{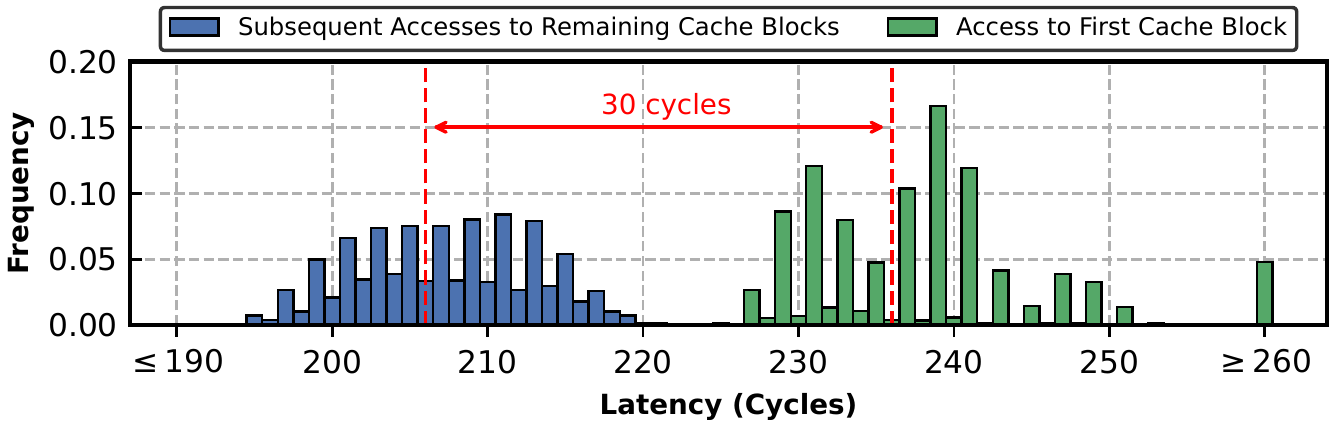}
    \caption{Histogram of the latency of the first and remaining cache block accesses to the same DRAM row.}
    \label{fig:real_system_access}
\end{figure}

{We observe that the {median} latency values of accessing the first cache block and the other cache blocks are 30 cycles apart.} Accessing the first cache block takes significantly {longer} than accessing other cache blocks. This happens because {the first access requires activation of the DRAM row but the remaining ones do not.} \revision{We conclude that, in the system we test, accessing consecutive cache blocks in {an activated} row causes the memory controller to keep the DRAM row open. Thus, existing memory controllers that behave similarly (e.g., using adaptive row buffer management policies{~\cite{IntelAPM, Awasthi2011Prediction, Rokicki2022Method, park2003history, Kahn2004method, Sander2005dynamic, xu2009prediction, mutlu2008parbs, rixner00, zuravleff1997controller}}) can facilitate future attacks leveraging RowPress.}

\section{Mitigating RowPress}
\label{sec:implications}

We examine four \om{potential} ways to mitigate RowPress bitflips: 1)~using error correcting codes (ECC), 2)~decoupling the row buffer from the opened DRAM row, 3)~limiting the maximum row-open time, and 4)~adapting existing RowHammer \om{mitigations} to account for RowPress. We believe the fourth way is the most effective among the four. {\secref{sec:ecc}, \secref{sec:rowbuffer_decoupling}, and \secref{sec:rowpolicy}} explain why the first three approaches are {either ineffective or undesirable} mitigations for RowPress. \secref{sec:graphere_para} describes and evaluates our proposed adaptations of RowHammer mitigations, using Graphene~\cite{park2020graphene} and PARA~\cite{kim2014flipping} as examples. Appendix~\secref{sec:extended_implications} provides detailed evaluation results with more benchmarks, {analyses}, and graphs.

\subsection{Error Correcting \om{Codes} (ECC)}
\label{sec:ecc}

{We} examine the {capability} of {ECC, which is widely used in {modern memory systems} to correct memory errors,} {in mitigating} RowPress{. W}e analyze the number of bitflips in \atb{every} 64-bit word for both single- and double-sided RowPress {for a} \DRAMTIMING{AggON} {of} \om{\SI{7.8}{\micro\second}}. To \atb{maximize the number of bitflips at this \DRAMTIMING{AggON}}, we activate the aggressor row(s) as many times as possible within \gf{\SI{60}{\milli\second}} at $80^{\circ}C$.
\figref{fig:ECC} is a box-and-whiskers plot that shows the \atb{distribution of the} number of {erroneous} 64-bit words {with}
1)~{at most two} bitflips {(1--2)}, 2)~{at least three} and {at most eight} bitflips {(3--8)}, and 3)~{more than eight bitflips {($>$8)} \atb{across all tested modules from every manufacturer (x-axis)}.} 
\begin{figure}[h]
    \centering
    \includegraphics[width=1.0\linewidth]{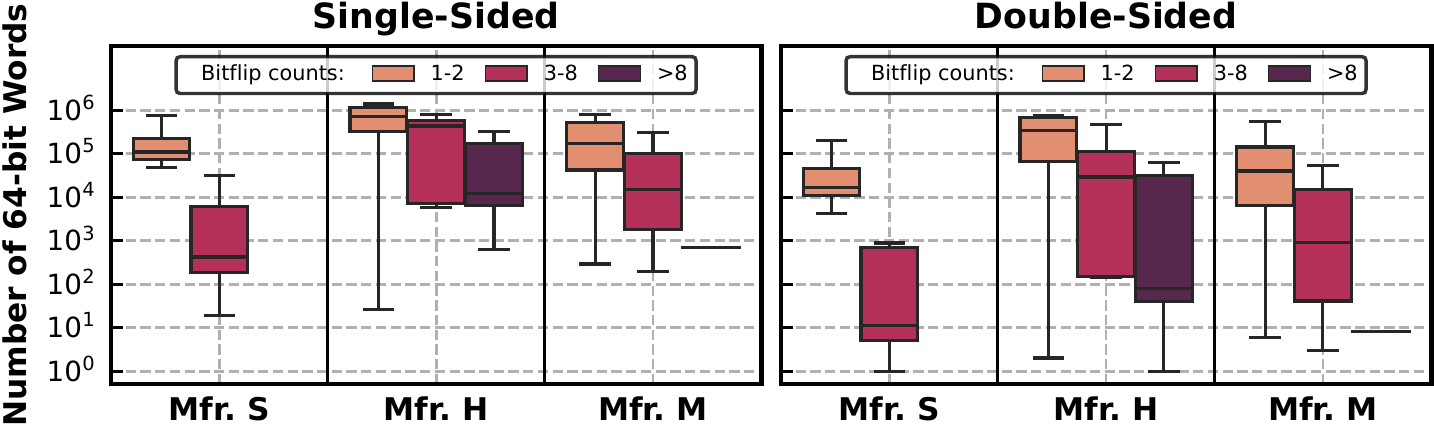}
    \caption{Number of 64-bit words with different bitflip counts for single-sided (left) and double-sided (right) RowPress.}
    \label{fig:ECC}
\end{figure}

\atb{We make two key observations from \hluo{our analysis}. First, there are up to \rb{\param{25}} RowPress bitflips {(not shown)} in a 64-bit \agy{data} word. ECC schemes that are widely used in memory systems (e.g., SECDED~\cite{hamming1950error} and Chipkill~\cite{dell1997white, locklear2000chipkill, memoryadvanced}\footnote{{Chipkill~\cite{dell1997white, locklear2000chipkill, memoryadvanced} can correct one-symbol errors and detect two-symbol errors. Because we observe up to 25 bitflips in a 64-bit data word, at least seven (four, two), symbols (i.e., data from seven, four, two DRAM chips, for x4, x8, and x16 chips, respectively) will be erroneous. Therefore, Chipkill \emph{cannot} provide guaranteed mitigation against RowPress.}}) cannot correct or detect \emph{all} RowPress bitflips {we observe}, which can lead to silent data corruption~\cite{fiala2012detection, George2021Demystifying, Michael2022Characterizing}. \om{Even a (7, 4) Hamming code (correcting one bitflip in a 4-bit data word)~\cite{hamming1950error} \agy{with} 75\% DRAM storage overhead (3 parity bits \yct{for }every 4 data bits), is not capable of correcting \rb{\param{25}} bitflips in a 64-bit data word. Other ECC schemes that can correct \emph{all} RowPress bitflips require prohibitively {large} storage overheads.} Thus, relying on ECC alone to prevent \emph{all} \yct{RowPress} bitflips is {a very expensive} solution.}
\atb{Second, for all three manufacturers \agy{(Mfrs. A, B, and C)}, a significant \agy{fraction (up to \rb{0.99\%, 35.77\%, and 10.08\%} \yct{for \DRAMTIMING{AggON} {$=$} \SI{7.8}{\micro\second}}, respectively)} of 64-bit \agy{data} words \agy{exhibit at least} three RowPress bitflips. This makes RowPress bitflips costly to prevent using techniques like \emph{memory page retirement} (where erroneous DRAM rows are {not used} in the system)~\cite{tang2006assessment, meza2015revisiting} {since such techniques} could render up to \rb{35.77}\% of storage capacity useless.}

{
\figref{fig:ECC_9REFI} shows the same distribution of the number of erroneous 64-bit words as \figref{fig:ECC} for \DRAMTIMING{AggON} $=$ \SI{70.2}{\micro\second}. We make similar observations and conclusions as for \figref{fig:ECC}.
}

\begin{figure}[h]
    \centering
    \includegraphics[width=1.0\linewidth]{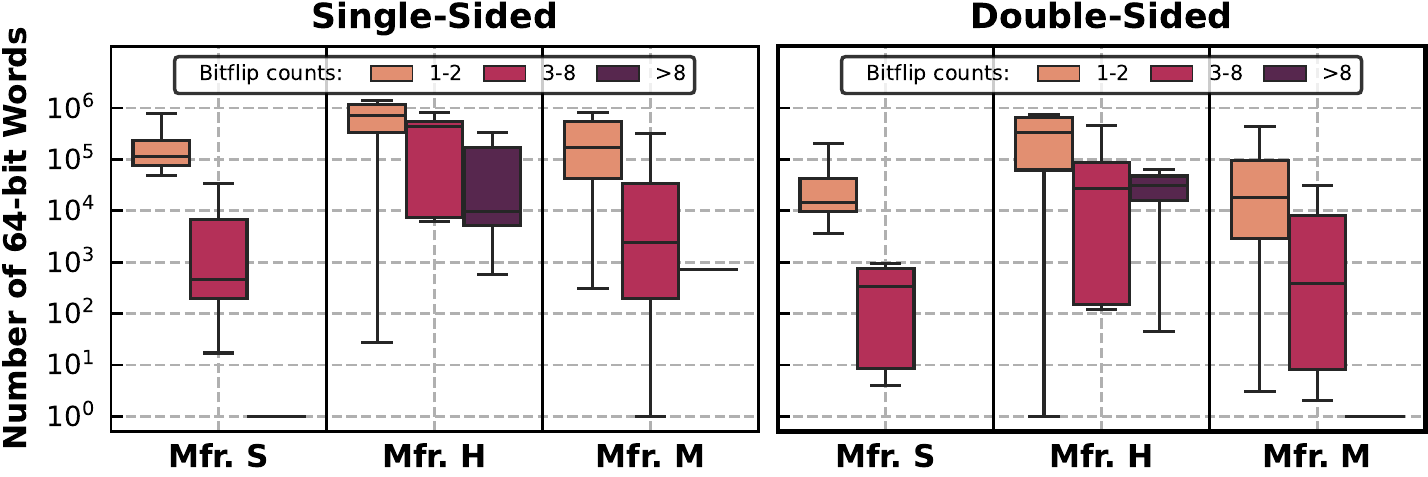}
    \caption{Number of 64-bit words with different bitflip counts for single-sided (left) and double-sided (right) RowPress when \DRAMTIMING{AggON} is \SI{70.2}{\micro\second}.}
    \label{fig:ECC_9REFI}
\end{figure}

\subsection{Decoupling the Row Buffer from the Row}
\label{sec:rowbuffer_decoupling}
\om{Prior works~\cite{O2014RowBuffer, Subramanian2018Closed} on improving DRAM performance and energy efficiency} propose to decouple the row buffer from the DRAM row \hluo{by disconnecting} the DRAM row from the row buffer {and de-asserting the wordline} once {the} charge restoration {process} is completed {after} row activation. Doing so can \emph{potentially} aid with RowPress mitigation because it limits \DRAMTIMING{AggON} to the minimum possible value (\DRAMTIMING{RAS}) regardless of the number of read requests sent to the DRAM row. {However, {there are at least three issues with this solution. First, it requires non-trivial changes in {cost-sensitive DRAM chips.} Second, to prevent write requests from increasing \DRAMTIMING{AggON}, the row needs to be reconnected to the row buffer ({by re-asserting} the wordline) only for the last write request, which further complicates {DRAM} chip design and memory controller request scheduling~\cite{Subramanian2018Closed}. Third,} row buffer decoupling does {\emph{not}} provide mitigation against {\emph{RowHammer}}. We leave {a} detailed evaluation of using row buffer decoupling to mitigate {RowPress} to future works.}

\revision{
\subsection{Limiting the Maximum Row-Open Time}
\label{sec:rowpolicy}
Since RowPress is caused by keeping a DRAM row open for a long period of time, limiting the \emph{maximum row-open time} ($t_{mro}$) by modifying the memory controller's row policy (i.e., forcing the closing of a row after $t_{mro}$ even if there are requests {in the memory controller} ready to be served from the opened row) 
may appear to be a mitigation for RowPress. However, it is \emph{not} {effective} because closing the row does \emph{not} mitigate the {read disturbance} already caused by the longer activation, unless $t_{mro}$ is set to its minimum possible value, \DRAMTIMING{RAS} (as we show in~\figref{fig:double_acmin_50C}, \gls{acmin} decreases as soon as \DRAMTIMING{AggON} is higher than \DRAMTIMING{RAS}). Having such a row policy that immediately closes an opened row after \DRAMTIMING{RAS} causes two issues. First, it may turn a benign workload with high row-buffer locality to a potential RowHammer attack as the same DRAM row {may} have to be activated more times. Second, it can cause large slowdown as it increases the average memory access latency by reducing {the} row buffer hit rate (up to $34.1\%$ single-core performance degradation, as we show in {Appendix}~\secref{sec:extended_rowpolicy}). {We show in \secref{sec:graphere_para} that mitigating RowPress is possible by co-designing a row policy that enforces $t_{mro}$ together with {an enhanced} RowHammer mitigation technique.}}

{Some existing row policy proposals adapt $t_{mro}$ based on row access patterns (e.g., keep the row open for longer when the row is predicted to be accessed soon in the future)~\cite{IntelAPM, Awasthi2011Prediction, Rokicki2022Method, park2003history, Kahn2004method, Sander2005dynamic, xu2009prediction}. Such row policies cannot mitigate RowPress as $t_{mro}$ can be controlled by an attacker to be set to larger values than \DRAMTIMING{RAS}, as we show in~\secref{sec:real}.
}

\subsection{\agy{\om{Adapting} Existing RowHammer \om{Mitigations}}}
\label{sec:graphere_para}
\noindent
\revision{\agyt{\textbf{Adaptation Methodology.}}
We propose a simple yet effective methodology to adapt existing RowHammer mitigation mechanisms to also mitigate RowPress with low \emph{additional} area overhead. 
The key idea is, based on device characterization (\secref{sec:characterization}, \secref{sec:sensitivity}), to
1)~quantify the worst-case (across different temperatures, access patterns, {and data patterns}) {read disturbance} caused by longer row-open time {and translate it} into an equivalent reduction in the RowHammer threshold ($T_{RH}$), {defined as the minimum number of aggressor row activations needed to cause a RowHammer bitflip,}
and 2)~{also} limit the maximum row-open time ($t_{mro}$).
For example, if the device characterization shows that for a \DRAMTIMING{AggON} of $X${ns}, \gls{acmin} reduces by {a maximum of} {$Y\%$}, then the adapted RowHammer mitigation mechanism will have $T^{'}_{RH} = (1-Y\%)T_{RH}$, and the memory controller must close the opened row after $X${ns} even if there are requests ready to be served by the row.}

\noindent
\revision{\agyt{\textbf{Security Analysis.}}
Assuming the original RowHammer mitigation is secure (i.e., it issues preventive {refreshes} to the victim rows before any DRAM row is activated $T_{RH}$ times within the refresh window) and the DRAM device is properly characterized {to uncover the worst-case RowPress vulnerability}, {our} adapted mitigation {mechanism} 1) still mitigates RowHammer because the adapted mitigation is more aggressive than the original mitigation (i.e., $T^{'}_{RH}$ is strictly smaller than $T_{RH}$), and  2)~mitigates RowPress {because} {the limited maximum row-open time ensures that at least $T^{'}_{RH}$ activations to a DRAM row are needed to induce RowPress bitflips, which the adapted mitigation already {properly} {prevents} (i.e., a preventive refresh is issued before a row is activated $T^{'}_{RH}$ times).}}

\noindent
\revision{\agyt{\textbf{Configuration and Evaluation.}}
\om{Our adaptation methodology is applicable to \agy{a wide range of} RowHammer mitigations. We demonstrate this by applying it to two} major ones: Graphene~\cite{park2020graphene}, a low performance overhead mechanism, and PARA~\cite{kim2014flipping}, a low area overhead mechanism. We denote the adapted versions of \agyh{Graphene and PARA} as Graphene\om{-RowPress (RP)} and PARA\om{-RowPress (RP)}, respectively. 
We use the characterization results of the 8Gb B-Die from Mfr. S to configure Graphene-RP and PARA-RP with a baseline $T_{RH}$ of 1K using the methodology provided in~\cite{park2020graphene, kim2014flipping}, as shown in Table~\ref{tab:gp_params}. We perform a sensitivity study of their {respective} performance overheads over Graphene and PARA\footnote{Measured by the weighted speedup~{\cite{snavely2000symbiotic, eyerman2008systemlevel}} of Graphene-RP (PARA-RP) normalized to Graphene (PARA).} with these configurations \gf{using} Ramulator~\cite{kim2016ramulator,ramulatorgithub} {with} a realistic baseline system {configuration}\footnote{\SI{4}{\giga\hertz} out-of-order core, dual-rank DDR4 DRAM~\cite{jedec2017ddr4}, FR-FCFS~\cite{rixner00, zuravleff1997controller} scheduling, open-row policy. 58 four-core multiprogrammed workloads from SPEC CPU2017~\cite{spec2017}, TPC-H~\cite{tpch}, and YCSB~\cite{ycsb}. {We find similar performance results for single-core workloads{, as shown in our extended version~\cite{rowpress-arxiv}}.}} and show the results in Table~\ref{tab:gp_params}.
}

\begin{table}[h!]
\centering
\setlength{\tabcolsep}{0.42em}
{
\caption{Graphene-RP and PARA-RP {evaluations}.}
\label{tab:gp_params}
\footnotesize
\begin{tabular}{ccccccc}
\toprule
$t_{mro}$ (ns)         & 36 (=\DRAMTIMING{RAS}) & 66  & 96  & 186   & 336    & 636    \\

\textbf{$T'_{RH}$}     & 1000 (=$T_{RH}$)  & 809    & 724   & 619   & 555    & 419    \\ 
\toprule

\textbf{Graphene-RP $T$}      & 333  & 269 & 241 & 206  & 185    & 139    \\ 
{Avg. Perf. Overhead} & 1.3\%   & -0.43\% & -0.63\% & -0.49\% & -0.14\% & 0.60\%    \\ 
{Max. Perf. Overhead} & 10.2\%  & 6.6\%   & 6.4\%   & 5.0\% & 5.0\%   & 4.6\%     \\ 
\midrule

\textbf{PARA-RP $p$}          & 0.034 & 0.042 & 0.047 & 0.054 & 0.061 & 0.079\\
{Avg. Perf. Overhead} & 3.2\%  & 3.6\%    & 4.5\%  & 6.0\% & 7.9\%  & 12.9\%     \\ 
{Max. Perf. Overhead} & 23.8\% & 13.4\%   & 13.1\% & 14.7\% & 19.4\% & 31.6\%   \\ 

\bottomrule

\end{tabular}}
\end{table}

\noindent
\agy{We make two {major} observations from the results. First,} Graphene-RP and PARA-RP can mitigate RowPress at low \emph{additional} performance overhead. Compared to Graphene (PARA), Graphene-RP (PARA-RP) has {an} average slowdown of only $-0.63\%$ ($3.2\%$) when $t_{mro}$ is $96$ns ($36$ns). {When $t_{mro}$ is $636$ns ($96$ns), Graphene-RP (PARA-RP) {causes} a maximum slowdown of only $4.6\%$ ($13.1\%$) over Graphene (PARA).} The reason for the small negative slowdowns (i.e., speedups) is that some $t_{mro}$ values improve fairness between cores in a way that increases weighted speedups {(similar to~\cite{mutlu2007stall, mutlu2008parbs})}.
\agyh{Second, the performance overheads of Graphene-RP and PARA-RP change differently with different $t_{mro}$ configurations.}
For Graphene-RP, having {a} $t_{mro}$ {value that is} either smaller or larger than $96$ns increases the performance overhead. 
\agyh{This is because row-buffer locality reduces at a {smaller} $t_{mro}$, and more preventive refreshes are issued at a {larger} $t_{mro}$. For PARA-RP, any $t_{mro}$ {value} larger than $36$ns increases the performance overhead. The reason is that PARA\agyh{'s performance overhead} does \agyh{\emph{not}} scale well with {smaller} $T'_{RH}$~\cite{kim2020revisiting, park2020graphene, yaglikci2021blockhammer}, {and} thus the benefit of longer row-open time is outweighed by the performance overhead of {more} preventive refreshes. We conclude that existing RowHammer mitigations can be {relatively} easily adapted to mitigate RowPress at low additional performance overhead. {We expect future work to introduce new mitigation mechanisms, as it has been happening analogously for RowHammer.}}

We provide {more} evaluations and {analyses} of our proposed mitigation {mechanisms} in Appendix~\secref{sec:extended_graphene_para}.

\section{Related Work}
\label{sec:related_work}

\hluo{To our knowledge, this is the first work to experimentally demonstrate and characterize RowPress, {\em a {widespread} read-disturb phenomenon in real DRAM chips}. Our analysis of RowPress {(especially in \secref{sec:relationship}, \secref{sec:sen_temperature} and \secref{sec:sen_acc_pattern}) shows} that RowPress is different from {RowHammer}. This section highlights the most relevant works.}

\noindent\textbf{RowHammer with Increased \DRAMTIMING{AggON}.} A recent experimental characterization of real DRAM chips~\cite{orosa2021deeper} and prior device-level studies~\cite{park2016experiments, yang2019trap} provide preliminary results on how increasing \DRAMTIMING{AggON} {\em by small amounts} affects RowHammer bitflips. These works {treat this phenomenon the same as RowHammer and} do \emph{not} {identify a DRAM read-disturb phenomenon \emph{different} from RowHammer} because they do \emph{not}: 1) test a wide range of \DRAMTIMING{AggON} values (only up to \gf{\SI{154.5}{\nano\second}} in~\cite{orosa2021deeper}, \gf{\SI{50}{\nano\second}} in~\cite{yang2019trap}, and \gf{\SI{72.5}{\nano\second}} in in~\cite{park2016experiments}, as opposed to up to \gf{\SI{30}{\milli\second}} in our work), 2) study {sensitivity of increased \DRAMTIMING{AggON} to} temperature and access pattern, and 3) study the properties of the bitflips they induce. As such, these works attribute the bitflips to RowHammer. {In contrast,} our work clearly shows that RowPress bitflips have almost no overlap with RowHammer bitflips {and thus RowPress is a different phenomenon from RowHammer}.

\noindent\textbf{RAS Clobber.} Two patents from {Micron}~\cite{ito2017Apparatus, Wolff2018wordline} \om{very briefly mention} a ``RAS Clobber'' effect similar to RowPress. \revision{They only describe RAS Clobber as ``the selected word line is driven to the active level continuously for a considerably long period''~\cite{ito2017Apparatus}, and ``stress applied to adjacent word lines by a word line being on for an extended duration''~\cite{Wolff2018wordline}. These patents do \emph{not} provide any evaluation, analysis or demonstration of this effect, and they do \emph{not} clearly distinguish this effect from RowHammer. We show through detailed real DRAM chip characterization that RowPress is different from RowHammer (\secref{sec:characterization}, \secref{sec:sensitivity}), and demonstrate that RowPress can be leveraged to induce bitflips in a real system  (\secref{sec:real}).}~\cite{ito2017Apparatus} describes a sampling-based {read disturbance} mitigation mechanism which they claim can handle both RowHammer and RAS Clobber. We introduce a general methodology {that adapts} existing RowHammer mitigation {mechanisms} to also mitigate RowPress (\secref{sec:graphere_para}).~\cite{Wolff2018wordline} proposes to {lower the wordline voltage after row activation and charge restoration to mitigate RAS Clobber. However, {it does} \emph{not} {demonstrate} that reduced wordline voltage {eliminates the {read disturbance} effect of increased \DRAMTIMING{AggON}}.}
Neither patent~\cite{ito2017Apparatus, Wolff2018wordline} evaluates or analyzes its proposed mitigation mechanisms {at the system-level}.

\noindent\textbf{{One-Location RowHammer.}} {A prior work~\cite{gruss2018another} proposes a single-sided RowHammer technique called ``One-Location Hammering'' that ``continuously {re-opens} the same DRAM row.'' However, it is unclear whether {the} bitflips {this work observes} are caused by increased \DRAMTIMING{AggON} or conventional single-sided RowHammer. {The} access pattern {in this work} does not consider on-die RowHammer mitigations (e.g., TRR~\cite{hassan2021utrr,frigo2020trrespass}), unlike our real-system demonstration (\secref{sec:real}).} 

\noindent\textbf{Other DRAM {Read Disturbance} {Mitigation Techniques}.}
Many works {(e.g., \cite{aichinger2015ddr, AppleRefInc, aweke2016anvil, kim2014flipping, kim2014architectural,son2017making, lee2019twice, you2019mrloc, seyedzadeh2018cbt, van2018guardion, konoth2018zebram, park2020graphene, yaglikci2021blockhammer, kang2020cattwo, bains2015row, bains2016distributed, bains2016row, brasser2017can, gomez2016dummy, jedec2017ddr4,hassan2019crow, devaux2021method, ryu2017overcoming, yang2016suppression, yang2017scanning, gautam2019row, yaglikci2021security, qureshi2021rethinking, greenfield2012throttling, saileshwar2022randomized, qureshi2022hydra, Wi2023shadow, Woo2023scalable})} propose \hluo{{techniques} to mitigate RowHammer bitflips}. None of these take RowPress into account.\footnote{{Two {recent} works~\cite{marazzi2022protrr, marazzi2023rega} discuss at a high level how to handle {increased vulnerability due to} small increases in \DRAMTIMING{AggON} {(as reported by~\cite{orosa2021deeper})} by modifying their proposed RowHammer mitigation mechanisms. However, {these works} do not evaluate {their} modified mechanisms.}} \hluo{We describe a methodology to adapt such {techniques} to mitigate both RowHammer and RowPress and evaluate it on two example {prior works}~\cite{kim2014flipping, park2020graphene} {(\secref{sec:graphere_para})}.}
\section{Conclusion}
\label{sec:conclusion}

We demonstrated and experimentally analyzed a widespread read-disturb phenomenon called RowPress in modern DRAM chips: keeping a row open for a long time disturbs physically nearby rows enough to cause bitflips. Our experimental characterization of 164 real DRAM chips reveals that RowPress 1) has a different underlying mechanism from the well-studied RowHammer {phenomenon}, 2) greatly amplifies DRAM's vulnerability to \gf{read} disturbance by reducing the number of activations to induce a bitflip by {one to two} orders of magnitude {(and in extreme cases to only a single activation)}, and 3) becomes worse as DRAM technology node size reduces. We demonstrate that a user-level program causes RowPress bitflips in a real system, even in the presence of in-DRAM read-disturb mitigation mechanisms, much more so than the bitflips RowHammer can induce. {We} describe a methodology to adapt existing read-disturb mitigation mechanisms {that only consider RowHammer {to also mitigate RowPress}}, enabling strong protection {against RowPress} with low \emph{additional performance overhead}. We hope that the findings reported in this work lead to further examination of and new solutions to the RowPress phenomenon at multiple levels of the computing stack. {To this end, we open source all our infrastructure, test programs, and raw data at~{\cite{rowpress-artifact-github}}.}

\section*{Acknowledgments}
{
We thank the anonymous reviewers of ISCA 2023 for feedback. We thank {the} SAFARI Research Group members for {valuable} feedback and the stimulating intellectual environment they provide. We acknowledge the generous gift funding provided by our industrial partners ({especially} Google, Huawei, Intel, Microsoft, VMware), which has been instrumental in enabling the decade-long research we have been conducting on read disturbance in DRAM in particular and memory systems in {general.} This work was in part supported by the {a Google Security and Privacy Research Award and the Microsoft Swiss Joint Research Center}.
}

\balance
{

  \let\OLDthebibliography\thebibliography
  \renewcommand\thebibliography[1]{
    \OLDthebibliography{#1}
    \setlength{\parskip}{0pt}
    \setlength{\itemsep}{0pt}
  }
  \bibliographystyle{IEEEtranS}

\bibliography{refs}

\begin{thebibliography}{100}
\providecommand{\url}[1]{#1}
\csname url@samestyle\endcsname
\providecommand{\newblock}{\relax}
\providecommand{\bibinfo}[2]{#2}
\providecommand{\BIBentrySTDinterwordspacing}{\spaceskip=0pt\relax}
\providecommand{\BIBentryALTinterwordstretchfactor}{4}
\providecommand{\BIBentryALTinterwordspacing}{\spaceskip=\fontdimen2\font plus
\BIBentryALTinterwordstretchfactor\fontdimen3\font minus
  \fontdimen4\font\relax}
\providecommand{\BIBforeignlanguage}[2]{{%
\expandafter\ifx\csname l@#1\endcsname\relax
\typeout{** WARNING: IEEEtranS.bst: No hyphenation pattern has been}%
\typeout{** loaded for the language `#1'. Using the pattern for}%
\typeout{** the default language instead.}%
\else
\language=\csname l@#1\endcsname
\fi
#2}}
\providecommand{\BIBdecl}{\relax}
\BIBdecl

\bibitem{aga2017good}
M.~T. Aga, Z.~B. Aweke, and T.~Austin, ``{When Good Protections Go Bad:
  Exploiting Anti-DoS Measures to Accelerate Rowhammer Attacks},'' in
  \emph{HOST}, 2017.

\bibitem{agarwal2018rowhammer}
S.~Agarwal, H.~Dixit, D.~Datta, M.~Tran, D.~Houssameddine, D.~Shum, and
  F.~Benistant, ``{Rowhammer for Spin Torque based Memory: Problem or not?}''
  in \emph{INTERMAG}, 2018.

\bibitem{aichinger2015ddr}
B.~Aichinger, ``{DDR Memory Errors Caused by Row Hammer},'' in \emph{HPEC},
  2015.

\bibitem{AppleRefInc}
{Apple Inc.}, ``{About the Security Content of Mac EFI Security Update
  2015-001},'' \url{https://support.apple.com/en-us/HT204934}, {2015}.

\bibitem{Awasthi2011Prediction}
M.~Awasthi, D.~W. Nellans, R.~Balasubramonian, and A.~Davis, ``{Prediction
  Based DRAM Row-Buffer Management in the Many-Core Era},'' in \emph{PACT},
  2011.

\bibitem{aweke2016anvil}
Z.~B. Aweke, S.~F. Yitbarek, R.~Qiao, R.~Das, M.~Hicks, Y.~Oren, and T.~Austin,
  ``{ANVIL: Software-Based Protection Against Next-Generation Rowhammer
  Attacks},'' in \emph{ASPLOS}, 2016.

\bibitem{bains2016row}
K.~Bains and J.~Halbert, ``{Row Hammer Monitoring Based on Stored Row Hammer
  Threshold Value},'' {U.S.}\ Patent 9384821, 2016.

\bibitem{bains2015row}
K.~Bains, J.~Halbert, C.~Mozak, T.~Schoenborn, and Z.~Greenfield, ``{Row Hammer
  Refresh Command},'' {U.S.}\ Patent 9117544, 2015.

\bibitem{bains2016distributed}
K.~S. Bains and J.~B. Halbert, ``{Distributed Row Hammer Tracking},'' {U.S.}\
  Patent 9299400, 2016.

\bibitem{barenghi2018software}
A.~Barenghi, L.~Breveglieri, N.~Izzo, and G.~Pelosi, ``{Software-Only Reverse
  Engineering of Physical DRAM Mappings for Rowhammer Attacks},'' in
  \emph{IVSW}, 2018.

\bibitem{baumann2005radiation}
R.~Baumann, ``{Radiation-Induced Soft Errors in Advanced Semiconductor
  Technologies},'' \emph{IEEE TDMR}, 2005.

\bibitem{bhattacharya2016curious}
S.~Bhattacharya and D.~Mukhopadhyay, ``{Curious Case of Rowhammer: Flipping
  Secret Exponent Bits Using Timing Analysis},'' in \emph{CHES}, 2016.

\bibitem{bhattacharya2018advanced}
{Bhattacharya, Sarani and Mukhopadhyay, Debdeep}, ``{Advanced Fault Attacks in
  Software: Exploiting the Rowhammer Bug},'' \emph{Fault Tolerant Architectures
  for Cryptography and Hardware Security}, 2018.

\bibitem{bosman2016dedup}
E.~Bosman, K.~Razavi, H.~Bos, and C.~Giuffrida, ``{Dedup Est Machina: Memory
  Deduplication as an Advanced Exploitation Vector},'' in \emph{S\&P}, 2016.

\bibitem{brasser2017can}
F.~Brasser, L.~Davi, D.~Gens, C.~Liebchen, and A.-R. Sadeghi, ``{Can't Touch
  This: Software-Only Mitigation Against Rowhammer Attacks Targeting Kernel
  Memory},'' in \emph{USENIX Security}, 2017.

\bibitem{burleson2016invited}
W.~Burleson, O.~Mutlu, and M.~Tiwari, ``{Invited: Who is the Major Threat to
  Tomorrow's Security? You, the Hardware Designer},'' in \emph{DAC}, 2016.

\bibitem{carre2018openssl}
S.~Carre, M.~Desjardins, A.~Facon, and S.~Guilley, ``{OpenSSL Bellcore's
  Protection Helps Fault Attack},'' in \emph{DSD}, 2018.

\bibitem{cohen2022hammerscope}
Y.~Cohen, K.~S. Tharayil, A.~Haenel, D.~Genkin, A.~D. Keromytis, Y.~Oren, and
  Y.~Yarom, ``{HammerScope: Observing DRAM Power Consumption Using
  Rowhammer},'' in \emph{CCS}, 2022.

\bibitem{cojocar2020rowhammer}
L.~Cojocar, J.~Kim, M.~Patel, L.~Tsai, S.~Saroiu, A.~Wolman, and O.~Mutlu,
  ``{Are We Susceptible to Rowhammer? An End-to-End Methodology for Cloud
  Providers},'' in \emph{S\&P}, 2020.

\bibitem{cojocar2019eccploit}
L.~Cojocar, K.~Razavi, C.~Giuffrida, and H.~Bos, ``{Exploiting Correcting
  Codes: On the Effectiveness of ECC Memory Against Rowhammer Attacks},'' in
  \emph{S\&P}, 2019.

\bibitem{ycsb}
B.~Cooper, A.~Silberstein, E.~Tam, R.~Ramakrishnan, and R.~Sears,
  ``{Benchmarking Cloud Serving Systems with {YCSB}},'' in \emph{SoCC}, 2010.

\bibitem{deridder2021smash}
F.~de~Ridder, P.~Frigo, E.~Vannacci, H.~Bos, C.~Giuffrida, and K.~Razavi,
  ``{SMASH}: {Synchronized} {Many-Sided} {Rowhammer} {Attacks} from
  {JavaScript},'' in \emph{{USENIX Security}}, 2021.

\bibitem{dell1997white}
T.~J. Dell, ``{A White Paper on the Benefits of Chipkill-Correct ECC for PC
  Server Main Memory},'' \emph{IBM Microelectronics Division}, 1997.

\bibitem{dennard1968dram}
R.~H. Dennard, ``{Field-Effect Transistor Memory},'' {U.S.}\ Patent 3387286,
  1968.

\bibitem{devaux2021method}
F.~Devaux and R.~Ayrignac, ``{Method and Circuit for Protecting a DRAM Memory
  Device from the Row Hammer Effect},'' {U.S.}\ Patent 10885966, 2021.

\bibitem{IntelAPM}
J.~M. Dodd, ``{Adaptive page management},'' {U.S.}\ Patent 7076617B2, 2005.

\bibitem{eyerman2008systemlevel}
S.~Eyerman and L.~Eeckhout, ``{System-Level Performance Metrics for
  Multiprogram Workloads},'' \emph{IEEE Micro}, 2008.

\bibitem{fahr2022frodo}
M.~Fahr~Jr, H.~Kippen, A.~Kwong, T.~Dang, J.~Lichtinger, D.~Dachman-Soled,
  D.~Genkin, A.~Nelson, R.~Perlner, A.~Yerukhimovich \emph{et~al.}, ``{When
  Frodo Flips: End-to-End Key Recovery on FrodoKEM via Rowhammer},''
  \emph{CCS}, 2022.

\bibitem{fiala2012detection}
D.~Fiala, F.~Mueller, C.~Engelmann, R.~Riesen, K.~Ferreira, and R.~Brightwell,
  ``{Detection and Correction of Silent Data Corruption for Large-Scale
  High-Performance Computing},'' in \emph{SC}, 2012.

\bibitem{fournaris2017exploiting}
A.~P. Fournaris, L.~Pocero~Fraile, and O.~Koufopavlou, ``{Exploiting Hardware
  Vulnerabilities to Attack Embedded System Devices: A Survey of Potent
  Microarchitectural Attacks},'' \emph{Electronics}, 2017.

\bibitem{frigo2018grand}
P.~Frigo, C.~Giuffrida, H.~Bos, and K.~Razavi, ``{Grand Pwning Unit:
  Accelerating Microarchitectural Attacks with the GPU},'' in \emph{S\&P},
  2018.

\bibitem{frigo2020trrespass}
P.~Frigo, E.~Vannacci, H.~Hassan, V.~van~der Veen, O.~Mutlu, C.~Giuffrida,
  H.~Bos, and K.~Razavi, ``{TRRespass: Exploiting the Many Sides of Target Row
  Refresh},'' in \emph{{S\&P}}, 2020.

\bibitem{gautam2019row}
S.~Gautam, S.~Manhas, A.~Kumar, M.~Pakala, and E.~Yieh, ``{Row Hammering
  Mitigation Using Metal Nanowire in Saddle Fin DRAM},'' \emph{IEEE TED}, 2019.

\bibitem{genssler2022reliability}
P.~R. Genssler, V.~M. van Santen, J.~Henkel, and H.~Amrouch, ``{On the
  Reliability of FeFET On-Chip Memory},'' \emph{TC}, 2022.

\bibitem{ghose2019demystifying}
S.~Ghose, T.~Li, N.~Hajinazar, D.~S. Cali, and O.~Mutlu, ``{Demystifying
  Complex Workload--DRAM Interactions: An Experimental Study},'' in
  \emph{{SIGMETRICS}}, 2019.

\bibitem{gomez2016dummy}
H.~{Gomez}, A.~{Amaya}, and E.~{Roa}, ``{{DRAM} Row-Hammer Attack Reduction
  Using Dummy Cells},'' in \emph{NORCAS}, 2016.

\bibitem{greenfield2012throttling}
Z.~Greenfield and T.~Levy, ``{Throttling Support for Row-Hammer Counters},''
  {U.S.\ Patent 9251885}, 2016.

\bibitem{gruss2018another}
D.~Gruss, M.~Lipp, M.~Schwarz, D.~Genkin, J.~Juffinger, S.~O'Connell,
  W.~Schoechl, and Y.~Yarom, ``{Another Flip in the Wall of Rowhammer
  Defenses},'' in \emph{S\&P}, 2018.

\bibitem{gruss2016rowhammer}
D.~Gruss, C.~Maurice, and S.~Mangard, ``{Rowhammer.js: A Remote
  Software-Induced Fault Attack in Javascript},'' arXiv:1507.06955 [cs.CR],
  2016.

\bibitem{rowhammer-js}
{Gruss, Daniel and Maurice, Clementine and Mangard, Stefan}, ``{Rowhammer.js: A
  Remote Software-Induced Fault Attack in JavaScript},'' \emph{arXiv:1507.06955
  [cs.CR]}, 2015.

\bibitem{hamming1950error}
R.~W. Hamming, ``{Error Detecting and Error Correcting Codes},'' \emph{The Bell
  System Technical Journal}, 1950.

\bibitem{hassan2019crow}
H.~{Hassan}, M.~{Patel}, J.~S. {Kim}, A.~G. {Ya\u{g}l{\i}k\c{c}{\i}},
  N.~{Vijaykumar}, N.~{Mansouri Ghiasi}, S.~{Ghose}, and O.~{Mutlu}, ``{CROW: A
  Low-Cost Substrate for Improving DRAM Performance, Energy Efficiency, and
  Reliability},'' in \emph{ISCA}, 2019.

\bibitem{hassan2021utrr}
H.~Hassan, Y.~C. Tugrul, J.~S. Kim, V.~v.~d. Veen, K.~Razavi, and O.~Mutlu,
  ``{Uncovering in-DRAM RowHammer Protection Mechanisms: A New Methodology,
  Custom RowHammer Patterns, and Implications},'' in \emph{MICRO}, 2021.

\bibitem{hassan2017softmc}
H.~Hassan, N.~Vijaykumar, S.~Khan, S.~Ghose, K.~Chang, G.~Pekhimenko, D.~Lee,
  O.~Ergin, and O.~Mutlu, ``{SoftMC: A Flexible and Practical Open-Source
  Infrastructure for Enabling Experimental DRAM Studies},'' in \emph{HPCA},
  2017.

\bibitem{hong2019terminal}
S.~Hong, P.~Frigo, Y.~Kaya, C.~Giuffrida, and T.~Dumitra\c{s}, ``{Terminal
  Brain Damage: Exposing the Graceless Degradation in Deep Neural Networks
  Under Hardware Fault Attacks},'' in \emph{USENIX Security}, 2019.

\bibitem{hong2023dsac}
S.~Hong, D.~Kim, J.~Lee, R.~Oh, C.~Yoo, S.~Hwang, and J.~Lee, ``{DSAC: Low-Cost
  Rowhammer Mitigation Using In-DRAM Stochastic and Approximate Counting
  Algorithm},'' arXiv:2302.03591, 2023.

\bibitem{horiguchi1997redundancy}
M.~Horiguchi, ``{Redundancy Techniques for High-Density DRAMs},'' in
  \emph{ISIS}, 1997.

\bibitem{intel-comet-lake}
{Intel}, ``{Intel Core i5-10400 Processor},''
  \url{https://ark.intel.com/content/www/us/en/ark/products/199271/intel-core-i510400-processor-12m-cache-up-to-4-30-ghz.html}.

\bibitem{intel-swd}
{{Intel}}, ``{Intel 64 and IA-32 Architectures Software Developer's Manual --
  Combined Volumes: 1, 2A, 2B, 2C, 2D, 3A, 3B, 3C, 3D and 4},''
  \url{https://www.intel.com/content/www/us/en/developer/articles/technical/intel-sdm.html},
  2022.

\bibitem{ito2017Apparatus}
Y.~Ito and Y.~He, ``{Apparatus and Methods for Refreshing Memory},'' {U.S.}\
  Patent 11062754B2, 2019.

\bibitem{itoh2013vlsi}
K.~Itoh, \emph{{VLSI Memory Chip Design}}.\hskip 1em plus 0.5em minus
  0.4em\relax Springer, 2001.

\bibitem{jang2017sgx}
Y.~Jang, J.~Lee, S.~Lee, and T.~Kim, ``{SGX-Bomb: Locking Down the Processor
  via Rowhammer Attack},'' in \emph{SOSP}, 2017.

\bibitem{jattke2022blacksmith}
P.~Jattke, V.~van~der Veen, P.~Frigo, S.~Gunter, and K.~Razavi, ``{Blacksmith:
  Scalable Rowhammering in the Frequency Domain},'' in \emph{SP}, 2022.

\bibitem{jedec2012ddr3}
{JEDEC}, \emph{{JESD79-3: DDR3 SDRAM Standard}}, 2012.

\bibitem{jedec2015lpddr4}
{{JEDEC}}, \emph{{JESD209-4B: Low Power Double Data Rate 4 (LPDDR4) Standard}},
  2017.

\bibitem{jedec2017ddr4}
{JEDEC}, \emph{{JESD79-4C: DDR4 SDRAM Standard}}, 2020.

\bibitem{jedec2020ddr5}
{{JEDEC}}, \emph{{JESD79-5: DDR5 SDRAM Standard}}, 2020.

\bibitem{ji2019pinpoint}
S.~Ji, Y.~Ko, S.~Oh, and J.~Kim, ``{Pinpoint Rowhammer: Suppressing Unwanted
  Bit Flips on Rowhammer Attacks},'' in \emph{ASIACCS}, 2019.

\bibitem{Kahn2004method}
O.~Kahn and J.~Wilcox, ``{Method for Dynamically Adjusting a Memory Page
  Closing Policy},'' 2004.

\bibitem{kang2020cattwo}
I.~Kang, E.~Lee, and J.~H. Ahn, ``{CAT-TWO: Counter-Based Adaptive Tree, Time
  Window Optimized for {DRAM} Row-Hammer Prevention},'' \emph{{IEEE} Access},
  2020.

\bibitem{keeth2001dram}
B.~Keeth and R.~Baker, \emph{{DRAM Circuit Design: A Tutorial}}.\hskip 1em plus
  0.5em minus 0.4em\relax Wiley, 2001.

\bibitem{khan2018analysis}
M.~N.~I. Khan and S.~Ghosh, ``{Analysis of Row Hammer Attack on STTRAM},'' in
  \emph{ICCD}, 2018.

\bibitem{khan2014efficacy}
S.~Khan, D.~Lee, Y.~Kim, A.~R. Alameldeen, C.~Wilkerson, and O.~Mutlu, ``{The
  Efficacy of Error Mitigation Techniques for DRAM Retention Failures: A
  Comparative Experimental Study},'' in \emph{SIGMETRICS}, 2014.

\bibitem{khan2016parbor}
S.~Khan, D.~Lee, and O.~Mutlu, ``{PARBOR: An Efficient System-Level Technique
  to Detect Data-Dependent Failures in DRAM},'' in \emph{DSN}, 2016.

\bibitem{khan2017detecting}
S.~Khan, C.~Wilkerson, Z.~Wang, A.~R. Alameldeen, D.~Lee, and O.~Mutlu,
  ``{Detecting and Mitigating Data-Dependent DRAM Failures by Exploiting
  Current Memory Content},'' in \emph{MICRO}, 2017.

\bibitem{kim2014architectural}
D.-H. Kim, P.~J. Nair, and M.~K. Qureshi, ``{Architectural Support for
  Mitigating Row Hammering in DRAM Memories},'' \emph{CAL}, 2015.

\bibitem{kim2020revisiting}
J.~S. Kim, M.~Patel, A.~G. Ya\u{g}l{\i}k\c{c}{\i}, H.~Hassan, R.~Azizi,
  L.~Orosa, and O.~Mutlu, ``{Revisiting RowHammer: An Experimental Analysis of
  Modern Devices and Mitigation Techniques},'' in \emph{ISCA}, 2020.

\bibitem{kim2014flipping}
Y.~{Kim}, R.~{Daly}, J.~{Kim}, C.~{Fallin}, J.~H. {Lee}, D.~{Lee},
  C.~{Wilkerson}, K.~{Lai}, and O.~{Mutlu}, ``{Flipping Bits in Memory Without
  Accessing Them: An Experimental Study of DRAM Disturbance Errors},'' in
  \emph{ISCA}, 2014.

\bibitem{kim2010atlas}
Y.~Kim, D.~Han, O.~Mutlu, and M.~Harchol-Balter, ``{ATLAS: A Scalable and
  High-Performance Scheduling Algorithm for Multiple Memory Controllers},'' in
  \emph{HPCA}, 2010.

\bibitem{kim2010thread}
Y.~Kim, M.~Papamichael, O.~Mutlu, and M.~Harchol-Balter, ``{Thread Cluster
  Memory Scheduling: Exploiting Differences in Memory Access Behavior},'' in
  \emph{MICRO}, 2010.

\bibitem{kim2016ramulator}
Y.~Kim, W.~Yang, and O.~Mutlu, ``{Ramulator: A Fast and Extensible DRAM
  Simulator},'' \emph{CAL}, 2016.

\bibitem{kogler2022half}
A.~Kogler, J.~Juffinger, S.~Qazi, Y.~Kim, M.~Lipp, N.~Boichat, E.~Shiu,
  M.~Nissler, and D.~Gruss, ``{Half-Double: Hammering From the Next Row
  Over},'' in \emph{USENIX Security}, 2022.

\bibitem{konoth2018zebram}
R.~K. Konoth, M.~Oliverio, A.~Tatar, D.~Andriesse, H.~Bos, C.~Giuffrida, and
  K.~Razavi, ``{ZebRAM: Comprehensive and Compatible Software Protection
  Against Rowhammer Attacks},'' in \emph{OSDI}, 2018.

\bibitem{kwong2020rambleed}
A.~Kwong, D.~Genkin, D.~Gruss, and Y.~Yarom, ``{RAMBleed: Reading Bits in
  Memory Without Accessing Them},'' in \emph{S\&P}, 2020.

\bibitem{Lantz1996Soft}
L.~Lantz, ``{Soft Errors Induced by Alpha Particles},'' in \emph{IEEE
  Transactions on Reliability}, 1996.

\bibitem{linux-kernel-540-131}
{Launchpad}, ``{linux 5.4.0-131.147 source package in Ubuntu},''
  \url{https://launchpad.net/ubuntu/+source/linux/5.4.0-131.147}, 2022.

\bibitem{lee2017design}
D.~Lee, S.~Khan, L.~Subramanian, S.~Ghose, R.~Ausavarungnirun, G.~Pekhimenko,
  V.~Seshadri, and O.~Mutlu, ``{Design-Induced Latency Variation in Modern DRAM
  Chips: Characterization, Analysis, and Latency Reduction Mechanisms},'' in
  \emph{SIGMETRICS}, 2017.

\bibitem{lee2019twice}
E.~Lee, I.~Kang, S.~Lee, G.~{Edward Suh}, and J.~{Ho Ahn}, ``{TWiCe: Preventing
  Row-Hammering by Exploiting Time Window Counters},'' in \emph{ISCA}, 2019.

\bibitem{li2014write}
H.~Li, H.-Y. Chen, Z.~Chen, B.~Chen, R.~Liu, G.~Qiu, P.~Huang, F.~Zhang,
  Z.~Jiang, B.~Gao, L.~Liu, X.~Liu, S.~Yu, H.-S.~P. Wong, and J.~Kang, ``{Write
  Disturb Analyses on Half-Selected Cells of Cross-Point RRAM Arrays},'' in
  \emph{IRPS}, 2014.

\bibitem{lim2017active}
C.~Lim, K.~Park, and S.~Baeg, ``{Active Precharge Hammering to Monitor
  Displacement Damage Using High-Energy Protons in 3x-nm SDRAM},'' \emph{TNS},
  2017.

\bibitem{lipp2018nethammer}
M.~Lipp, M.~T. Aga, M.~Schwarz, D.~Gruss, C.~Maurice, L.~Raab, and L.~Lamster,
  ``{Nethammer: Inducing Rowhammer Faults Through Network Requests},''
  arXiv:1805.04956 [cs.CR], 2018.

\bibitem{liu2013experimental}
J.~Liu, B.~Jaiyen, Y.~Kim, C.~Wilkerson, O.~Mutlu, J.~Liu, B.~Jaiyen, Y.~Kim,
  C.~Wilkerson, and O.~Mutlu, ``{An Experimental Study of Data Retention
  Behavior in Modern DRAM Devices},'' in \emph{ISCA}, 2013.

\bibitem{liu2012raidr}
J.~Liu, B.~Jaiyen, R.~Veras, and O.~Mutlu, ``{RAIDR: Retention-Aware
  Intelligent DRAM Refresh},'' in \emph{ISCA}, 2012.

\bibitem{liu2022generating}
L.~Liu, Y.~Guo, Y.~Cheng, Y.~Zhang, and J.~Yang, ``{Generating Robust DNN with
  Resistance to Bit-Flip based Adversarial Weight Attack},'' \emph{IEEE
  Transactions on Computers}, 2022.

\bibitem{locklear2000chipkill}
D.~Locklear, ``{Chipkill Correct Memory Architecture},'' \emph{{Dell Enterprise
  Systems Group, Technology Brief}}, 2000.

\bibitem{rowpress-arxiv}
H.~Luo, A.~Olgun, A.~G. Ya\u{g}l{\i}k\c{c}{\i}, Y.~C. Tu\u{g}rul, S.~Rhyner,
  M.~B. Cavlak, J.~Lindegger, M.~Sadrosadati, and O.~Mutlu, ``{RowPress:
  Amplifying Read Disturbance in Modern DRAM Chips},'' \emph{arXiv}, 2023.

\bibitem{marazzi2022protrr}
M.~Marazzi, P.~Jattke, F.~Solt, and K.~Razavi, ``{ProTRR}: {Principled} yet
  {Optimal} {In-DRAM} {Target Row Refresh},'' in \emph{{S\&P}}, 2022.

\bibitem{marazzi2023rega}
M.~Marazzi, F.~Solt, P.~Jattke, K.~Takashi, and K.~Razavi, ``{REGA}: {Scalable}
  {Rowhammer} {Mitigation} with {Refresh-Generating} {Activations},'' in
  \emph{{S\&P}}, 2023.

\bibitem{maxwellFT200}
{Maxwell}, ``{FT20X},''
  \url{https://www.maxwell-fa.com/upload/files/base/8/m/311.pdf}.

\bibitem{may1979alpha}
T.~May and M.~Woods, ``{Alpha-Particle-Induced Soft Errors in Dynamic
  Memories},'' in \emph{IEEE Transactions on Electron Devices}, 1979.

\bibitem{memoryadvanced}
I.~C. Memory, ``{Advanced ECC Memory for the IBM Netfinity 7000 M10},''
  \emph{Enhancing IBM Nethnity Server Reliability}.

\bibitem{meza2015revisiting}
J.~Meza, Q.~Wu, S.~Kumar, and O.~Mutlu, ``{Revisiting Memory Errors in
  Large-Scale Production Data Centers: Analysis and Modeling of New Trends from
  the Field},'' in \emph{{DSN}}, 2015.

\bibitem{moscibroda2007memory}
T.~Moscibroda and O.~Mutlu, ``{Memory Performance Attacks: Denial of Memory
  Service in Multi-Core Systems},'' in \emph{USENIX Security}, 2007.

\bibitem{mutlu2017rowhammer}
O.~Mutlu, ``{The RowHammer Problem and Other Issues We May Face as Memory
  Becomes Denser},'' in \emph{DATE}, 2017.

\bibitem{mutlu2019rowhammer}
O.~Mutlu and J.~S. Kim, ``{RowHammer: A Retrospective},'' \emph{TCAD}, 2019.

\bibitem{mutlu2007stall}
O.~Mutlu and T.~Moscibroda, ``{Stall-Time Fair Memory Access Scheduling for
  Chip Multiprocessors},'' in \emph{MICRO}, 2007.

\bibitem{mutlu2008parbs}
------, ``{Parallelism-Aware Batch Scheduling: Enhancing Both Performance and
  Fairness of Shared DRAM Systems},'' in \emph{ISCA}, 2008.

\bibitem{mutlu2022fundamentally}
O.~Mutlu, A.~Olgun, and A.~G. Ya{\u{g}}l{\i}k{\c{c}}{\i}, ``{Fundamentally
  Understanding and Solving RowHammer},'' in \emph{ASP-DAC}, 2023.

\bibitem{ni2018write}
K.~Ni, X.~Li, J.~A. Smith, M.~Jerry, and S.~Datta, ``{Write Disturb in
  Ferroelectric FETs and Its Implication for 1T-FeFET AND Memory Arrays},''
  \emph{IEEE EDL}, 2018.

\bibitem{Gorman1994cosmic}
T.~O'Gorman, in \emph{{The Effect of Cosmic Rays on the Soft Error Rate of a
  DRAM at Ground Level}}, 1994.

\bibitem{olgun2022drambender}
A.~Olgun, H.~Hassan, A.~G. Yaglikci, Y.~C. Tugrul, L.~Orosa, H.~Luo, M.~Patel,
  O.~Ergin, and O.~Mutlu, ``{DRAM Bender: An Extensible and Versatile
  FPGA-based Infrastructure to Easily Test State-of-the-art DRAM Chips},''
  arXiv:2211.05838, 2022.

\bibitem{orosa2022spyhammer}
L.~Orosa, U.~R{\"u}hrmair, A.~G. Yaglikci, H.~Luo, A.~Olgun, P.~Jattke,
  M.~Patel, J.~Kim, K.~Razavi, and O.~Mutlu, ``{SpyHammer: Using RowHammer to
  Remotely Spy on Temperature},'' arXiv:2210.04084, 2022.

\bibitem{orosa2021deeper}
L.~Orosa, A.~G. Ya{\u{g}}l{\i}k{\c{c}}{\i}, H.~Luo, A.~Olgun, J.~Park,
  H.~Hassan, M.~Patel, J.~S. Kim, and O.~Mutlu, ``{A Deeper Look into
  RowHammer's Sensitivities: Experimental Analysis of Real DRAM Chips and
  Implications on Future Attacks and Defenses},'' in \emph{MICRO}, 2021.

\bibitem{George2021Demystifying}
G.~Papadimitriou and D.~Gizopoulos, ``Demystifying the system vulnerability
  stack: Transient fault effects across the layers,'' in \emph{ISCA}, 2021.

\bibitem{park2014active}
K.~Park, S.~Baeg, S.~Wen, and R.~Wong, ``{Active-Precharge Hammering on a
  Row-Induced Failure in DDR3 SDRAMs Under 3x nm Technology},'' in \emph{IIRW},
  2014.

\bibitem{park2016experiments}
K.~Park, C.~Lim, D.~Yun, and S.~Baeg, ``{Experiments and Root Cause Analysis
  for Active-Precharge Hammering Fault in DDR3 SDRAM under 3xnm Technology},''
  \emph{Microelectronics Reliability}, 2016.

\bibitem{park2016statistical}
K.~Park, D.~Yun, and S.~Baeg, ``{Statistical Distributions of Row-Hammering
  Induced Failures in DDR3 Components},'' \emph{Microelectronics Reliability},
  2016.

\bibitem{park2003history}
S.-I. Park and I.-C. Park, ``{History-Based Memory Mode Prediction For
  Improving Memory Performance},'' in \emph{ISCAS}, 2003.

\bibitem{park2020graphene}
Y.~Park, W.~Kwon, E.~Lee, T.~J. Ham, J.~H. Ahn, and J.~W. Lee, ``{Graphene:
  Strong yet Lightweight Row Hammer Protection},'' in \emph{MICRO}, 2020.

\bibitem{patel2020beer}
M.~Patel, J.~Kim, T.~Shahroodi, H.~Hassan, and O.~Mutlu, ``{Bit-Exact ECC
  Recovery (BEER): Determining DRAM On-Die ECC Functions by Exploiting DRAM
  Data Retention Characteristics (Best Paper)},'' in \emph{{MICRO}}, 2020.

\bibitem{patel2017reaper}
M.~Patel, J.~S. Kim, and O.~Mutlu, ``{The Reach Profiler (REAPER): Enabling the
  Mitigation of DRAM Retention Failures via Profiling at Aggressive
  Conditions},'' in \emph{ISCA}, 2017.

\bibitem{pessl2016drama}
P.~Pessl, D.~Gruss, C.~Maurice, M.~Schwarz, and S.~Mangard, ``{DRAMA:
  Exploiting DRAM Addressing for Cross-CPU Attacks},'' in \emph{USENIX
  Security}, 2016.

\bibitem{poddebniak2018attacking}
D.~Poddebniak, J.~Somorovsky, S.~Schinzel, M.~Lochter, and P.~R{\"o}sler,
  ``{Attacking Deterministic Signature Schemes using Fault Attacks},'' in
  \emph{EuroS\&P}, 2018.

\bibitem{qiao2016new}
R.~Qiao and M.~Seaborn, ``{A New Approach for RowHammer Attacks},'' in
  \emph{HOST}, 2016.

\bibitem{qureshi2021rethinking}
M.~Qureshi, ``{Rethinking ECC in the Era of Row-Hammer},'' \emph{{DRAMSec}},
  2021.

\bibitem{qureshi2022hydra}
M.~Qureshi, A.~Rohan, G.~Saileshwar, and P.~J. Nair, ``{Hydra: Enabling
  Low-Overhead Mitigation of Row-Hammer at Ultra-Low Thresholds via Hybrid
  Tracking},'' in \emph{ISCA}, 2022.

\bibitem{rakin2022deepsteal}
A.~S. Rakin, M.~H.~I. Chowdhuryy, F.~Yao, and D.~Fan, ``{DeepSteal: Advanced
  Model Extractions Leveraging Efficient Weight Stealing in Memories},'' in
  \emph{SP}, 2022.

\bibitem{razavi2016flip}
K.~Razavi, B.~Gras, E.~Bosman, B.~Preneel, C.~Giuffrida, and H.~Bos, ``{Flip
  Feng Shui: Hammering a Needle in the Software Stack},'' in \emph{USENIX
  Security}, 2016.

\bibitem{rixner00}
S.~Rixner, W.~J. Dally, U.~J. Kapasi, P.~Mattson, and J.~D. Owens, ``{Memory
  Access Scheduling},'' in \emph{ISCA}, 2000.

\bibitem{Rokicki2022Method}
T.~G. Rokicki, ``{Method and computer system for speculatively closing pages in
  memory},'' {U.S.}\ Patent 6389514B1, 2002.

\bibitem{ryu2017overcoming}
S.-W. Ryu, K.~Min, J.~Shin, H.~Kwon, D.~Nam, T.~Oh, T.-S. Jang, M.~Yoo, Y.~Kim,
  and S.~Hong, ``{Overcoming the Reliability Limitation in the Ultimately
  Scaled DRAM using Silicon Migration Technique by Hydrogen Annealing},'' in
  \emph{IEDM}, 2017.

\bibitem{drambendergithub}
{SAFARI Research Group}, ``{DRAM Bender --- GitHub Repository},''
  \url{https://github.com/CMU-SAFARI/DRAM-Bender}.

\bibitem{ramulatorgithub}
{{SAFARI Research Group}}, ``{Ramulator --- GitHub Repository},''
  \url{https://github.com/CMU-SAFARI/ramulator}.

\bibitem{rowhammergithub}
{SAFARI Research Group}, ``{RowHammer --- GitHub Repository},''
  \url{https://github.com/CMU-SAFARI/rowhammer}.

\bibitem{rowpress-artifact-github}
{{SAFARI Research Group}}, ``{RowPress Artifact --- GitHub Repository},''
  \url{https://github.com/CMU-SAFARI/RowPress}.

\bibitem{softmcgithub}
------, ``{SoftMC --- GitHub Repository},''
  \url{https://github.com/CMU-SAFARI/softmc}.

\bibitem{rowpress-artifact}
{SAFARI Research Group}, ``{RowPress Artifact --- Zenodo Repository},''
  \url{https://doi.org/10.5281/zenodo.7750890}, 2023.

\bibitem{saileshwar2022randomized}
G.~Saileshwar, B.~Wang, M.~Qureshi, and P.~J. Nair, ``{Randomized Row-Swap:
  Mitigating Row Hammer by Breaking Spatial Correlation Between Aggressor and
  Victim Rows},'' in \emph{ASPLOS}, 2022.

\bibitem{samsung-real-datasheet}
{Samsung Electronics}, ``{288pin Unbuffered DIMM based on 8Gb C-die},''
  \url{https://download.semiconductor.samsung.com/resources/data-sheet/DDR4_8Gb_C_die_Unbuffered_DIMM_Rev1.4_Apr.18.pdf}.

\bibitem{Sander2005dynamic}
B.~Sander, P.~Madrid, and G.~Samus, ``{Dynamic Idle Counter Threshold Value for
  Use in Memory Paging Policy},'' 2005.

\bibitem{google-project-zero}
M.~Seaborn and T.~Dullien, ``{Exploiting the DRAM Rowhammer Bug to Gain Kernel
  Privileges},''
  \url{http://googleprojectzero.blogspot.com.tr/2015/03/exploiting-dram-rowhammer-bug-to-gain.html},
  2015.

\bibitem{seaborn2015exploiting}
{Seaborn, Mark and Dullien, Thomas}, ``{Exploiting the DRAM Rowhammer Bug to
  Gain Kernel Privileges},'' \emph{Black Hat}, 2015.

\bibitem{O2014RowBuffer}
O.~Seongil, Y.~H. Son, N.~S. Kim, and J.~H. Ahn, ``{Row-buffer Decoupling: A
  Case for Low-Latency DRAM Microarchitecture},'' in \emph{ISCA}, 2014.

\bibitem{seshadri2015gather}
V.~Seshadri, T.~Mullins, A.~Boroumand, O.~Mutlu, P.~B. Gibbons, M.~A. Kozuch,
  and T.~C. Mowry, ``{Gather-Scatter DRAM: In-DRAM Address Translation to
  Improve the Spatial Locality of Non-Unit Strided Accesses},'' in
  \emph{MICRO}, 2015.

\bibitem{seyedzadeh2018cbt}
S.~M. {Seyedzadeh}, A.~K. {Jones}, and R.~{Melhem}, ``{Mitigating Wordline
  Crosstalk Using Adaptive Trees of Counters},'' in \emph{ISCA}, 2018.

\bibitem{smith1981laser}
R.~T. Smith, J.~D. Chlipala, J.~F.~M. Bindels, R.~G. Nelson, F.~H. Fischer, and
  T.~F. Mantz, ``{Laser Programmable Redundancy and Yield Improvement in a 64K
  DRAM},'' \emph{JSSC}, 1981.

\bibitem{snavely2000symbiotic}
A.~Snavely and D.~M. Tullsen, ``{Symbiotic Job Scheduling for A Simultaneous
  Multithreaded Processor},'' in \emph{{ASPLOS}}, 2000.

\bibitem{son2017making}
M.~Son, H.~Park, J.~Ahn, and S.~Yoo, ``{Making DRAM Stronger Against Row
  Hammering},'' in \emph{DAC}, 2017.

\bibitem{spec2006}
{{Standard Performance Evaluation Corp.}}, ``{SPEC CPU 2006},''
  \url{http://www.spec.org/cpu2006/}.

\bibitem{spec2017}
{Standard Performance Evaluation Corp.}, ``{SPEC CPU 2017},''
  \url{http://www.spec.org/cpu2017/}.

\bibitem{subramanian2016bliss}
L.~Subramanian, D.~Lee, V.~Seshadri, H.~Rastogi, and O.~Mutlu, ``{BLISS:
  Balancing Performance, Fairness and Complexity in Memory Access
  Scheduling},'' \emph{TPDS}, 2016.

\bibitem{Subramanian2018Closed}
L.~Subramanian, K.~Vaidyanathan, A.~Nori, S.~Subramoney, T.~Karnik, and
  H.~Wang, ``{Closed yet Open DRAM: Achieving Low Latency and High Performance
  in DRAM Memory Systems},'' in \emph{DAC}, 2018.

\bibitem{Michael2022Characterizing}
M.~B. Sullivan, N.~R. Saxena, M.~O’Connor, D.~Lee, P.~Racunas, S.~Hukerikar,
  T.~Tsai, S.~K.~S. Hari, and S.~W. Keckler, ``{Characterizing and Mitigating
  Soft Errors in GPU DRAM},'' \emph{IEEE Micro}, 2022.

\bibitem{tang2006assessment}
D.~Tang, P.~Carruthers, Z.~Totari, and M.~Shapiro, ``{Assessment of the Effect
  of Memory Page Retirement on System RAS Against Hardware Faults},'' in
  \emph{DSN}, 2006.

\bibitem{tatar2018defeating}
A.~Tatar, C.~Giuffrida, H.~Bos, and K.~Razavi, ``{Defeating Software
  Mitigations Against Rowhammer: A Surgical Precision Hammer},'' in
  \emph{RAID}, 2018.

\bibitem{tatar2018throwhammer}
A.~Tatar, R.~K. Konoth, E.~Athanasopoulos, C.~Giuffrida, H.~Bos, and K.~Razavi,
  ``{Throwhammer: {Rowhammer} {Attacks} Over the {Network} and {Defenses}},''
  in \emph{{USENIX} {ATC}}, 2018.

\bibitem{hugepage-linux}
{The Linux Kernel Archives}, ``{Summary of Hugetlbpage Support},''
  \url{https://www.kernel.org/doc/Documentation/vm/hugetlbpage.txt}, 2022.

\bibitem{tobah2022spechammer}
Y.~Tobah, A.~Kwong, I.~Kang, D.~Genkin, and K.~G. Shin, ``{SpecHammer:
  Combining Spectre and Rowhammer for New Speculative Attacks},'' in \emph{SP},
  2022.

\bibitem{tol2022toward}
M.~C. Tol, S.~Islam, B.~Sunar, and Z.~Zhang, ``{Toward Realistic Backdoor
  Injection Attacks on DNNs using RowHammer},'' {arXiv:2110.07683v2 [cs.LG]},
  2022.

\bibitem{tpch}
{Transaction Processing Performance Council}, ``{TPC-H},''
  \url{https://www.tpc.org/tpch}.

\bibitem{tullsen2001handling}
D.~Tullsen and J.~Brown, ``{Handling Long-Latency Loads in a Simultaneous
  Multithreading Processor},'' in \emph{MICRO}, 2001.

\bibitem{vandegoor2002address}
A.~van~de Goor and I.~Schanstra, ``{Address and Data Scrambling: Causes and
  Impact on Memory Tests},'' in \emph{DELTA}, 2002.

\bibitem{van2016drammer}
V.~van~der Veen, Y.~Fratantonio, M.~Lindorfer, D.~Gruss, C.~Maurice, G.~Vigna,
  H.~Bos, K.~Razavi, and C.~Giuffrida, ``{Drammer: Deterministic Rowhammer
  Attacks on Mobile Platforms},'' in \emph{CCS}, 2016.

\bibitem{van2018guardion}
V.~van~der Veen, M.~Lindorfer, Y.~Fratantonio, H.~P. Pillai, G.~Vigna,
  C.~Kruegel, H.~Bos, and K.~Razavi, ``{GuardION: Practical Mitigation of
  DMA-Based Rowhammer Attacks on ARM},'' in \emph{{DIMVA}}, 2018.

\bibitem{walker2021ondramrowhammer}
A.~J. Walker, S.~Lee, and D.~Beery, ``{On DRAM RowHammer and the Physics on
  Insecurity},'' \emph{IEEE TED}, 2021.

\bibitem{weissman2020jackhammer}
Z.~Weissman, T.~Tiemann, D.~Moghimi, E.~Custodio, T.~Eisenbarth, and B.~Sunar,
  ``{JackHammer: Efficient Rowhammer on Heterogeneous FPGA--CPU Platforms},''
  arXiv:1912.11523 [cs.CR], 2020.

\bibitem{Wi2023shadow}
M.~Wi, J.~Park, S.~Ko, M.~J. Kim, N.~Sung~Kim, E.~Lee, and J.~H. Ahn,
  ``{SHADOW: Preventing Row Hammer in DRAM with Intra-Subarray Row
  Shuffling},'' in \emph{HPCA}, 2023.

\bibitem{Wolff2018wordline}
G.~D. Wolff, ``{Word Line Cache Mode},'' {U.S.}\ Patent 10366733B1, 2019.

\bibitem{Woo2023scalable}
J.~Woo, G.~Saileshwar, and P.~J. Nair, ``{Scalable and Secure Row-Swap:
  Efficient and Safe Row Hammer Mitigation in Memory Systems},'' in
  \emph{HPCA}, 2023.

\bibitem{xiao2016one}
Y.~Xiao, X.~Zhang, Y.~Zhang, and R.~Teodorescu, ``{One Bit Flips, One Cloud
  Flops: Cross-VM Row Hammer Attacks and Privilege Escalation},'' in
  \emph{USENIX Security}, 2016.

\bibitem{alveou200}
Xilinx, ``{Xilinx Alveo U200 FPGA Board},''
  \url{https://www.xilinx.com/products/boards-and-kits/alveo/u200.html}, 2021.

\bibitem{xu2009prediction}
Y.~Xu, A.~Agarwal, and B.~Davis, ``{Prediction in Dynamic SDRAM Controller
  Policies},'' in \emph{SAMOS}, 2009.

\bibitem{yaglikci2021security}
A.~G. Ya{\u{g}}l{\i}k{\c{c}}{\i}, J.~S. Kim, F.~Devaux, and O.~Mutlu,
  ``{Security Analysis of the Silver Bullet Technique for RowHammer
  Prevention},'' {arXiv:2106.07084 [cs.CR]}, 2021.

\bibitem{yaglikci2022understanding}
A.~G. Ya{\u{g}}l{\i}k{\c{c}}{\i}, H.~Luo, G.~F. Oliveira, A.~Olgun, M.~Patel,
  J.~Park, H.~Hassan, J.~S. Kim, L.~Orosa, and O.~Mutlu, ``{Understanding
  RowHammer Under Reduced Wordline Voltage: An Experimental Study Using Real
  DRAM Devices},'' in \emph{DSN}, 2022.

\bibitem{yaglikci2022hira}
A.~G. Ya{\u{g}}lik{\c{c}}i, A.~Olgun, M.~Patel, H.~Luo, H.~Hassan, L.~Orosa,
  O.~Ergin, and O.~Mutlu, ``{HiRA: Hidden Row Activation for Reducing Refresh
  Latency of Off-the-Shelf DRAM Chips},'' in \emph{MICRO}, 2022.

\bibitem{yaglikci2021blockhammer}
A.~G. Ya{\u{g}}l{\i}k{\c{c}}{\i}, M.~Patel, J.~S. Kim, R.~Azizibarzoki,
  A.~Olgun, L.~Orosa, H.~Hassan, J.~Park, K.~Kanellopoullos, T.~Shahroodi,
  S.~Ghose, and O.~Mutlu, ``{BlockHammer: Preventing RowHammer at Low Cost by
  Blacklisting Rapidly-Accessed DRAM Rows},'' in \emph{HPCA}, 2021.

\bibitem{yang2016suppression}
C.~Yang, C.~K. Wei, Y.~J. Chang, T.~C. Wu, H.~P. Chen, and C.~S. Lai,
  ``{Suppression of RowHammer Effect by Doping Profile Modification in
  Saddle-Fin Array Devices for Sub-30-nm DRAM Technology},'' \emph{IEEE
  Transactions on Device and Materials Reliability}, 2016.

\bibitem{yang2017scanning}
C.-M. Yang, C.-K. Wei, H.-P. Chen, J.-S. Luo, Y.~J. Chang, T.-C. Wu, and C.-S.
  Lai, ``{Scanning Spreading Resistance Microscopy for Doping Profile in
  Saddle-Fin Devices},'' \emph{IEEE Transactions on Nanotechnology}, 2017.

\bibitem{yang2019trap}
T.~Yang and X.-W. Lin, ``{Trap-Assisted DRAM Row Hammer Effect},'' \emph{EDL},
  2019.

\bibitem{yao2020deephammer}
F.~Yao, A.~S. Rakin, and D.~Fan, ``{DeepHammer: Depleting the Intelligence of
  Deep Neural Networks Through Targeted Chain of Bit Flips},'' in \emph{USENIX
  Security}, 2020.

\bibitem{you2019mrloc}
J.~M. You and J.-S. Yang, ``{MRLoc: Mitigating Row-Hammering Based on Memory
  Locality},'' in \emph{DAC}, 2019.

\bibitem{yun2018study}
D.~Yun, M.~Park, C.~Lim, and S.~Baeg, ``{Study of TID Effects on One Row
  Hammering using Gamma in DDR4 SDRAMs},'' in \emph{IRPS}, 2018.

\bibitem{zhang2018triggering}
Z.~Zhang, Z.~Zhan, D.~Balasubramanian, X.~Koutsoukos, and G.~Karsai,
  ``{Triggering Rowhammer Hardware Faults on ARM: A Revisit},'' in
  \emph{ASHES}, 2018.

\bibitem{zhang2020pthammer}
Z.~Zhang, Y.~Cheng, D.~Liu, S.~Nepal, Z.~Wang, and Y.~Yarom, ``{PThammer:
  Cross-User-Kernel-Boundary Rowhammer through Implicit Accesses},'' in
  \emph{MICRO}, 2020.

\bibitem{zhang2022implicit}
Z.~Zhang, W.~He, Y.~Cheng, W.~Wang, Y.~Gao, D.~Liu, K.~Li, S.~Nepal, A.~Fu, and
  Y.~Zou, ``{Implicit Hammer: Cross-Privilege-Boundary Rowhammer through
  Implicit Accesses},'' \emph{IEEE Transactions on Dependable and Secure
  Computing}, 2022.

\bibitem{zheng2022trojvit}
M.~Zheng, Q.~Lou, and L.~Jiang, ``{TrojViT: Trojan Insertion in Vision
  Transformers},'' arXiv:2208.13049, 2022.

\bibitem{zuravleff1997controller}
W.~K. Zuravleff and T.~Robinson, ``{Controller for a Synchronous DRAM That
  Maximizes Throughput by Allowing Memory Requests and Commands to Be Issued
  Out of Order},'' {U.S.}\ Patent 5630096, 1997.

\end{thebibliography}
}

\clearpage
\appendix
\nobalance

\section{Artifact {Description} Appendix}

  \titleformat{\subsubsection}%
    {\normalfont\normalsize\bfseries}{\thesubsubsection}{1em}{}%
      \titlespacing*{\subsubsection}{0pt}{\baselineskip}{0.3em}
  \NoIndentAfterEnv{subsubsection} 
  
\subsection{Abstract}

Our artifact{~\cite{rowpress-artifact, rowpress-artifact-github}} contains the data, source code, and scripts needed to reproduce our results, including all figures in the paper. We provide: 1) original characterization data from our real-chip characterization~(\secref{sec:characterization},~\secref{sec:sensitivity}) and source code of the DRAM Bender~{\cite{olgun2022drambender, drambendergithub}} program used to perform the characterization, 2) the source code of our real-system demonstration~(\secref{sec:real}), and 3) the source code of the Ramulator~\cite{kim2016ramulator, ramulatorgithub} implementation of our proposed RowPress mitigation~(\secref{sec:graphere_para}). We provide Python scripts and Jupyter Notebooks to analyze and plot the results for all three parts (referred to as \emph{Characterization}, \emph{Demonstration}, and \emph{Mitigation}, respectively).

\subsection{{Artifact Check-list (Meta-information)}}

\begin{table}[h]
\centering
{\fontsize{8}{10}\selectfont
\label{tab:appendix-checklist}
\resizebox{\columnwidth}{!}{%
\begin{tabular}{ll}
\textbf{Parameter}        & \textbf{Value}                     \\ \hline
Program &
  \begin{tabular}[c]{@{}l@{}}C++ program \\ Python3 scripts/Jupyter Notebooks\\ Shell scripts\end{tabular} \\ \hline
Compilation               & C++17 compiler (tested with GCC 9) \\ \hline
Run-time environment &
\begin{tabular}[c]{@{}l@{}}Ubuntu 20.04 (or similar) Linux \\ Ubuntu 18.04 (with Linux kernel 5.4.0-131-generic~\cite{linux-kernel-540-131}), used for \\ reproducing \emph{Demonstration} results \\ Python 3.9+\\ DRAM Bender~\cite{olgun2022drambender}\\ Boost 1.71+\\ Xilinx Vivado 2020.2+\\ Slurm 20+\end{tabular} \\ \hline
Hardware &
  \begin{tabular}[c]{@{}l@{}}x86 machine w/ PCIe 3.0 x16 slot\\ FPGA development board supported by DRAM Bender\\ (e.g., Xilinx Alveo U200)\\ Temperature control setup for DRAM modules under test\\ (e.g., Maxwell FT200)\end{tabular} \\ \hline
Output                    & Data and execution logs in plain text and plots in pdf and png format    \\ \hline
Experiment workflow &
  \begin{tabular}[c]{@{}l@{}}Perform characterizations (simulations), aggregate results, and \\ run analysis scripts on the results\end{tabular} \\ \hline
Experiment Customization  & Possible. See~\secref{sec:characterization-custom} for details      \\ \hline
Disk space requirement    & $\approx$ 1TB                 \\ \hline
Workflow preparation time & $\approx$ 1 day                    \\ \hline
Experiment completion time &
  \begin{tabular}[c]{@{}l@{}}$\approx$ 3 hours (Reproduce characterization figures with provided raw data) \\ {3 to 4} weeks per DRAM module (Replicate characterization results)\\  $\approx$ 5 days (Demonstration)\\ $\approx$ 1 day (Mitigation)\end{tabular} \\ \hline
Publicly available?       &  \begin{tabular}[c]{@{}l@{}} Zenodo (\url{https://doi.org/10.5281/zenodo.7750890}) \\ {Github (https://github.com/CMU-SAFARI/RowPress)}\end{tabular}       \\ \hline
Code licenses             & MIT                                \\ \hline
\end{tabular}%
}
}
\end{table}

\subsection{Description}

\subsubsection{How to Access}
The artifact is available on Zenodo with DOI \url{https://doi.org/10.5281/zenodo.7750890}. The live repository is at \url{https://github.com/CMU-SAFARI/RowPress}.

\subsubsection{Hardware Dependencies}
\label{sec:a-hwdependency}

\noindent{\textbf{Characterization. }} 
{To reproduce our real-DRAM characterization results (figures) using the provided raw data from our experiments, a Linux workstation with 1TB free disk space is required (the data size is about 800GB before compression). To replicate our results, the reader needs a similar setup as shown in~\figref{fig:hostmachine}:
\begin{itemize} 
    \item A host x86 machine with a PCIe 3.0 x16 slot.
    \item An FPGA board with a DIMM/SODIMM slot supported by DRAM Bender~\cite{olgun2022drambender, drambendergithub} (e.g., Xilinx Alveo U200~\cite{alveou200}).
    \item Heater pads attached to the DRAM module under test.
    \item A temperature controller (e.g., MaxWell FT200~\cite{maxwellFT200}) connected to the heater pads and programmable by the host machine.
\end{itemize}
}
\noindent{\textbf{Demonstration. }} 
{To reproduce our real-system demonstration of RowPress, the reader needs a system with an Intel Core i5 10400 (Comet Lake-S)~\cite{intel-comet-lake} processor and a Samsung \texttt{M378A2K43CB1-CTD} DDR4 DRAM module with {the 8Gb C-Dies from Mfr. S} (\texttt{K4A8G085WC-BCTD})~\cite{samsung-real-datasheet}. {We describe how to adapt our demonstration program to replicate our results on systems with a different processor and DRAM module in~\secref{sec:demo-custom}.}}

\noindent{\textbf{Mitigation. }} 
 The Ramulator~{\cite{kim2016ramulator, ramulatorgithub}} implementation of our proposed RowPress mitigation can be run on a {Linux} workstation. We recommend using a machine or a compute cluster with many CPU cores and large main memory to parallelize the simulation tasks.

\subsubsection{Software Dependencies}
\begin{itemize}
\item \texttt{GNU Make}, \texttt{CMake 3.10+}
\item \texttt{C++17} build toolchain (tested with \texttt{GCC 9})
\item \texttt{boost 1.71+}
\item  Xilinx \texttt{Vivado 2020.2+}
\item {\texttt{pigz} for fast decompression of raw characterization {data}}
\item \texttt{Python 3.9+} with Jupyter Notebook
\item \texttt{pip} packages: \texttt{pandas}, \texttt{scipy}, \texttt{matplotlib}, and \texttt{seaborn}
\item Slurm 20+
\item {Ubuntu 18.04 (Linux kernel 5.4.0-131-generic~\cite{linux-kernel-540-131}) for reproducing \emph{Demonstration}}

\end{itemize}

\subsection{Installation}
{To reproduce our results, no system-level installation is needed for \emph{Characterization} and \emph{Mitigation}. For \emph{Demonstration}, {1GB} hugepage {support} is required to simplify the process of finding neighboring DRAM rows in a real system.}

{To replicate our real-DRAM characterization, please follow the instructions in DRAM Bender's Github repository~\cite{drambendergithub} to install all dependencies to run DRAM Bender programs.}
\subsection{Experiment Workflow}
\label{sec:a-workflow}
\subsubsection{Characterization (Reproducing Figures)} 
{We describe how to reproduce all figures related to our real-DRAM characterization using the raw data from the artifact. For readers who wish to replicate our characterization results using their own infrastructure and DRAM modules, please see~\secref{sec:characterization-custom} for details.}

\begin{enumerate}
\item Extract raw characterization data ($\approx$ 800GB):
    \shellcmd{tar -I pigz -pxvf rowpress\_characterization\_data.tar.gz}

\item Process the raw data into pandas dataframes:
    \shellcmd{cd characterization/analysis/scripts}
    \shellcmd{DATA\_ROOT=<path-to-data>}            
    \shellcmd{./process\_data\_slurm.sh \$\{DATA\_ROOT\} }
\end{enumerate}

The processed characterization data will be placed at \texttt{characterization/analysis/processed\_data/}. To reproduce all figures related to \emph{Characterization}, open \texttt{characterization/analysis/plots/paper\_plots.ipynb} and run all {code blocks}. We use Markdown blocks in the notebook to clearly mark and explain all figures. The generated figures can be viewed both in the notebook and in \texttt{characterization/analysis/plots/output/}.

\subsubsection{Demonstration}
\begin{enumerate}
\item Build the demonstration program:
            \shellcmd{cd demonstration/}
            \shellcmd{make}
\item Run the program with root privilege (required only for {accessing the} hugepage) and analyze the bitflip results:
            \shellcmd{sudo ./mount\_hugepage.sh {\color{CadetBlue} \# Should print 1 if successful}}
            \shellcmd{sudo demo -{}-num\_victims 1500 > bitflips.txt}
            \shellcmd{python3 analyze.py bitflips.txt > parsed\_results.txt}
        
Open \texttt{real\_system\_bitflips.ipynb} and run all {code blocks} to analyze the results and reproduce~\figref{fig:real_rowpress_bitflips}.
\item Verify that \DRAMTIMING{AggON} increases (\secref{sec:real_verify}):
            \shellcmd{sudo ./disable\_prefetching.sh}
            \shellcmd{sudo demo -{}-verify}
        
Open \texttt{real\_system\_access.ipynb} and run all {code blocks} to reproduce \figref{fig:real_system_access}.
\end{enumerate}

\subsubsection{Mitigation}
Our artifact contains: 1) a modified version of Ramulator where we implement our proposed RowPress mitigation, 2) traces used to form workloads, and 3) {scripts} that automatically {generate} simulation configurations. The following instructions assume the reader is using Slurm to schedule a large number of parallelizable simulation jobs. Alternatively, readers can find the command lines for individual simulation jobs in the form of \texttt{mitigation/run\_cmds/<config>-<workload>.sh} after executing step 2 to be used for their own job scheduler.
\begin{enumerate}
\item Build ramulator:
            \shellcmd{cd mitigation/ramulator/}
            \shellcmd{./build.sh}
\item Generate simulation configurations and submit jobs:
            \shellcmd{python3 gen\_jobs.py}
            \shellcmd{./run.sh}
\end{enumerate}

Executing the above generates Ramulator statistics files from the simulations in \texttt{mitigation/results}. The reader can then open the \texttt{mitigation/analyze.ipynb} Jupyter notebook and run all {code blocks} to reproduce our results in~\tabref{tab:gp_params}.
\subsection{Evaluation and Expected Results}
Running each of the experiments described in~\secref{sec:a-workflow} is sufficient to reproduce all of 1) our real-chip characterization results (\figref{fig:hcf_intro},~\figref{fig:acmin_characterization} to~\figref{fig:taggonmin_temp_sweep},~\figref{fig:double_acmin_50C} to ~\figref{fig:dp_example_double},~\figref{fig:ftber}, and~\figref{fig:ECC}), 2) real-system demonstration of RowPress (\figref{fig:real_rowpress_bitflips} and~\figref{fig:real_system_access}), and 3) simulation results of our proposed RowPress mitigation (\tabref{tab:gp_params}).
\subsection{Experiment Customization}
\label{sec:custom}

\subsubsection{Characterization}
\label{sec:characterization-custom}
{The source code of our RowPress characterization program is at \texttt{characterization/DRAM-Bender/sources/apps/RowPress/}. {A} python script \texttt{characterization/run.py} {automates} the experiments. Note that this script is tightly coupled to our internal DRAM testing infrastructure to provide ad-hoc functionalities (e.g., experiment and infrastructure status book-keeping, communicating with the temperature controller). Readers who wish to replicate our characterization {on their own infrastructure} can modify \texttt{characterization/run\_bare.py}, which includes the infrastructure-independent experiment parameters, with \texttt{characterization/run.py} as a reference to perform the experiments on their own testing infrastructure. Performing all experiments for a single DRAM module takes about {three to four} weeks.}

Our RowPress characterization program is highly configurable to test different DRAM modules, data and access patterns, aggressor row activation counts, \DRAMTIMING{AggON}/\DRAMTIMING{AggOFF} values, etc. {Note that it is the responsibility of the reader's own DRAM testing infrastructure, not our characterization program, to control the temperature of the DRAM chips.} We explain {some key options} in~\tabref{tab:characterization-custom}, and encourage the reader to refer to the help messages of the program {for all options} and their explanations.

\begin{table}[h!]
\centering
\vspace{1em}
\caption{Key Options of RowPress Characterization Program}
\label{tab:characterization-custom}
\resizebox{\columnwidth}{!}{%
\begin{tabular}{@{}ll@{}}
\textbf{{Option}}        & \textbf{{Explanation}}                                                \\ \toprule
-{}-help & Print all available options and their explanations.                       \\ \midrule

\multirow{4}{*}{-{}-experiment} & 0 (Bitflips for given access pattern and activation count)                                                                                 \\
                & 1 (ACmin for given access pattern)                                 \\
                & 3 (Retention failures for given refresh-idle time)                 \\
                & 5 (Bitflips for given RowPress-ONOFF pattern and activation count) \\ \midrule
-{}-pattern\_file               & \begin{tabular}[c]{@{}l@{}}Path to a file specifying the data pattern and \\ spatial layout of the aggressor and victim rows.\end{tabular} \\ \midrule
-{}-hammer\_count & The number of activations per aggressor row.                       \\ \midrule
-{}-RAS\_scale    & The increase in \DRAMTIMING{AggON} beyond \DRAMTIMING{RAS} (1 unit = 30ns).                \\ \midrule
-{}-extra\_cycles               & \begin{tabular}[c]{@{}l@{}}$\Delta$ \DRAMTIMING{A2A} for the RowPress-ONOFF pattern {(1 unit = 6ns)}.\end{tabular}         \\ \midrule
-{}-RAS\_ratio    & Fraction of $\Delta$ \DRAMTIMING{A2A} that contributes to \DRAMTIMING{AggON}                   \\ \bottomrule
\end{tabular}%
}
\end{table}

\subsubsection{Demonstration}
\label{sec:demo-custom}
{On} the system described in~\secref{sec:a-hwdependency}, the reader can change the number of victim rows to be tested {using} the demonstration program with the command line option \texttt{-{}-num\_victims}. The number of cache blocks accessed per aggressor row activation can be configured by modifying the \texttt{no\_reads\_arr} array in line 635 of \texttt{main.cpp}.

To {successfully run} {the} demonstration program on a different system {(i.e., different processor and/or DRAM module)} {from that} described in~\secref{sec:a-hwdependency}, the reader needs {to perform} the {following}:

\begin{enumerate}
    \item Reverse engineer the DRAM address mapping of the memory controller of the processor.
    \item Obtain a baseline access pattern (e.g., using U-TRR~\cite{hassan2021utrr}) that can bypass the existing on-die RowHammer mitigation mechanism.
    \item Profile the system to obtain a threshold memory access latency that can be used to decide whether a DRAM refresh is happening (used to synchronize the access pattern with DRAM refresh).
\end{enumerate}

\noindent We explain {these steps} and how to modify the {demonstration} program in \texttt{demonstration/README.md}. 

\subsubsection{Mitigation}
The provided configurations can be evaluated with {user-provided} Ramulator traces. To include more traces in the job generation script, please modify the list of traces in \texttt{mitigation/gen\_jobs.py}.

\clearpage
\onecolumn
\begin{landscape}
\section{{\textbf{Summary Tables of RowPress and RowHammer Characteristics of All Tested DRAM Modules}}}
\label{sec:appendix}

\begin{table}[h]
\centering
\caption{{Summary of all tested DDR4 modules and their RowHammer/RowPress vulnerabilities {in terms of \gls{acmin} (\DRAMTIMING{AggON}$_{min}$).} We report the smaller \gls{acmin} (\DRAMTIMING{AggON}$_{min}$) we observe from single-sided and double-sided versions of RowPress and RowHammer.} \prearxiv{Recall that \gls{acmin} is the minimum number of total aggressor row activations needed to induce at least one bitflip, and \DRAMTIMING{AggON}$_{min}$ is the {\emph{minimum aggressor row on time (i.e., \DRAMTIMING{AggON})}} to induce at least one bitflip for a given aggressor row activation count.}}
\label{tab:extended-module-info}
\scriptsize
\setlength{\tabcolsep}{2pt}
\renewcommand*{\arraystretch}{1.3}
\begin{tabular}{@{}ccccccc|c|cccccccccc|@{}}
\toprule
 &
   &
   &
   &
   &
   &
   &
   &
  \multicolumn{1}{c|}{\textbf{\begin{tabular}[c]{@{}c@{}}RowHammer\\ Vulnerability\end{tabular}}} &
  \multicolumn{2}{c|}{\textbf{\begin{tabular}[c]{@{}c@{}}RowPress\\ Vulnerability\end{tabular}}} &
  \multicolumn{1}{c|}{\textbf{\begin{tabular}[c]{@{}c@{}}RowHammer\\ Vulnerability\end{tabular}}} &
  \multicolumn{6}{c|}{\textbf{\begin{tabular}[c]{@{}c@{}}RowPress\\ Vulnerability\end{tabular}}} \\ \cmidrule(l){9-18} 
 &
   &
   &
   &
   &
   &
   &
   &
  \multicolumn{6}{c|}{\textbf{\begin{tabular}[c]{@{}c@{}}\gls{acmin} @ Representative \DRAMTIMING{AggON}\\ Avg. (Min.)\end{tabular}}} &
  \multicolumn{4}{c|}{\textbf{\begin{tabular}[c]{@{}c@{}}\DRAMTIMING{AggON}$_{min}$ (@ Representative AC)\\ Avg. (Min.)\end{tabular}}} \\ \cmidrule(l){9-18} 
 &
   &
   &
   &
   &
   &
   &
   &
  \multicolumn{3}{c|}{\textbf{50$^{\circ}$C}} &
  \multicolumn{3}{c|}{\textbf{80$^{\circ}$C}} &
  \multicolumn{2}{c|}{\textbf{50$^{\circ}$C}} &
  \multicolumn{2}{c|}{\textbf{80$^{\circ}$C}} \\ \cmidrule(l){9-18} 
  \multirow{-7}{*}{\textbf{Mfr.}} &
  \multirow{-7}{*}{\textbf{DIMM Part}} &
  \multirow{-7}{*}{\textbf{DRAM Part}} &
  \multirow{-7}{*}{\textbf{\begin{tabular}[c]{@{}c@{}}Die\\ Rev.\tablefootnote{We report the die revision marked on the DRAM chip package (if available). A die revision of ``X'' means the original markings on the DRAM chips package are removed by the DRAM module vendor, and thus the die revision could not be identified.}\end{tabular}}} &
  \multirow{-7}{*}{\textbf{\begin{tabular}[c]{@{}c@{}}Die\\ Density\end{tabular}}} &
  \multirow{-7}{*}{\textbf{DQ}} &
  \multirow{-7}{*}{\textbf{\begin{tabular}[c]{@{}c@{}}Date\\ Code\tablefootnote{\prearxiv{In most cases, we report the date code of a DRAM module in the WW-YY format (i.e., 20-53 means the module is manufactured in the $53^{\text{rd}}$ week of year 2020) as marked on the label of the module. We report the date codes of modules S6 and S7 in the ``month-year'' format because it is the only date marked on the labels. We report ``N/A'' if no date is marked on the label of a module.}}\end{tabular}}} &
  \multirow{-7}{*}{\textbf{ID}} &
  \multicolumn{1}{c|}{\textbf{\begin{tabular}[c]{@{}c@{}}\DRAMTIMING{AggON}=36ns\\ (tRAS)\end{tabular}}} &
  \multicolumn{1}{c|}{\textbf{\begin{tabular}[c]{@{}c@{}}\DRAMTIMING{AggON}=7.8us\\ (tREFI)\end{tabular}}} &
  \multicolumn{1}{c|}{\textbf{\begin{tabular}[c]{@{}c@{}}\DRAMTIMING{AggON}=70.2us\\ (9xtREFI)\end{tabular}}} &
  \multicolumn{1}{c|}{\textbf{\begin{tabular}[c]{@{}c@{}}\DRAMTIMING{AggON}=36ns\\ (tRAS)\end{tabular}}} &
  \multicolumn{1}{c|}{\textbf{\begin{tabular}[c]{@{}c@{}}\DRAMTIMING{AggON}=7.8us\\ (tREFI)\end{tabular}}} &
  \multicolumn{1}{c|}{\textbf{\begin{tabular}[c]{@{}c@{}}\DRAMTIMING{AggON}=70.2us\\ (9xtREFI)\end{tabular}}} &
  \multicolumn{1}{c|}{\textbf{AC=1}} &
  \multicolumn{1}{c|}{\textbf{AC=10K}} &
  \multicolumn{1}{c|}{\textbf{AC=1}} &
  \textbf{AC=10K} \\ \midrule
 &
   &
   &
   &
   &
   &
   &
  \cellcolor[HTML]{DAE8FC}S0 &
  \multicolumn{1}{c|}{\cellcolor[HTML]{DAE8FC}279K (47K)} &
  \multicolumn{1}{c|}{\cellcolor[HTML]{DAE8FC}6.1K (1.6K)} &
  \multicolumn{1}{c|}{\cellcolor[HTML]{DAE8FC}682 (176)} &
  \multicolumn{1}{c|}{\cellcolor[HTML]{DAE8FC}295K (46K)} &
  \multicolumn{1}{c|}{\cellcolor[HTML]{DAE8FC}3.9K (776)} &
  \multicolumn{1}{c|}{\cellcolor[HTML]{DAE8FC}427 (87)} &
  \multicolumn{1}{c|}{\cellcolor[HTML]{DAE8FC}47.3 (12.4) ms} &
  \multicolumn{1}{c|}{\cellcolor[HTML]{DAE8FC}4.7 (1.3) us} &
  \multicolumn{1}{c|}{\cellcolor[HTML]{DAE8FC}24.8 (6.2) ms} &
  \cellcolor[HTML]{DAE8FC}2.5 (0.6) us \\
 &
  \multirow{-2}{*}{M393A1K43BB1-CTD} &
  \multirow{-2}{*}{K4A8G085WB-BCTD} &
  \multirow{-2}{*}{B} &
  \multirow{-2}{*}{8 Gb} &
  \multirow{-2}{*}{x8} &
  \multirow{-2}{*}{20-53} &
  S1 &
  \multicolumn{1}{c|}{262K (38K)} &
  \multicolumn{1}{c|}{6.3K (1.7K)} &
  \multicolumn{1}{c|}{700 (187)} &
  \multicolumn{1}{c|}{284K (41K)} &
  \multicolumn{1}{c|}{4.5K (808)} &
  \multicolumn{1}{c|}{486 (89)} &
  \multicolumn{1}{c|}{49.4 (14.1) ms} &
  \multicolumn{1}{c|}{4.9 (1.4) us} &
  \multicolumn{1}{c|}{29.0 (6.6) ms} &
  2.9 (0.8) us \\ \cmidrule(l){2-18} 
 &
  M378A2K43CB1-CTD &
  K4A8G085WC-BCTD &
  C &
  8 Gb &
  x8 &
  N/A &
  \cellcolor[HTML]{DAE8FC}S2 &
  \multicolumn{1}{c|}{\cellcolor[HTML]{DAE8FC}110K (24K)} &
  \multicolumn{1}{c|}{\cellcolor[HTML]{DAE8FC}6.4K (1.6K)} &
  \multicolumn{1}{c|}{\cellcolor[HTML]{DAE8FC}708 (179)} &
  \multicolumn{1}{c|}{\cellcolor[HTML]{DAE8FC}108K (23K)} &
  \multicolumn{1}{c|}{\cellcolor[HTML]{DAE8FC}5.3K (1.0K)} &
  \multicolumn{1}{c|}{\cellcolor[HTML]{DAE8FC}590 (107)} &
  \multicolumn{1}{c|}{\cellcolor[HTML]{DAE8FC}49.1 (13.0) ms} &
  \multicolumn{1}{c|}{\cellcolor[HTML]{DAE8FC}4.9 (1.3) us} &
  \multicolumn{1}{c|}{\cellcolor[HTML]{DAE8FC}33.9 (7.9) ms} &
  \cellcolor[HTML]{DAE8FC}3.4 (0.8) us \\ \cmidrule(l){2-18} 
 &
   &
   &
   &
   &
   &
   &
  S3 &
  \multicolumn{1}{c|}{41K (12K)} &
  \multicolumn{1}{c|}{5.7K (1.3K)} &
  \multicolumn{1}{c|}{627 (147)} &
  \multicolumn{1}{c|}{43K (15K)} &
  \multicolumn{1}{c|}{4.0K (835)} &
  \multicolumn{1}{c|}{447 (79)} &
  \multicolumn{1}{c|}{40.7 (11.4) ms} &
  \multicolumn{1}{c|}{4.1 (1.2) us} &
  \multicolumn{1}{c|}{23.4 (6.8) ms} &
  2.4 (0.7) us \\
 &
   &
   &
   &
   &
   &
   &
  \cellcolor[HTML]{DAE8FC}S4 &
  \multicolumn{1}{c|}{\cellcolor[HTML]{DAE8FC}42K (13K)} &
  \multicolumn{1}{c|}{\cellcolor[HTML]{DAE8FC}5.5K (1.0K)} &
  \multicolumn{1}{c|}{\cellcolor[HTML]{DAE8FC}606 (107)} &
  \multicolumn{1}{c|}{\cellcolor[HTML]{DAE8FC}42K (13K)} &
  \multicolumn{1}{c|}{\cellcolor[HTML]{DAE8FC}4.5K (721)} &
  \multicolumn{1}{c|}{\cellcolor[HTML]{DAE8FC}493 (81)} &
  \multicolumn{1}{c|}{\cellcolor[HTML]{DAE8FC}38.7 (9.6) ms} &
  \multicolumn{1}{c|}{\cellcolor[HTML]{DAE8FC}3.9 (1.0) us} &
  \multicolumn{1}{c|}{\cellcolor[HTML]{DAE8FC}26.9 (6.6) ms} &
  \cellcolor[HTML]{DAE8FC}2.7 (0.7) us \\
 &
  \multirow{-3}{*}{M378A1K43DB2-CTD} &
  \multirow{-3}{*}{K4A8G085WD-BCTD} &
  \multirow{-3}{*}{D} &
  \multirow{-3}{*}{8 Gb} &
  \multirow{-3}{*}{x8} &
  \multirow{-3}{*}{21-10} &
  S5 &
  \multicolumn{1}{c|}{41K (15K)} &
  \multicolumn{1}{c|}{5.5K (932)} &
  \multicolumn{1}{c|}{607 (98)} &
  \multicolumn{1}{c|}{43K (13K)} &
  \multicolumn{1}{c|}{4.2K (712)} &
  \multicolumn{1}{c|}{470 (77)} &
  \multicolumn{1}{c|}{38.7 (9.2) ms} &
  \multicolumn{1}{c|}{3.9 (1.0) us} &
  \multicolumn{1}{c|}{24.4 (5.5) ms} &
  2.5 (0.6) us \\ \cmidrule(l){2-18} 
 &
   &
   &
   &
   &
   &
   &
  \cellcolor[HTML]{DAE8FC}S6 &
  \multicolumn{1}{c|}{\cellcolor[HTML]{DAE8FC}116K (20K)} &
  \multicolumn{1}{c|}{\cellcolor[HTML]{DAE8FC}6.4K (1.4K)} &
  \multicolumn{1}{c|}{\cellcolor[HTML]{DAE8FC}703 (147)} &
  \multicolumn{1}{c|}{\cellcolor[HTML]{DAE8FC}117K (21K)} &
  \multicolumn{1}{c|}{\cellcolor[HTML]{DAE8FC}3.0K (450)} &
  \multicolumn{1}{c|}{\cellcolor[HTML]{DAE8FC}328 (51)} &
  \multicolumn{1}{c|}{\cellcolor[HTML]{DAE8FC}48.5 (15.0) ms} &
  \multicolumn{1}{c|}{\cellcolor[HTML]{DAE8FC}4.9 (1.7) us} &
  \multicolumn{1}{c|}{\cellcolor[HTML]{DAE8FC}17.7 (5.7) ms} &
  \cellcolor[HTML]{DAE8FC}1.8 (0.7) us \\
\multirow{-10}{*}{\textbf{\begin{tabular}[c]{@{}c@{}}Samsung\\ (Mfr. S)\end{tabular}}} &
  \multirow{-2}{*}{\begin{tabular}[c]{@{}c@{}}(G.Skill)\\ F4-2400C17S-8GNT\end{tabular}} &
  \multirow{-2}{*}{K4A4G085WF-BCTD} &
  \multirow{-2}{*}{F} &
  \multirow{-2}{*}{4 Gb} &
  \multirow{-2}{*}{x8} &
  \multirow{-2}{*}{Mar. 21} &
  S7 &
  \multicolumn{1}{c|}{129K (22K)} &
  \multicolumn{1}{c|}{5.9K (1.5K)} &
  \multicolumn{1}{c|}{651 (166)} &
  \multicolumn{1}{c|}{130K (22K)} &
  \multicolumn{1}{c|}{2.6K (682)} &
  \multicolumn{1}{c|}{291 (75)} &
  \multicolumn{1}{c|}{41.8 (13.5) ms} &
  \multicolumn{1}{c|}{4.2 (1.4) us} &
  \multicolumn{1}{c|}{14.4 (4.8) ms} &
  1.5 (0.5) us \\ \midrule
 &
   &
   &
   &
   &
   &
   &
  \cellcolor[HTML]{DAE8FC}H0 &
  \multicolumn{1}{c|}{\cellcolor[HTML]{DAE8FC}119K (21K)} &
  \multicolumn{1}{c|}{\cellcolor[HTML]{DAE8FC}6.1K (1.8K)} &
  \multicolumn{1}{c|}{\cellcolor[HTML]{DAE8FC}680 (200)} &
  \multicolumn{1}{c|}{\cellcolor[HTML]{DAE8FC}112K (26K)} &
  \multicolumn{1}{c|}{\cellcolor[HTML]{DAE8FC}1.7K (380)} &
  \multicolumn{1}{c|}{\cellcolor[HTML]{DAE8FC}190 (43)} &
  \multicolumn{1}{c|}{\cellcolor[HTML]{DAE8FC}46.2 (14.3) ms} &
  \multicolumn{1}{c|}{\cellcolor[HTML]{DAE8FC}4.6 (1.4) us} &
  \multicolumn{1}{c|}{\cellcolor[HTML]{DAE8FC}10.0 (3.0) ms} &
  \cellcolor[HTML]{DAE8FC}1.0 (0.3) us \\
 &
  \multirow{-2}{*}{HMAA4GU6AJR8N-XN} &
  \multirow{-2}{*}{H5ANAG8NAJR-XN} &
  \multirow{-2}{*}{A} &
  \multirow{-2}{*}{16 Gb} &
  \multirow{-2}{*}{x8} &
  \multirow{-2}{*}{20-51} &
  H1 &
  \multicolumn{1}{c|}{115K (24K)} &
  \multicolumn{1}{c|}{6.9K (2.4K)} &
  \multicolumn{1}{c|}{759 (268)} &
  \multicolumn{1}{c|}{108K (25K)} &
  \multicolumn{1}{c|}{2.7K (527)} &
  \multicolumn{1}{c|}{299 (65)} &
  \multicolumn{1}{c|}{53.5 (28.2) ms} &
  \multicolumn{1}{c|}{5.4 (2.9) us} &
  \multicolumn{1}{c|}{15.9 (5.6) ms} &
  1.6 (0.6) us \\ \cmidrule(l){2-18} 
 &
   &
   &
   &
   &
   &
   &
  \cellcolor[HTML]{DAE8FC}H2 &
  \multicolumn{1}{c|}{\cellcolor[HTML]{DAE8FC}77K (14K)} &
  \multicolumn{1}{c|}{\cellcolor[HTML]{DAE8FC}6.7K (2.8K)} &
  \multicolumn{1}{c|}{\cellcolor[HTML]{DAE8FC}736 (316)} &
  \multicolumn{1}{c|}{\cellcolor[HTML]{DAE8FC}75K (17K)} &
  \multicolumn{1}{c|}{\cellcolor[HTML]{DAE8FC}3.8K (959)} &
  \multicolumn{1}{c|}{\cellcolor[HTML]{DAE8FC}422 (105)} &
  \multicolumn{1}{c|}{\cellcolor[HTML]{DAE8FC}51.9 (25.4) ms} &
  \multicolumn{1}{c|}{\cellcolor[HTML]{DAE8FC}5.1 (2.5) us} &
  \multicolumn{1}{c|}{\cellcolor[HTML]{DAE8FC}22.0 (7.5) ms} &
  \cellcolor[HTML]{DAE8FC}2.3 (0.8) us \\
 &
  \multirow{-2}{*}{HMAA4GU7CJR8N-XN} &
  \multirow{-2}{*}{H5ANAG8NCJR-XN} &
  \multirow{-2}{*}{C} &
  \multirow{-2}{*}{16 Gb} &
  \multirow{-2}{*}{x8} &
  \multirow{-2}{*}{21-36} &
  H3 &
  \multicolumn{1}{c|}{78K (17K)} &
  \multicolumn{1}{c|}{6.7K (1.3K)} &
  \multicolumn{1}{c|}{7.8 (135)} &
  \multicolumn{1}{c|}{76K (16K)} &
  \multicolumn{1}{c|}{3.9K (739)} &
  \multicolumn{1}{c|}{426 (87)} &
  \multicolumn{1}{c|}{51.3 (9.8) ms} &
  \multicolumn{1}{c|}{5.1 (1.0) us} &
  \multicolumn{1}{c|}{22.6 (6.0) ms} &
  2.3 (0.6) us \\ \cmidrule(l){2-18} 
 &
  \begin{tabular}[c]{@{}c@{}}(Kingston)\\ KVR24R17S8/4\end{tabular} &
  H5AN4G8NAFR-UHC &
  A &
  4 Gb &
  x8 &
  19-46 &
  \cellcolor[HTML]{DAE8FC}H4 &
  \multicolumn{1}{c|}{\cellcolor[HTML]{DAE8FC}382K (83K)} &
  \multicolumn{1}{c|}{\cellcolor[HTML]{DAE8FC}No Bitflip} &
  \multicolumn{1}{c|}{\cellcolor[HTML]{DAE8FC}No Bitflip} &
  \multicolumn{1}{c|}{\cellcolor[HTML]{DAE8FC}373K (85K)} &
  \multicolumn{1}{c|}{\cellcolor[HTML]{DAE8FC}6.5K (2.7K)} &
  \multicolumn{1}{c|}{\cellcolor[HTML]{DAE8FC}719 (305)} &
  \multicolumn{1}{c|}{\cellcolor[HTML]{DAE8FC}No Bitflip} &
  \multicolumn{1}{c|}{\cellcolor[HTML]{DAE8FC}No Bitflip} &
  \multicolumn{1}{c|}{\cellcolor[HTML]{DAE8FC}50.8 (28.2) ms} &
  \cellcolor[HTML]{DAE8FC}5.1 (2.8) us \\ \cmidrule(l){2-18} 
\multirow{-9}{*}{\textbf{\begin{tabular}[c]{@{}c@{}}SK Hynix\\ (Mfr. H)\end{tabular}}} &
  \begin{tabular}[c]{@{}c@{}}(Corsair)\\ CMV4GX4M1A2133C15\end{tabular} &
  N/A &
  X &
  4 Gb &
  x8 &
  N/A &
  H5 &
  \multicolumn{1}{c|}{119K (20K)} &
  \multicolumn{1}{c|}{6.8K (2.4K)} &
  \multicolumn{1}{c|}{754 (259)} &
  \multicolumn{1}{c|}{116K (21K)} &
  \multicolumn{1}{c|}{2.3K (469)} &
  \multicolumn{1}{c|}{259 (53)} &
  \multicolumn{1}{c|}{53.5 (21.8) ms} &
  \multicolumn{1}{c|}{5.3 (2.2) us} &
  \multicolumn{1}{c|}{13.9 (4.1) ms} &
  1.4 (0.5) us \\ \midrule
 &
  MTA18ASF2G72PZ-2G3B1 &
  MT40A2G4WE-083E:B &
  B &
  8 Gb &
  x4 &
  N/A &
  \cellcolor[HTML]{DAE8FC}M0 &
  \multicolumn{1}{c|}{\cellcolor[HTML]{DAE8FC}386K (87K)} &
  \multicolumn{1}{c|}{\cellcolor[HTML]{DAE8FC}No Bitflip} &
  \multicolumn{1}{c|}{\cellcolor[HTML]{DAE8FC}No Bitflip} &
  \multicolumn{1}{c|}{\cellcolor[HTML]{DAE8FC}367K (80K)} &
  \multicolumn{1}{c|}{\cellcolor[HTML]{DAE8FC}No Bitflip} &
  \multicolumn{1}{c|}{\cellcolor[HTML]{DAE8FC}No Bitflip} &
  \multicolumn{1}{c|}{\cellcolor[HTML]{DAE8FC}No Bitflip} &
  \multicolumn{1}{c|}{\cellcolor[HTML]{DAE8FC}No Bitflip} &
  \multicolumn{1}{c|}{\cellcolor[HTML]{DAE8FC}No Bitflip} &
  \cellcolor[HTML]{DAE8FC}No Bitflip \\ \cmidrule(l){2-18} 
 &
   &
   &
   &
   &
   &
   &
  M1 &
  \multicolumn{1}{c|}{114K (24K)} &
  \multicolumn{1}{c|}{7.1K (3.7K)} &
  \multicolumn{1}{c|}{784 (403)} &
  \multicolumn{1}{c|}{105K (23K)} &
  \multicolumn{1}{c|}{6.3K (2.4K)} &
  \multicolumn{1}{c|}{689 (259)} &
  \multicolumn{1}{c|}{55.0 (35.2) ms} &
  \multicolumn{1}{c|}{5.5 (3.4) us} &
  \multicolumn{1}{c|}{44.5 (21.8) ms} &
  4.5 (1.8) us \\
 &
  \multirow{-2}{*}{MTA4ATF1G64HZ-3G2B2} &
  \multirow{-2}{*}{MT40A1G16RC-062E:B} &
  \multirow{-2}{*}{B} &
  \multirow{-2}{*}{16 Gb} &
  \multirow{-2}{*}{x16} &
  \multirow{-2}{*}{21-26} &
  \cellcolor[HTML]{DAE8FC}M2 &
  \multicolumn{1}{c|}{\cellcolor[HTML]{DAE8FC}118K (26K)} &
  \multicolumn{1}{c|}{\cellcolor[HTML]{DAE8FC}7.0K (5.5K)} &
  \multicolumn{1}{c|}{\cellcolor[HTML]{DAE8FC}785 (621)} &
  \multicolumn{1}{c|}{\cellcolor[HTML]{DAE8FC}110K (22K)} &
  \multicolumn{1}{c|}{\cellcolor[HTML]{DAE8FC}7.0K (3.5K)} &
  \multicolumn{1}{c|}{\cellcolor[HTML]{DAE8FC}781 (379)} &
  \multicolumn{1}{c|}{\cellcolor[HTML]{DAE8FC}58.4 (56.8) ms} &
  \multicolumn{1}{c|}{\cellcolor[HTML]{DAE8FC}5.9 (5.8) us} &
  \multicolumn{1}{c|}{\cellcolor[HTML]{DAE8FC}55.0 (28.2) us} &
  \cellcolor[HTML]{DAE8FC}5.5 (2.8) us \\ \cmidrule(l){2-18} 
 &
  MTA36ASF8G72PZ-2G9E1 &
  MT40A4G4JC-062E:E &
  E &
  16 Gb &
  x4 &
  20-14 &
  M3 &
  \multicolumn{1}{c|}{41K (10K)} &
  \multicolumn{1}{c|}{7.2K (2.4K)} &
  \multicolumn{1}{c|}{770 (310)} &
  \multicolumn{1}{c|}{39K (11K)} &
  \multicolumn{1}{c|}{4.8K (815)} &
  \multicolumn{1}{c|}{545 (91)} &
  \multicolumn{1}{c|}{53.3 (28.1) ms} &
  \multicolumn{1}{c|}{5.3 (2.8) us} &
  \multicolumn{1}{c|}{28.3 (9.8) ms} &
  2.8 (1.0) us \\ \cmidrule(l){2-18} 
 &
   &
   &
   &
   &
   &
   &
  \cellcolor[HTML]{DAE8FC}M4 &
  \multicolumn{1}{c|}{\cellcolor[HTML]{DAE8FC}36K (12K)} &
  \multicolumn{1}{c|}{\cellcolor[HTML]{DAE8FC}7.0K (2.2K)} &
  \multicolumn{1}{c|}{\cellcolor[HTML]{DAE8FC}746 (245)} &
  \multicolumn{1}{c|}{\cellcolor[HTML]{DAE8FC}34K (10K)} &
  \multicolumn{1}{c|}{\cellcolor[HTML]{DAE8FC}4.1K (925)} &
  \multicolumn{1}{c|}{\cellcolor[HTML]{DAE8FC}468 (102)} &
  \multicolumn{1}{c|}{\cellcolor[HTML]{DAE8FC}52.3 (17.1) ms} &
  \multicolumn{1}{c|}{\cellcolor[HTML]{DAE8FC}5.2 (1.7) us} &
  \multicolumn{1}{c|}{\cellcolor[HTML]{DAE8FC}25.1 (7.2) ms} &
  \cellcolor[HTML]{DAE8FC}2.5 (0.8) us \\
 &
  \multirow{-2}{*}{MTA4ATF1G64HZ-3G2E1} &
  \multirow{-2}{*}{MT40A1G16KD-062E:E} &
  \multirow{-2}{*}{E} &
  \multirow{-2}{*}{16 Gb} &
  \multirow{-2}{*}{x16} &
  \multirow{-2}{*}{20-46} &
  M5 &
  \multicolumn{1}{c|}{40K (11K)} &
  \multicolumn{1}{c|}{5.6K (1.2K)} &
  \multicolumn{1}{c|}{610 (127)} &
  \multicolumn{1}{c|}{37K (11K)} &
  \multicolumn{1}{c|}{2.6K (616)} &
  \multicolumn{1}{c|}{289 (67)} &
  \multicolumn{1}{c|}{34.6 (9.0) ms} &
  \multicolumn{1}{c|}{3.5 (0.9) us} &
  \multicolumn{1}{c|}{15.8 (4.6) ms} &
  1.6 (0.5) us \\ \cmidrule(l){2-18} 
\multirow{-10}{*}{\textbf{\begin{tabular}[c]{@{}c@{}}Micron\\ (Mfr. M)\end{tabular}}} &
  MTA4ATF1G64HZ-3G2F1 &
  MT40A1G16TB-062E:F &
  F &
  16 Gb &
  x16 &
  21-50 &
  \cellcolor[HTML]{DAE8FC}M6 &
  \multicolumn{1}{c|}{\cellcolor[HTML]{DAE8FC}31K (8.7K)} &
  \multicolumn{1}{c|}{\cellcolor[HTML]{DAE8FC}6.7K (1.4K)} &
  \multicolumn{1}{c|}{\cellcolor[HTML]{DAE8FC}737 (181)} &
  \multicolumn{1}{c|}{\cellcolor[HTML]{DAE8FC}30K (8.2K)} &
  \multicolumn{1}{c|}{\cellcolor[HTML]{DAE8FC}3.4K (611)} &
  \multicolumn{1}{c|}{\cellcolor[HTML]{DAE8FC}381 (67)} &
  \multicolumn{1}{c|}{\cellcolor[HTML]{DAE8FC}50.9 (17.9) ms} &
  \multicolumn{1}{c|}{\cellcolor[HTML]{DAE8FC}5.1 (0.1) us} &
  \multicolumn{1}{c|}{\cellcolor[HTML]{DAE8FC}18.9 (6.4) us} &
  \cellcolor[HTML]{DAE8FC}1.9 (0.1) us \\ \bottomrule
\end{tabular}
\end{table}
\end{landscape}
\clearpage
\onecolumn
\begin{landscape}

\begin{table}[h]
\centering
\caption{{
Summary of all tested DDR4 modules and their RowHammer/RowPress vulnerabilities {in terms of maximum bit error rate.} We report the maximum bit error rate at representative \DRAMTIMING{AggON} values with the maximum activation count within 60ms.}}
\label{tab:extended-module-info-ber}
\scriptsize
\setlength{\tabcolsep}{2pt}
\renewcommand*{\arraystretch}{1.3}
\begin{tabular}{@{}ccccccc|c|cccccccccc|@{}}
\toprule
 &
   &
   &
   &
   &
   &
   &
   &
  \multicolumn{1}{c|}{\textbf{\begin{tabular}[c]{@{}c@{}}RowHammer\\ Vulnerability\end{tabular}}} &
  \multicolumn{2}{c|}{\textbf{\begin{tabular}[c]{@{}c@{}}RowPress\\ Vulnerability\end{tabular}}} &
  \multicolumn{1}{c|}{\textbf{\begin{tabular}[c]{@{}c@{}}RowHammer\\ Vulnerability\end{tabular}}} &
  \multicolumn{2}{c|}{\textbf{\begin{tabular}[c]{@{}c@{}}RowPress\\ Vulnerability\end{tabular}}} \\ \cmidrule(l){9-14} 
 &
   &
   &
   &
   &
   &
   &
   &
  \multicolumn{6}{c|}{\textbf{\begin{tabular}[c]{@{}c@{}}Maximum Bit Error Rate @ Representative \DRAMTIMING{AggON} and Maximum Activation Count\\ Single-Sided (Double-Sided)\end{tabular}}} \\ \cmidrule(l){9-14} 
 &
   &
   &
   &
   &
   &
   &
   &
  \multicolumn{3}{c|}{\textbf{50$^{\circ}$C}} &
  \multicolumn{3}{c|}{\textbf{80$^{\circ}$C}} \\ \cmidrule(l){9-14} 
  \multirow{-7}{*}{\textbf{Mfr.}} &
  \multirow{-7}{*}{\textbf{DIMM Part}} &
  \multirow{-7}{*}{\textbf{DRAM Part}} &
  \multirow{-7}{*}{\textbf{\begin{tabular}[c]{@{}c@{}}Die\\ Rev.\tablefootnote{We report the die revision marked on the DRAM chip package (if available). A die revision of ``X'' means the original markings on the DRAM chips package are removed by the DRAM module vendor, and thus the die revision could not be identified.}\end{tabular}}} &
  \multirow{-7}{*}{\textbf{\begin{tabular}[c]{@{}c@{}}Die\\ Density\end{tabular}}} &
  \multirow{-7}{*}{\textbf{DQ}} &
  \multirow{-7}{*}{\textbf{\begin{tabular}[c]{@{}c@{}}Date\\ Code\tablefootnote{\prearxiv{In most cases, we report the date code of a DRAM module in the WW-YY format (i.e., 20-53 means the module is manufactured in the $53^{\text{rd}}$ week of year 2020) as marked on the label of the module. We report the date codes of modules S6 and S7 in the ``month-year'' format because it is the only date marked on the labels. We report ``N/A'' if no date is marked on the label of a module.}}\end{tabular}}} &
  \multirow{-7}{*}{\textbf{ID}} &
  \multicolumn{1}{c|}{\textbf{\begin{tabular}[c]{@{}c@{}}\DRAMTIMING{AggON}=36ns\\ (tRAS)\end{tabular}}} &
  \multicolumn{1}{c|}{\textbf{\begin{tabular}[c]{@{}c@{}}\DRAMTIMING{AggON}=7.8us\\ (tREFI)\end{tabular}}} &
  \multicolumn{1}{c|}{\textbf{\begin{tabular}[c]{@{}c@{}}\DRAMTIMING{AggON}=70.2us\\ (9xtREFI)\end{tabular}}} &
  \multicolumn{1}{c|}{\textbf{\begin{tabular}[c]{@{}c@{}}\DRAMTIMING{AggON}=36ns\\ (tRAS)\end{tabular}}} &
  \multicolumn{1}{c|}{\textbf{\begin{tabular}[c]{@{}c@{}}\DRAMTIMING{AggON}=7.8us\\ (tREFI)\end{tabular}}} &
  \textbf{\begin{tabular}[c]{@{}c@{}}\DRAMTIMING{AggON}=70.2us\\ (9xtREFI)\end{tabular}} \\ \midrule
 &
   &
   &
   &
   &
   &
   &
  \cellcolor[HTML]{DAE8FC}S0 &
  \multicolumn{1}{c|}{\cellcolor[HTML]{DAE8FC}0.1\% (3.8\%)} &
  \multicolumn{1}{c|}{\cellcolor[HTML]{DAE8FC}0.009\% (0.005\%)} &
  \multicolumn{1}{c|}{\cellcolor[HTML]{DAE8FC}0.009\% (0.005\%)} &
  \multicolumn{1}{c|}{\cellcolor[HTML]{DAE8FC}0.1\% (3.6\%)} &
  \multicolumn{1}{c|}{\cellcolor[HTML]{DAE8FC}0.09\% (0.04\%)} &
  \cellcolor[HTML]{DAE8FC}0.09\% (0.04\%) \\
 &
  \multirow{-2}{*}{M393A1K43BB1-CTD} &
  \multirow{-2}{*}{K4A8G085WB-BCTD} &
  \multirow{-2}{*}{B} &
  \multirow{-2}{*}{8 Gb} &
  \multirow{-2}{*}{x8} &
  \multirow{-2}{*}{20-53} &
  S1 &
  \multicolumn{1}{c|}{0.2\% (4.3\%)} &
  \multicolumn{1}{c|}{\cellcolor[HTML]{FFFFFF}0.008\% (0.003\%)} &
  \multicolumn{1}{c|}{\cellcolor[HTML]{FFFFFF}0.008\% (0.003\%)} &
  \multicolumn{1}{c|}{0.2\% (3.8\%)} &
  \multicolumn{1}{c|}{\cellcolor[HTML]{FFFFFF}0.07\% (0.03\%)} &
  \cellcolor[HTML]{FFFFFF}0.07\% (0.03\%) \\ \cmidrule(l){2-14} 
 &
  M378A2K43CB1-CTD &
  K4A8G085WC-BCTD &
  C &
  8 Gb &
  x8 &
  N/A &
  \cellcolor[HTML]{DAE8FC}S2 &
  \multicolumn{1}{c|}{\cellcolor[HTML]{DAE8FC}0.7\% (9.5\%)} &
  \multicolumn{1}{c|}{\cellcolor[HTML]{DAE8FC}0.02\% (0.003\%)} &
  \multicolumn{1}{c|}{\cellcolor[HTML]{DAE8FC}0.02\% (0.003\%)} &
  \multicolumn{1}{c|}{\cellcolor[HTML]{DAE8FC}0.8\% (9.0\%)} &
  \multicolumn{1}{c|}{\cellcolor[HTML]{DAE8FC}0.1\% (0.02\%)} &
  \cellcolor[HTML]{DAE8FC}0.1\% (0.02\%) \\ \cmidrule(l){2-14} 
 &
   &
   &
   &
   &
   &
   &
  S3 &
  \multicolumn{1}{c|}{7.7\% (33.1\%)} &
  \multicolumn{1}{c|}{0.07\% (0.01\%)} &
  \multicolumn{1}{c|}{0.08\% (0.01\%)} &
  \multicolumn{1}{c|}{8.2\% (32.5\%)} &
  \multicolumn{1}{c|}{0.6\% (0.2\%)} &
  0.6\% (0.1\%) \\
 &
   &
   &
   &
   &
   &
   &
  \cellcolor[HTML]{DAE8FC}S4 &
  \multicolumn{1}{c|}{\cellcolor[HTML]{DAE8FC}5.2\% (30.0\%)} &
  \multicolumn{1}{c|}{\cellcolor[HTML]{DAE8FC}0.04\% (0.02\%)} &
  \multicolumn{1}{c|}{\cellcolor[HTML]{DAE8FC}0.04\% (0.02\%)} &
  \multicolumn{1}{c|}{\cellcolor[HTML]{DAE8FC}5.9\% (30.0\%)} &
  \multicolumn{1}{c|}{\cellcolor[HTML]{DAE8FC}0.2\% (0.05\%)} &
  \cellcolor[HTML]{DAE8FC}0.2\% (0.04\%) \\
 &
  \multirow{-3}{*}{M378A1K43DB2-CTD} &
  \multirow{-3}{*}{K4A8G085WD-BCTD} &
  \multirow{-3}{*}{D} &
  \multirow{-3}{*}{8 Gb} &
  \multirow{-3}{*}{x8} &
  \multirow{-3}{*}{21-10} &
  S5 &
  \multicolumn{1}{c|}{7.8\% (33.9\%)} &
  \multicolumn{1}{c|}{0.06\% (0.01\%)} &
  \multicolumn{1}{c|}{0.06\% (0.01\%)} &
  \multicolumn{1}{c|}{8.0\% (33.0\%)} &
  \multicolumn{1}{c|}{0.3\% (0.07\%)} &
  0.3\% (0.07\%) \\ \cmidrule(l){2-14} 
 &
   &
   &
   &
   &
   &
   &
  \cellcolor[HTML]{DAE8FC}S6 &
  \multicolumn{1}{c|}{\cellcolor[HTML]{DAE8FC}0.5\% (7.9\%)} &
  \multicolumn{1}{c|}{\cellcolor[HTML]{DAE8FC}0.02\% (0.01\%)} &
  \multicolumn{1}{c|}{\cellcolor[HTML]{DAE8FC}0.02\% (0.01\%)} &
  \multicolumn{1}{c|}{\cellcolor[HTML]{DAE8FC}0.6\% (7.6\%)} &
  \multicolumn{1}{c|}{\cellcolor[HTML]{DAE8FC}0.7\% (0.3\%)} &
  \cellcolor[HTML]{DAE8FC}0.8\% (0.3\%) \\
\multirow{-10}{*}{\textbf{\begin{tabular}[c]{@{}c@{}}Samsung\\ (Mfr. S)\end{tabular}}} &
  \multirow{-2}{*}{\begin{tabular}[c]{@{}c@{}}(G.Skill)\\ F4-2400C17S-8GNT\end{tabular}} &
  \multirow{-2}{*}{K4A4G085WF-BCTD} &
  \multirow{-2}{*}{F} &
  \multirow{-2}{*}{4 Gb} &
  \multirow{-2}{*}{x8} &
  \multirow{-2}{*}{Mar. 21} &
  S7 &
  \multicolumn{1}{c|}{\cellcolor[HTML]{FFFFFF}0.5\% (7.6\%)} &
  \multicolumn{1}{c|}{\cellcolor[HTML]{FFFFFF}0.03\% (0.01\%)} &
  \multicolumn{1}{c|}{\cellcolor[HTML]{FFFFFF}0.02\% (0.01\%)} &
  \multicolumn{1}{c|}{\cellcolor[HTML]{FFFFFF}0.6\% (7.2\%)} &
  \multicolumn{1}{c|}{\cellcolor[HTML]{FFFFFF}0.9\% (0.3\%)} &
  \cellcolor[HTML]{FFFFFF}1.0\% (0.3\%) \\ \midrule
 &
   &
   &
   &
   &
   &
   &
  \cellcolor[HTML]{DAE8FC}H0 &
  \multicolumn{1}{c|}{\cellcolor[HTML]{DAE8FC}1.0\% (9.3\%)} &
  \multicolumn{1}{c|}{\cellcolor[HTML]{DAE8FC}0.03\% (0.01\%)} &
  \multicolumn{1}{c|}{\cellcolor[HTML]{DAE8FC}0.03\% (0.01\%)} &
  \multicolumn{1}{c|}{\cellcolor[HTML]{DAE8FC}2.8\% (10.7\%)} &
  \multicolumn{1}{c|}{\cellcolor[HTML]{DAE8FC}9.4\% (5.7\%)} &
  \cellcolor[HTML]{DAE8FC}9.4\% (5.7\%) \\
 &
  \multirow{-2}{*}{HMAA4GU6AJR8N-XN} &
  \multirow{-2}{*}{H5ANAG8NAJR-XN} &
  \multirow{-2}{*}{A} &
  \multirow{-2}{*}{16 Gb} &
  \multirow{-2}{*}{x8} &
  \multirow{-2}{*}{20-51} &
  H1 &
  \multicolumn{1}{c|}{1.1\% (9.6\%)} &
  \multicolumn{1}{c|}{0.008\% (0.002\%)} &
  \multicolumn{1}{c|}{0.006\% (0.002\%)} &
  \multicolumn{1}{c|}{1.7\% (10.8\%)} &
  \multicolumn{1}{c|}{3.9\% (1.4\%)} &
  3.9\% (1.3\%) \\ \cmidrule(l){2-14} 
 &
   &
   &
   &
   &
   &
   &
  \cellcolor[HTML]{DAE8FC}H2 &
  \multicolumn{1}{c|}{\cellcolor[HTML]{DAE8FC}2.2\% (14.0\%)} &
  \multicolumn{1}{c|}{\cellcolor[HTML]{DAE8FC}0.002\% (0.002\%)} &
  \multicolumn{1}{c|}{\cellcolor[HTML]{DAE8FC}0.002\% (0.002\%)} &
  \multicolumn{1}{c|}{\cellcolor[HTML]{DAE8FC}2.6\% (15.0\%)} &
  \multicolumn{1}{c|}{\cellcolor[HTML]{DAE8FC}0.5\% (0.1\%)} &
  \cellcolor[HTML]{DAE8FC}0.5\% (0.1\%) \\
 &
  \multirow{-2}{*}{HMAA4GU7CJR8N-XN} &
  \multirow{-2}{*}{H5ANAG8NCJR-XN} &
  \multirow{-2}{*}{C} &
  \multirow{-2}{*}{16 Gb} &
  \multirow{-2}{*}{x8} &
  \multirow{-2}{*}{21-36} &
  H3 &
  \multicolumn{1}{c|}{2.0\% (13.0\%)} &
  \multicolumn{1}{c|}{0.003\% (0.002\%)} &
  \multicolumn{1}{c|}{0.003\% (0.002\%)} &
  \multicolumn{1}{c|}{2.0\% (14.0\%)} &
  \multicolumn{1}{c|}{0.4\% (0.1\%)} &
  0.4\% (0.1\%) \\ \cmidrule(l){2-14} 
 &
  \begin{tabular}[c]{@{}c@{}}(Kingston)\\ KVR24R17S8/4\end{tabular} &
  H5AN4G8NAFR-UHC &
  A &
  4 Gb &
  x8 &
  19-46 &
  \cellcolor[HTML]{DAE8FC}H4 &
  \multicolumn{1}{c|}{\cellcolor[HTML]{DAE8FC}0.2\% (1.1\%)} &
  \multicolumn{1}{c|}{\cellcolor[HTML]{DAE8FC}No Bitflip} &
  \multicolumn{1}{c|}{\cellcolor[HTML]{DAE8FC}No Bitflip} &
  \multicolumn{1}{c|}{\cellcolor[HTML]{DAE8FC}0.2\% (1.2\%)} &
  \multicolumn{1}{c|}{\cellcolor[HTML]{DAE8FC}0.003\% (0.002\%)} &
  \cellcolor[HTML]{DAE8FC}0.003\% (0.002\%) \\ \cmidrule(l){2-14} 
\multirow{-9}{*}{\textbf{\begin{tabular}[c]{@{}c@{}}SK Hynix\\ (Mfr. H)\end{tabular}}} &
  \begin{tabular}[c]{@{}c@{}}(Corsair)\\ CMV4GX4M1A2133C15\end{tabular} &
  N/A &
  X &
  4 Gb &
  x8 &
  N/A &
  H5 &
  \multicolumn{1}{c|}{0.9\% (9.0\%)} &
  \multicolumn{1}{c|}{0.005\% (0.002\%)} &
  \multicolumn{1}{c|}{0.005\% (0.002\%)} &
  \multicolumn{1}{c|}{1.7\% (9.8\%)} &
  \multicolumn{1}{c|}{4.0\% (1.6\%)} &
  3.8\% (1.5\%) \\ \midrule
 &
  MTA18ASF2G72PZ-2G3B1 &
  MT40A2G4WE-083E:B &
  B &
  8 Gb &
  x4 &
  N/A &
  \cellcolor[HTML]{DAE8FC}M0 &
  \multicolumn{1}{c|}{\cellcolor[HTML]{DAE8FC}0.3\% (2.6\%)} &
  \multicolumn{1}{c|}{\cellcolor[HTML]{DAE8FC}No Bitflip} &
  \multicolumn{1}{c|}{\cellcolor[HTML]{DAE8FC}No Bitflip} &
  \multicolumn{1}{c|}{\cellcolor[HTML]{DAE8FC}0.3\% (3.0\%)} &
  \multicolumn{1}{c|}{\cellcolor[HTML]{DAE8FC}No Bitflip} &
  \cellcolor[HTML]{DAE8FC}No Bitflip \\ \cmidrule(l){2-14} 
 &
   &
   &
   &
   &
   &
   &
  M1 &
  \multicolumn{1}{c|}{1.2\% (12.0\%)} &
  \multicolumn{1}{c|}{0.005\% (0.002\%)} &
  \multicolumn{1}{c|}{0.005\% (0.002\%)} &
  \multicolumn{1}{c|}{1.7\% (13.2\%)} &
  \multicolumn{1}{c|}{0.03\% (0.006\%)} &
  0.03\% (0.005\%) \\
 &
  \multirow{-2}{*}{MTA4ATF1G64HZ-3G2B2} &
  \multirow{-2}{*}{MT40A1G16RC-062E:B} &
  \multirow{-2}{*}{B} &
  \multirow{-2}{*}{16 Gb} &
  \multirow{-2}{*}{x16} &
  \multirow{-2}{*}{21-26} &
  \cellcolor[HTML]{DAE8FC}M2 &
  \multicolumn{1}{c|}{\cellcolor[HTML]{DAE8FC}1.3\% (12.0\%)} &
  \multicolumn{1}{c|}{\cellcolor[HTML]{DAE8FC}0.002\% (No Bitflip)} &
  \multicolumn{1}{c|}{\cellcolor[HTML]{DAE8FC}0.002\% (No Bitflip)} &
  \multicolumn{1}{c|}{\cellcolor[HTML]{DAE8FC}1.6\% (12.8\%)} &
  \multicolumn{1}{c|}{\cellcolor[HTML]{DAE8FC}0.003\% (0.002\%)} &
  \cellcolor[HTML]{DAE8FC}0.005\% (0.002\%) \\ \cmidrule(l){2-14} 
 &
  MTA36ASF8G72PZ-2G9E1 &
  MT40A4G4JC-062E:E &
  E &
  16 Gb &
  x4 &
  20-14 &
  M3 &
  \multicolumn{1}{c|}{7.4\% (39.2\%)} &
  \multicolumn{1}{c|}{0.003\% (0.009\%)} &
  \multicolumn{1}{c|}{0.003\% (0.002\%)} &
  \multicolumn{1}{c|}{9.3\% (41.3\%)} &
  \multicolumn{1}{c|}{0.1\% (0.04\%)} &
  0.1\% (0.03\%) \\ \cmidrule(l){2-14} 
 &
   &
   &
   &
   &
   &
   &
  \cellcolor[HTML]{DAE8FC}M4 &
  \multicolumn{1}{c|}{\cellcolor[HTML]{DAE8FC}9.0\% (41.0\%)} &
  \multicolumn{1}{c|}{\cellcolor[HTML]{DAE8FC}0.001\% (0.001\%)} &
  \multicolumn{1}{c|}{\cellcolor[HTML]{DAE8FC}0.009\% (0.005\%)} &
  \multicolumn{1}{c|}{\cellcolor[HTML]{DAE8FC}11.3\% (43.7\%)} &
  \multicolumn{1}{c|}{\cellcolor[HTML]{DAE8FC}0.4\% (0.1\%)} &
  \cellcolor[HTML]{DAE8FC}0.4\% (0.06\%) \\
 &
  \multirow{-2}{*}{MTA4ATF1G64HZ-3G2E1} &
  \multirow{-2}{*}{MT40A1G16KD-062E:E} &
  \multirow{-2}{*}{E} &
  \multirow{-2}{*}{16 Gb} &
  \multirow{-2}{*}{x16} &
  \multirow{-2}{*}{20-46} &
  M5 &
  \multicolumn{1}{c|}{8.6\% (39.8\%)} &
  \multicolumn{1}{c|}{0.06\% (0.02\%)} &
  \multicolumn{1}{c|}{0.06\% (0.01\%)} &
  \multicolumn{1}{c|}{11.4\% (43.2\%)} &
  \multicolumn{1}{c|}{2.6\% (1.2\%)} &
  2.6\% (1.0\%) \\ \cmidrule(l){2-14} 
\multirow{-10}{*}{\textbf{\begin{tabular}[c]{@{}c@{}}Micron\\ (Mfr. M)\end{tabular}}} &
  MTA4ATF1G64HZ-3G2F1 &
  MT40A1G16TB-062E:F &
  F &
  16 Gb &
  x16 &
  21-50 &
  \cellcolor[HTML]{DAE8FC}M6 &
  \multicolumn{1}{c|}{\cellcolor[HTML]{DAE8FC}7.1\% (23.2\%)} &
  \multicolumn{1}{c|}{\cellcolor[HTML]{DAE8FC}0.01\% (0.02\%)} &
  \multicolumn{1}{c|}{\cellcolor[HTML]{DAE8FC}0.01\% (0.001\%)} &
  \multicolumn{1}{c|}{\cellcolor[HTML]{DAE8FC}8.6\% (24.3\%)} &
  \multicolumn{1}{c|}{\cellcolor[HTML]{DAE8FC}1.0\% (0.4\%)} &
  \cellcolor[HTML]{DAE8FC}1.0\% (0.3\%) \\ \bottomrule
\end{tabular}
\end{table}
\end{landscape}
\clearpage
\twocolumn
\section{{\textbf{Extended Characterization Results}}}
\label{sec:extended_characterization_results}

\subsection{{\textbf{Extended Characterization Results of the RowPress-ONOFF Pattern}}}
\label{sec:appendix-allonoff}
{We plot the bit error rate (\gls{ber}) for both single-sided (top row of plots) and double-sided (bottom row of plots) RowPress-ONOFF pattern for all die revisions using the same methodology in~\secref{sec:sen_ftber} in the following figures{~(i.e.,~\figref{fig:Mfr.S_4GbF-Die} to \figref{fig:Mfr.M_16GbF-Die})}. We sweep $\Delta \DRAMTIMING{A2A}$ (different {lines} in each {plot}) and the percentage of $\Delta \DRAMTIMING{A2A}$ that contributes to \DRAMTIMING{AggON} (x-axis) at $50^{\circ}C$ ({left} column) and $80^{\circ}C$ ({right} column). {The error band shows the standard deviation of \gls{ber}.} }
\begin{figure}[h]
    \centering
    \includegraphics[width=1.0\linewidth]{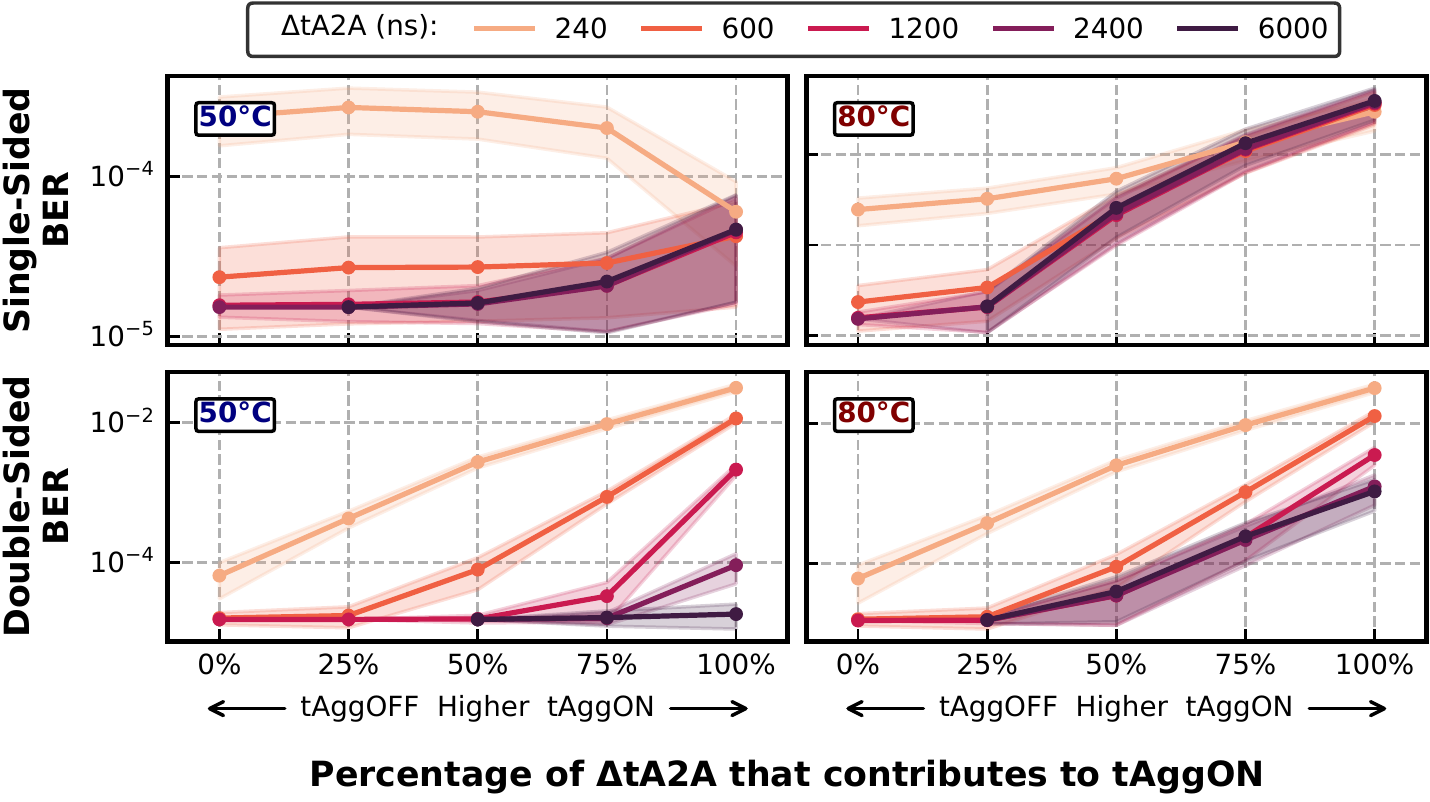}
    \caption{Mfr. S 4Gb F-Die.}
    \label{fig:Mfr.S_4GbF-Die}
\end{figure}

\begin{figure}[h]
    \centering
    \includegraphics[width=1.0\linewidth]{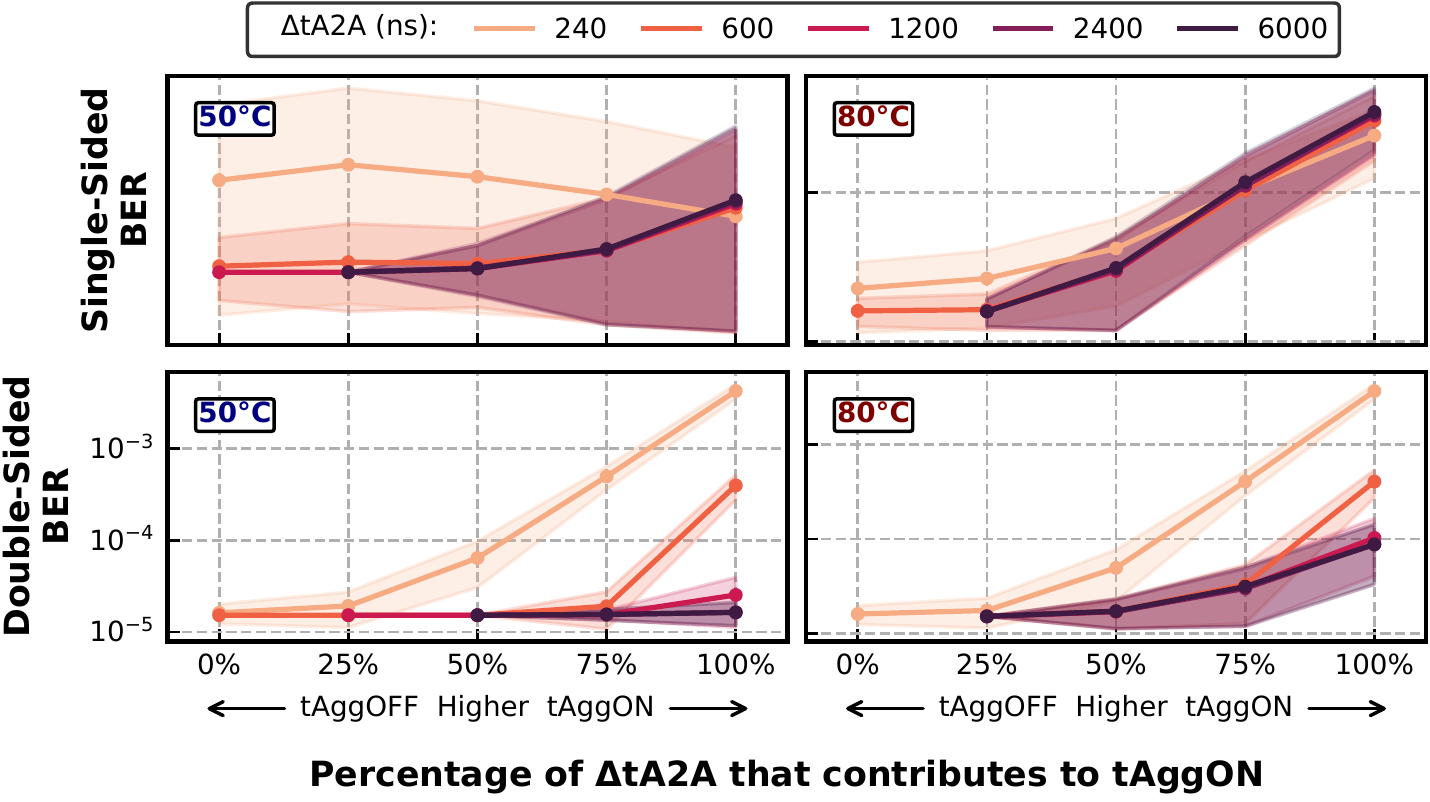}
    \caption{Mfr. S 8Gb B-Die.}
    \label{fig:Mfr.S_8GbB-Die}
\end{figure}

\begin{figure}[h]
    \centering
    \includegraphics[width=1.0\linewidth]{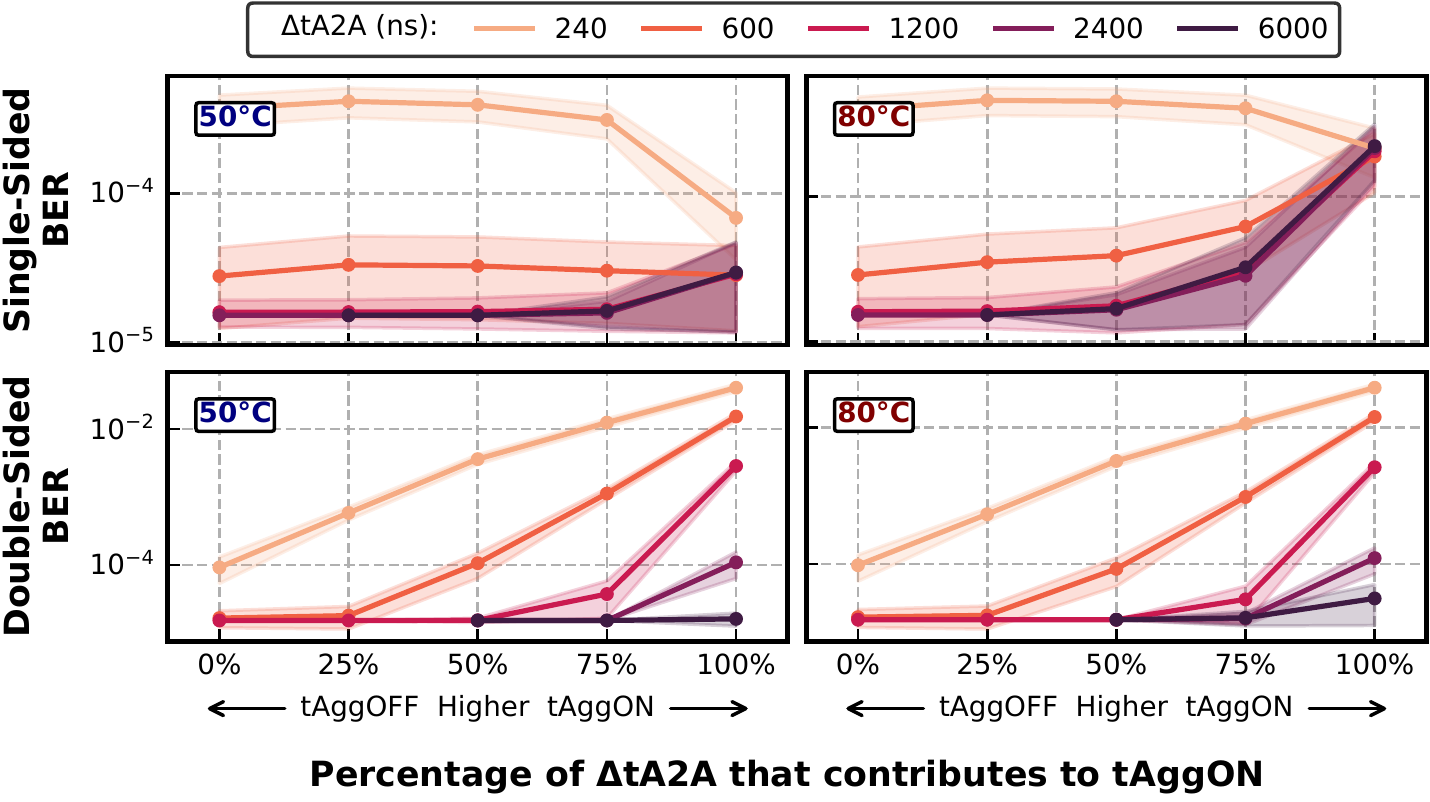}
    \caption{Mfr. S 8Gb C-Die.}
    \label{fig:Mfr.S_8GbC-Die}
\end{figure}

\begin{figure}[h]
    \centering
    \includegraphics[width=1.0\linewidth]{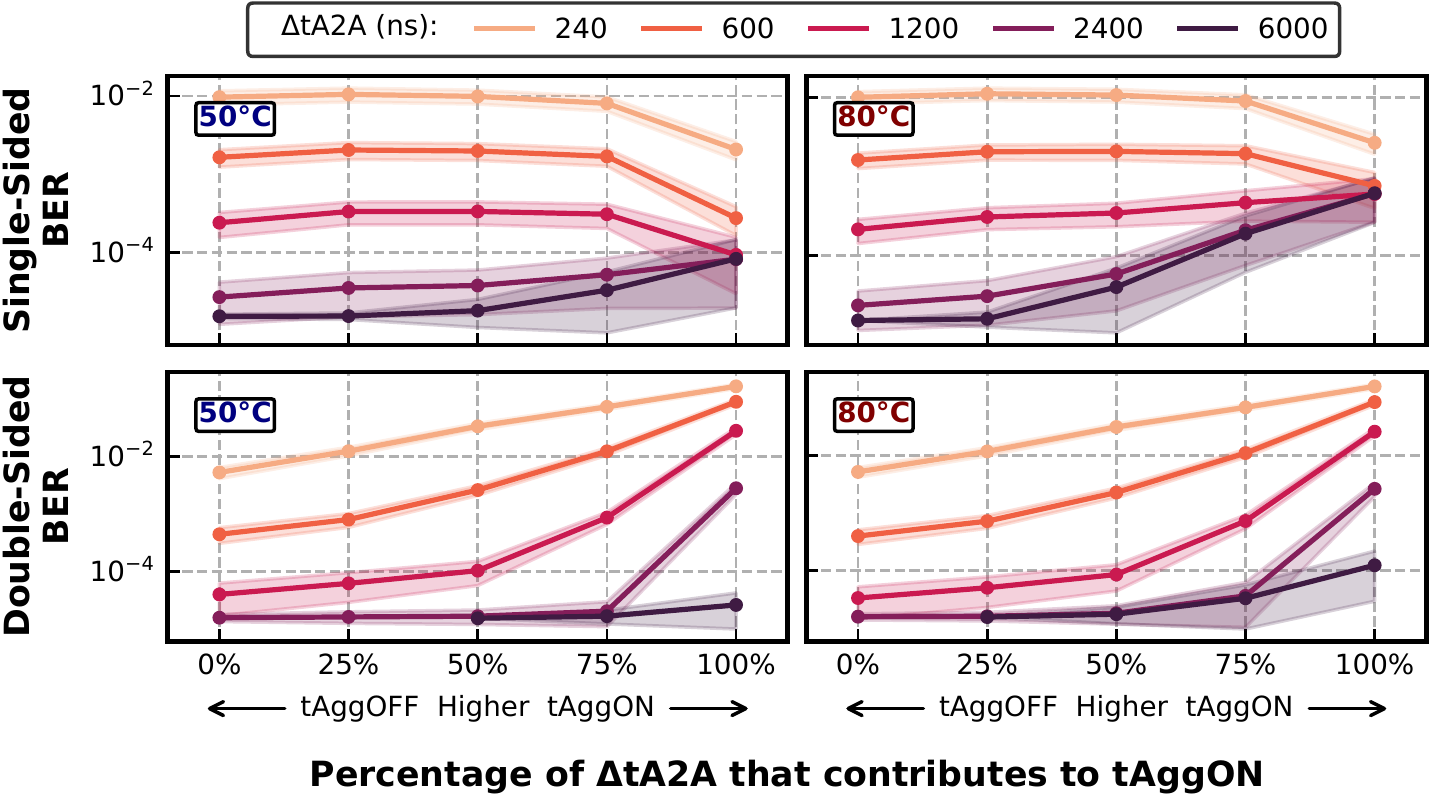}
    \caption{Mfr. S 8Gb D-Die.}
    \label{fig:Mfr.S_8GbD-Die}
\end{figure}

\begin{figure}[h]
    \centering
    \includegraphics[width=1.0\linewidth]{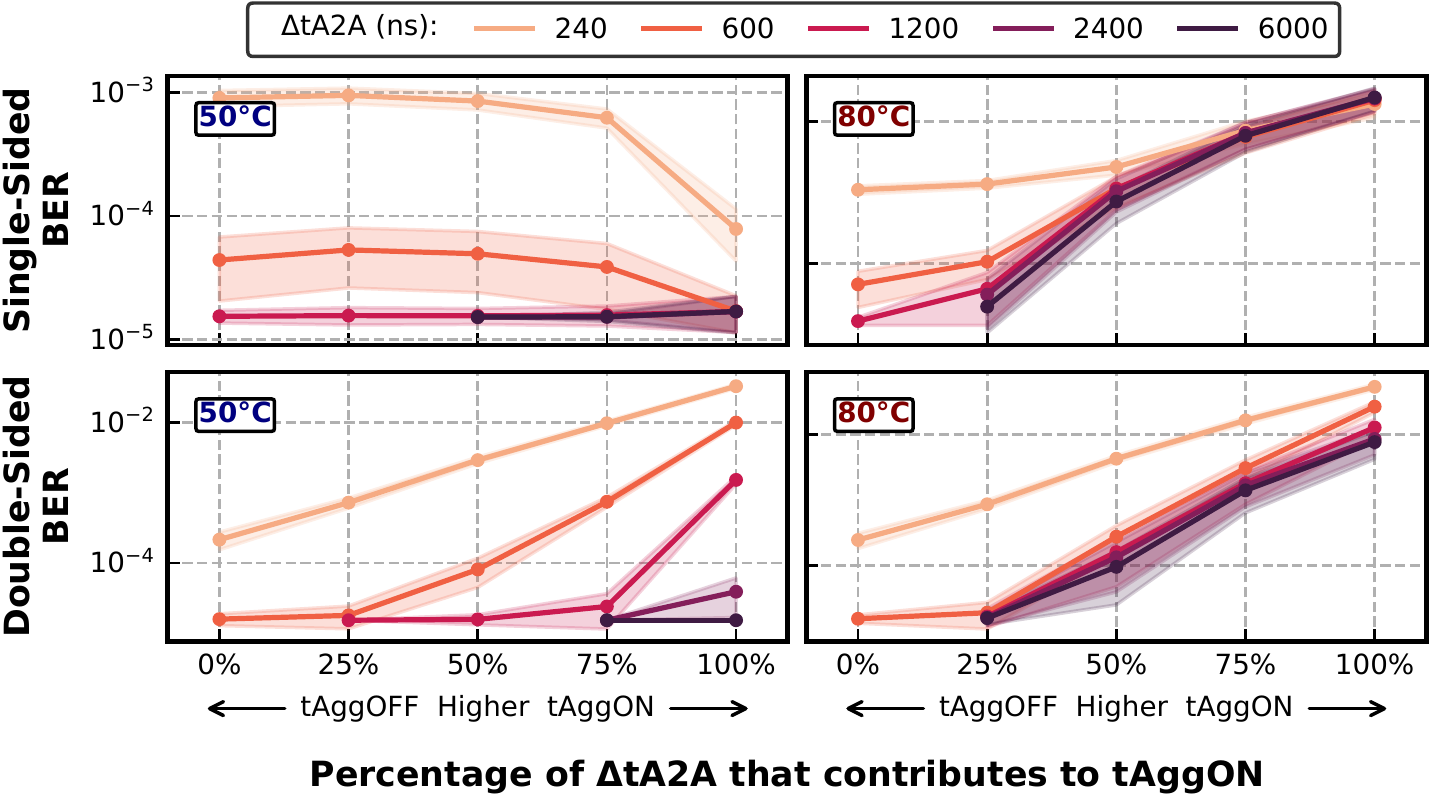}
    \caption{Mfr. H 4Gb X-Die.}
    \label{fig:Mfr.H_4GbX-Die}
\end{figure}

\begin{figure}[h]
    \centering
    \includegraphics[width=1.0\linewidth]{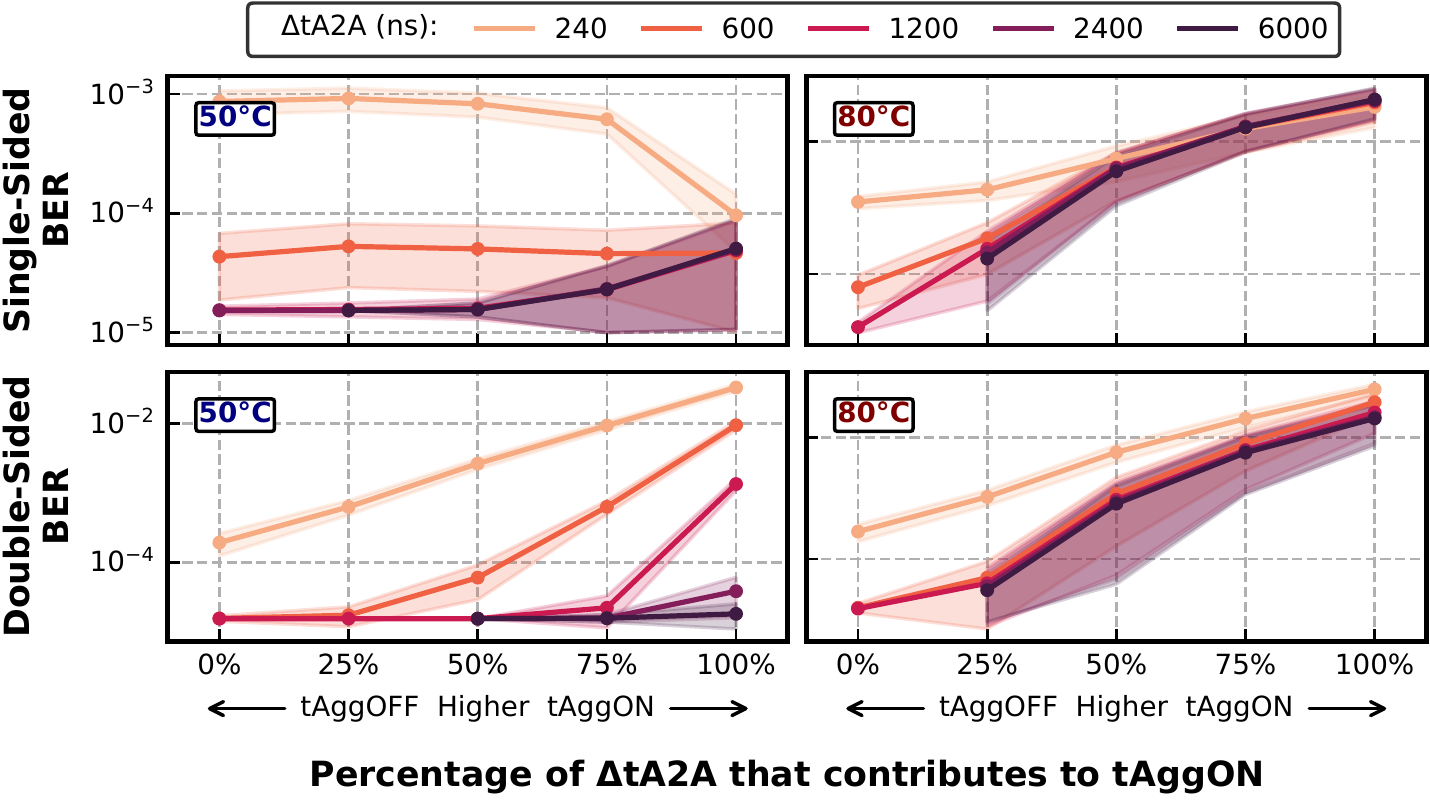}
    \caption{Mfr. H 16Gb A-Die.}
    \label{fig:Mfr.H_16GbA-Die}
\end{figure}

\begin{figure}[h]
    \centering
    \includegraphics[width=1.0\linewidth]{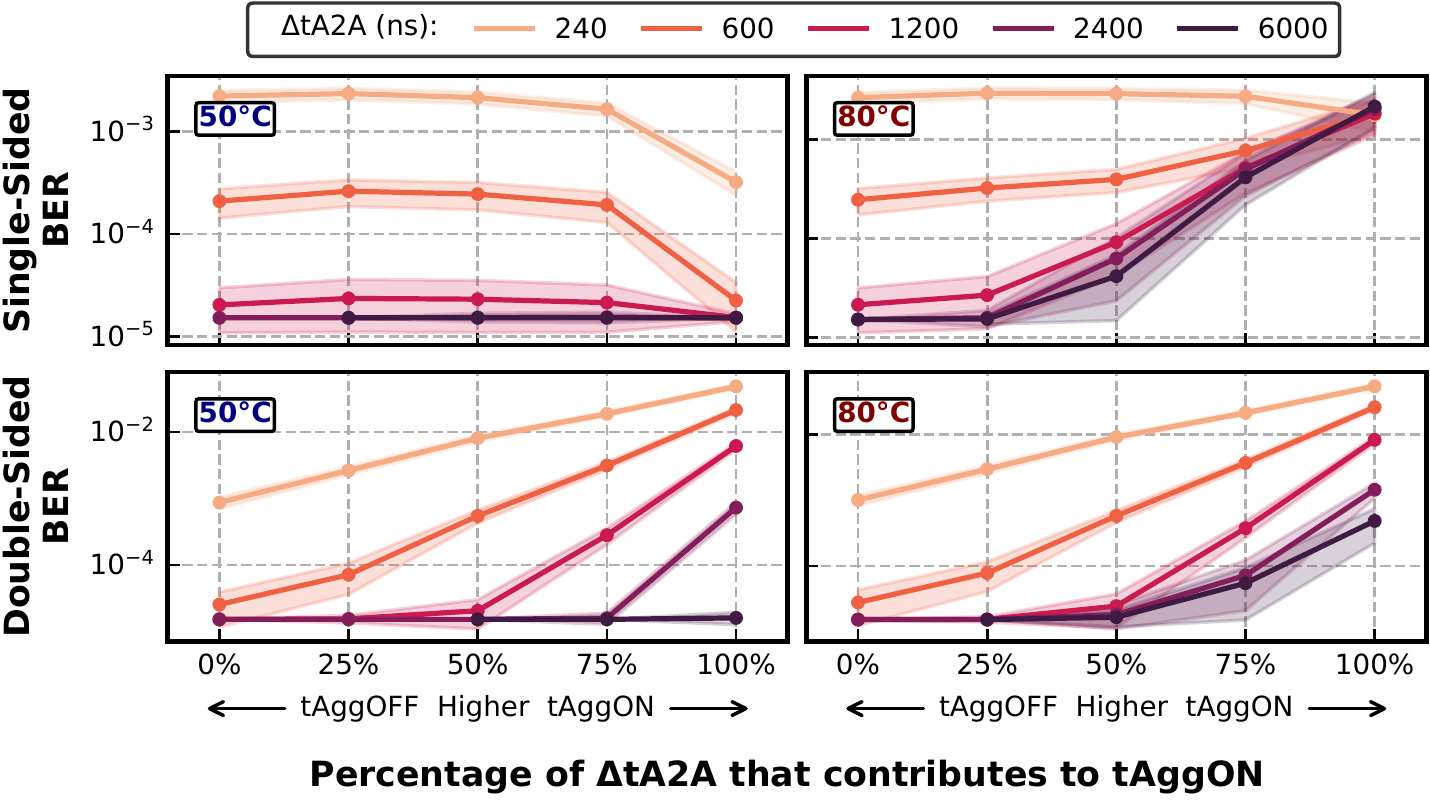}
    \caption{Mfr. H 16Gb C-Die.}
    \label{fig:Mfr.H_16GbC-Die}
\end{figure}

\begin{figure}[h]
    \centering
    \includegraphics[width=1.0\linewidth]{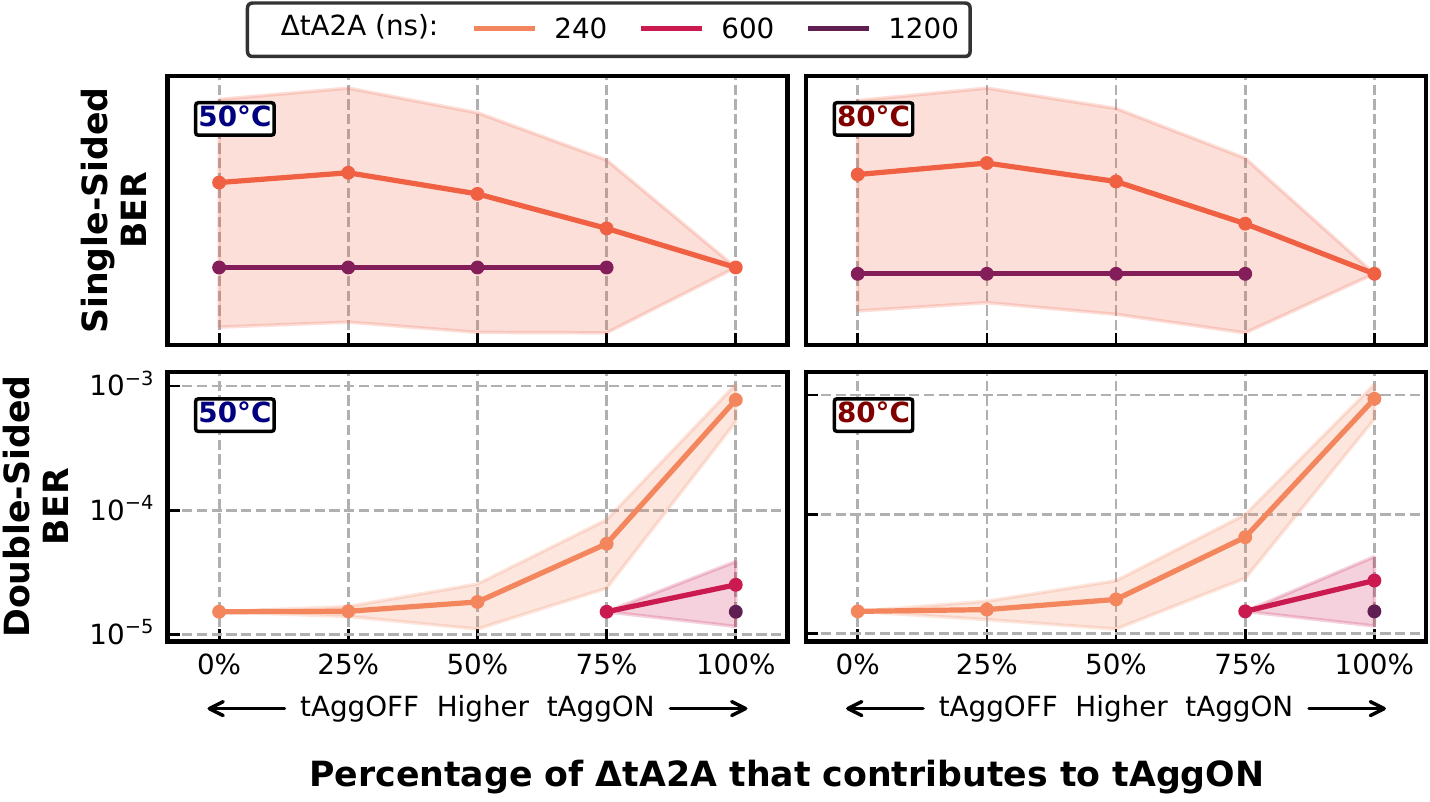}
    \caption{Mfr. M 8Gb B-Die.}
    \label{fig:Mfr.M_8GbB-Die}
\end{figure}

\begin{figure}[h]
    \centering
    \includegraphics[width=1.0\linewidth]{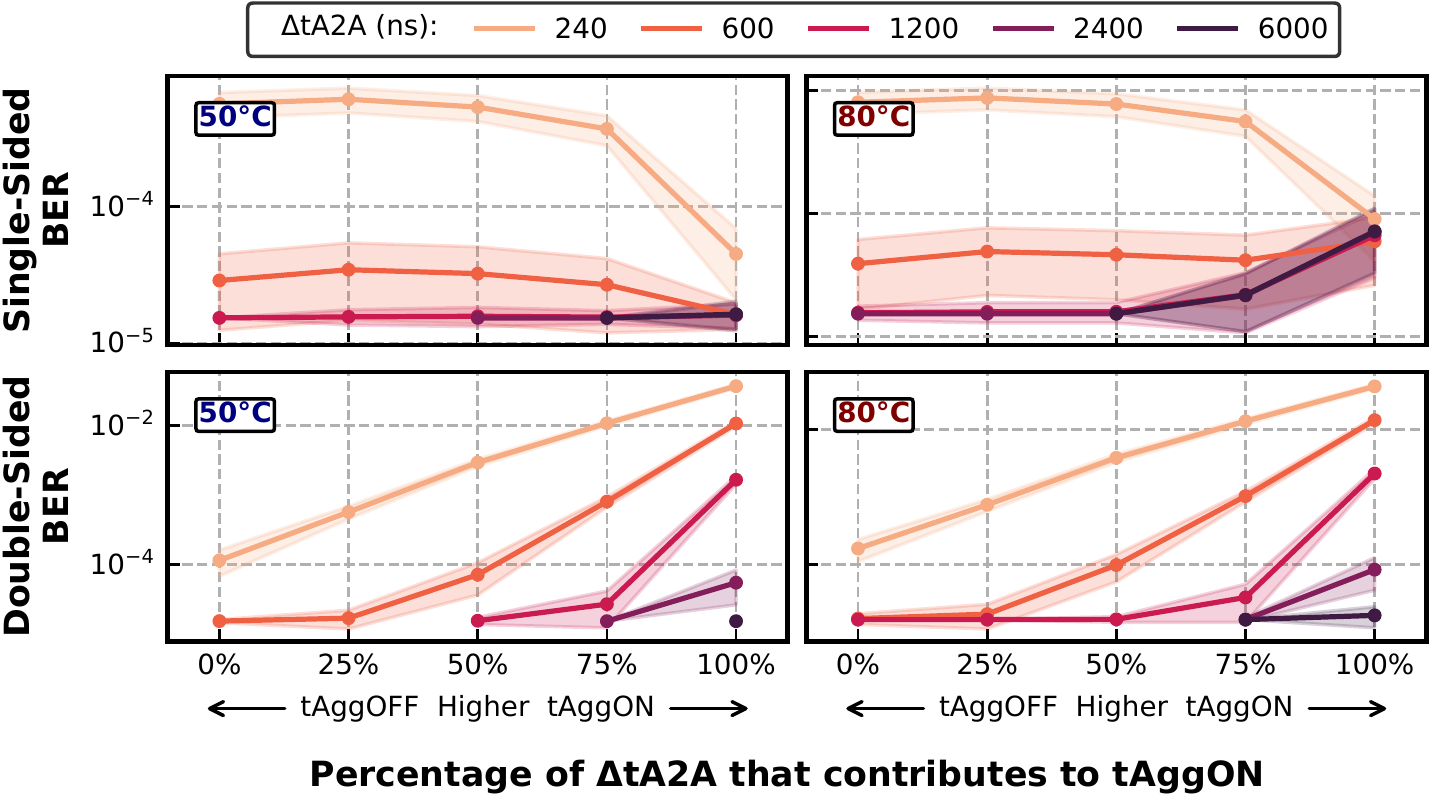}
    \caption{Mfr. M 16Gb B-Die.}
    \label{fig:Mfr.M_16GbB-Die}
\end{figure}

\begin{figure}[h]
    \centering
    \includegraphics[width=1.0\linewidth]{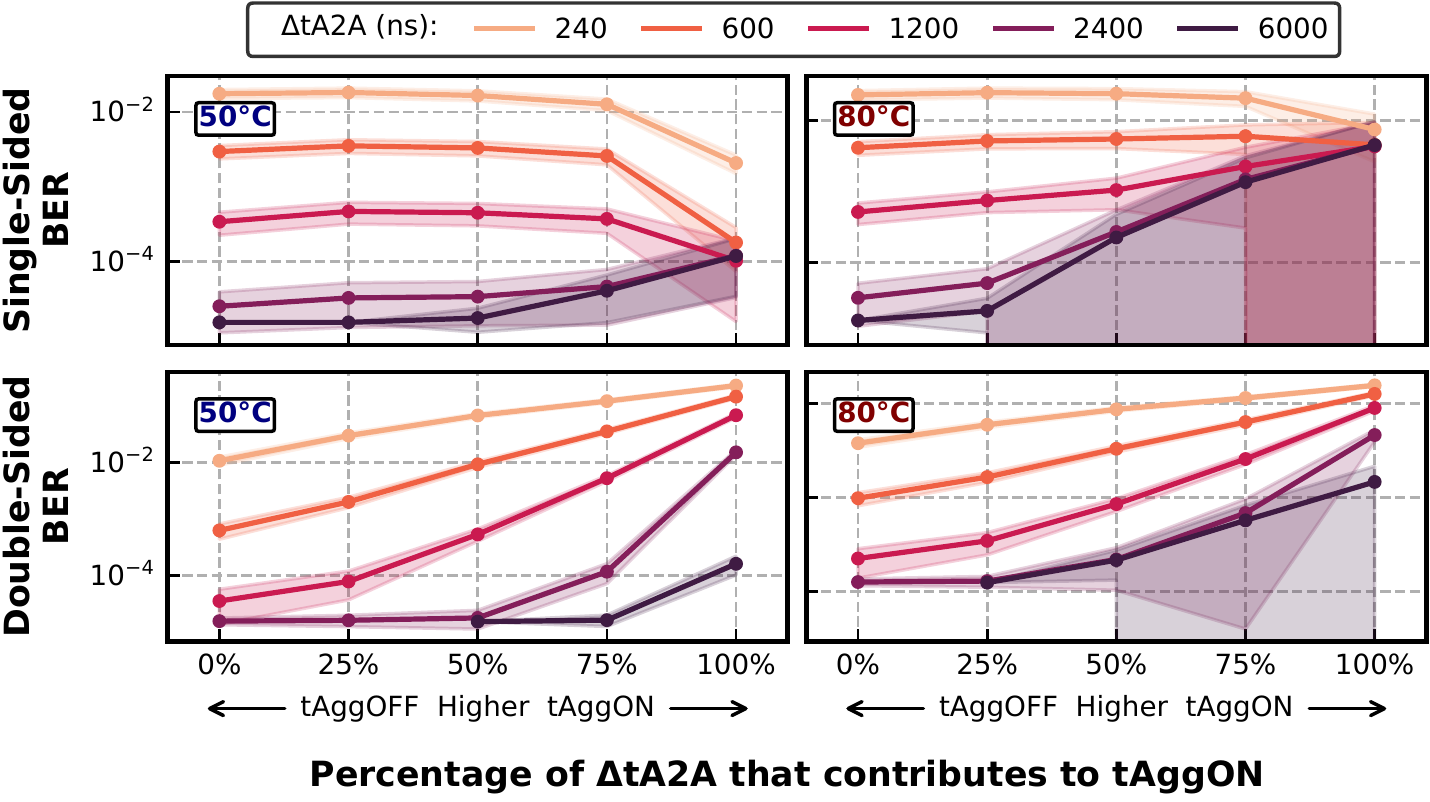}
    \caption{Mfr. M 16Gb E-Die.}
    \label{fig:Mfr.M_16GbE-Die}
\end{figure}

\begin{figure}[h]
    \centering
    \includegraphics[width=1.0\linewidth]{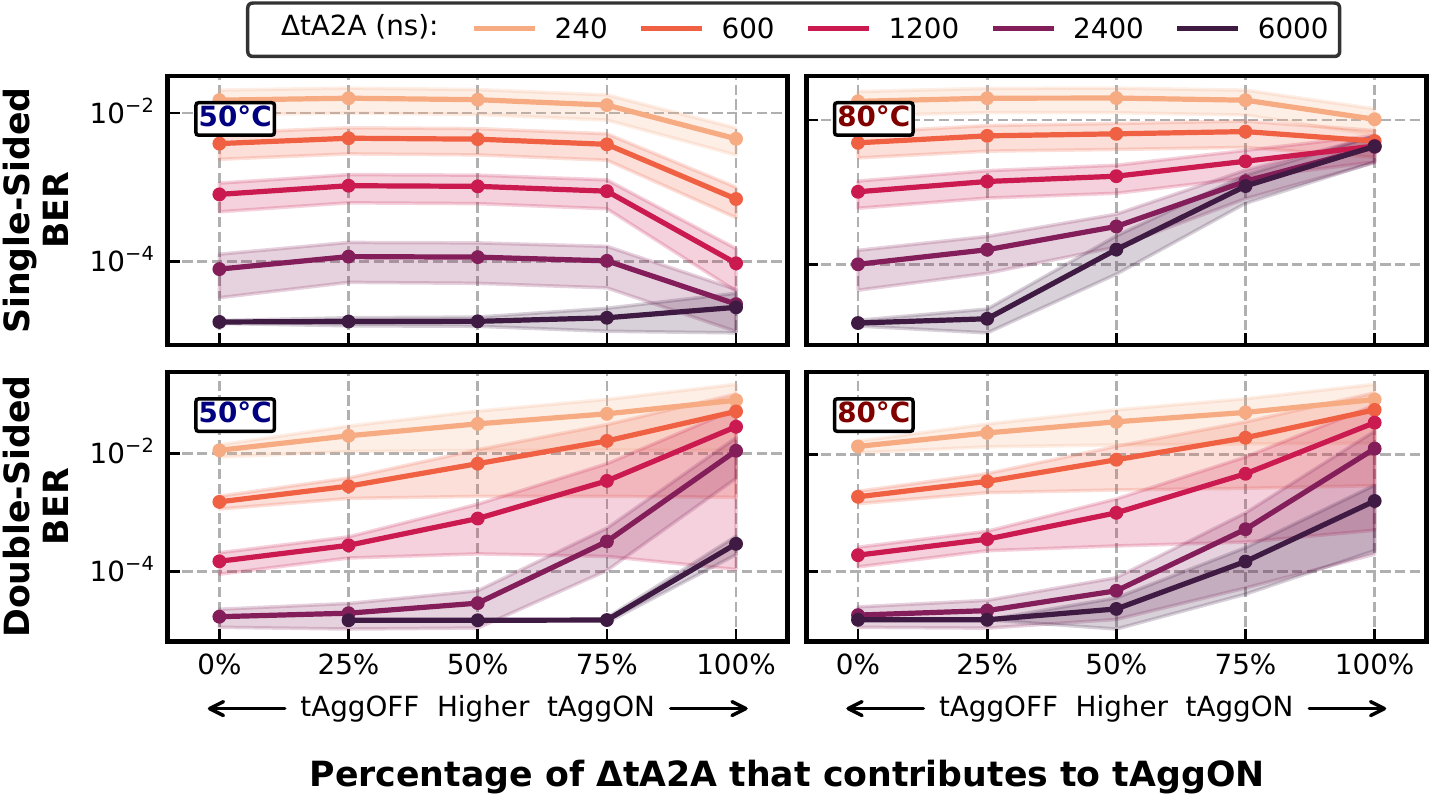}
    \caption{Mfr. M 16Gb F-Die.}
    \label{fig:Mfr.M_16GbF-Die}
\end{figure}
\clearpage
\section{{\textbf{Extended Evaluation Results}}}
\label{sec:extended_implications}

\subsection{{Limiting the Maximum Row-Open Time}}
\label{sec:extended_rowpolicy}

We evaluate 61 workloads from SPEC CPU2006~\cite{spec2006}, SPEC CPU-2017~\cite{spec2017}, TPC-H~\cite{tpch}, and YCSB~\cite{ycsb} using Ramulator~\cite{kim2016ramulator,ramulatorgithub} {with} a realistic baseline system {configuration}, as shown in Table~\ref{tab:system_config}. We compare the increase in the number of activations to each individual DRAM row in a \SI{64}{\milli\second} (\DRAMTIMING{REFW}) time window and the performance overhead with the minimally-open-row policy to those of the {baseline open-row policy~\cite{rixner00}}.

\begin{table}[h]
\centering
\scriptsize
\caption{Simulated system configuration.}
\label{tab:system_config}
\begin{tabular}{@{}cc@{}}
\toprule
\textbf{Processor} &
  \begin{tabular}[c]{@{}c@{}}4 GHz Out-of-Order core, \\ 4-wide issue, 128-entry instruction window.\\ 8 MSHRs per core, 2MiB LLC per core\end{tabular} \\ \midrule
\textbf{Memory Controller} &
  \begin{tabular}[c]{@{}c@{}}64-entry read/write request buffer, \\ FR-FCFS request scheduling~\cite{rixner00, zuravleff1997controller}, {open-row policy~\cite{rixner00}}\end{tabular} \\ \midrule
\textbf{DRAM} &
  \begin{tabular}[c]{@{}c@{}}DDR4~\cite{jedec2017ddr4} 3200MT/s, \\ 1 channel, 2 rank, 4 bankgroups, 4 banks per bankgroup\\ JEDEC DDR4-3200W Speedbin~\cite{jedec2017ddr4}\end{tabular} \\ \bottomrule
\end{tabular}
\end{table}

\figref{fig:increase_act} shows the maximum increase in the number of activations to each individual DRAM row within \DRAMTIMING{REFW}. For clarity, we only plot the workloads with a maximum increase over $50\times$.

\begin{figure}[h]
    \centering
    \includegraphics[width=\linewidth]{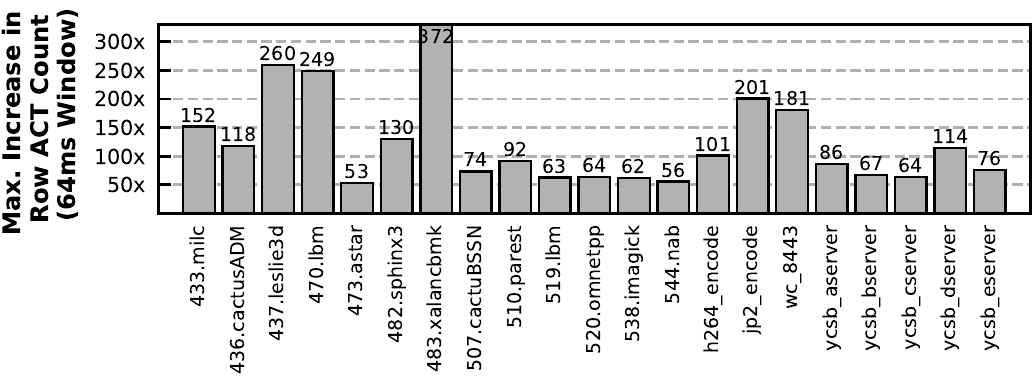}
    \caption{Maximum increase in the number of activations to each individual DRAM row in \DRAMTIMING{REFW} with the minimally-open-row policy, compared to an open-row policy baseline.}
    \label{fig:increase_act}
\end{figure}

We observe that using a minimally-open-row policy significantly increases the number of activations to a single DRAM row within \DRAMTIMING{REFW} for a large group of workloads (i.e., 21 out of 58 workloads have at least $50\times$ increase), by up to $372\times$ (from only 1 to 372 activations for 483.xalancbmk). We also observe that, across all the rows accessed by the workloads, using a minimally-open-row policy significantly increases the number of activations to the most activated DRAM row. For example, the most activated row in 510.parest is only activated 497 times within \DRAMTIMING{REFW} for the open-row policy, but this increases to 3808 times for the minimally-open-row policy. We find that 436.cactusADM, jp2\_decode, 505.mcf, 471.omnetpp, and 483.xalancbmk also have their most-activated row activation count increased from less than 1000 to over 1000, which is used by many prior works~{(e.g., \cite{kim2020revisiting, park2020graphene, yaglikci2021blockhammer, qureshi2022hydra, yaglikci2022hira})} as a projected RowHammer threshold ($T_{RH}$), \agyh{defined as the minimum number of aggressor row activations needed to cause a RowHammer bitflip,} in the near future. We conclude that using a minimally-open-row policy can potentially turn benign workloads into a RowHammer attack~{\exploitingRowHammerAllCitations}.

{
\figref{fig:max_slowdown} shows the instruction per cycle (IPC) of the workloads with the minimally-open-row policy, normalized to the open-row policy baseline. For clarity, we only plot the workloads with a normalized IPC smaller than 0.95. {We do \emph{not} observe any workload with a normalized IPC higher than 1.0.}
}

\begin{figure}[h]
    \centering
    \includegraphics[width=\linewidth]{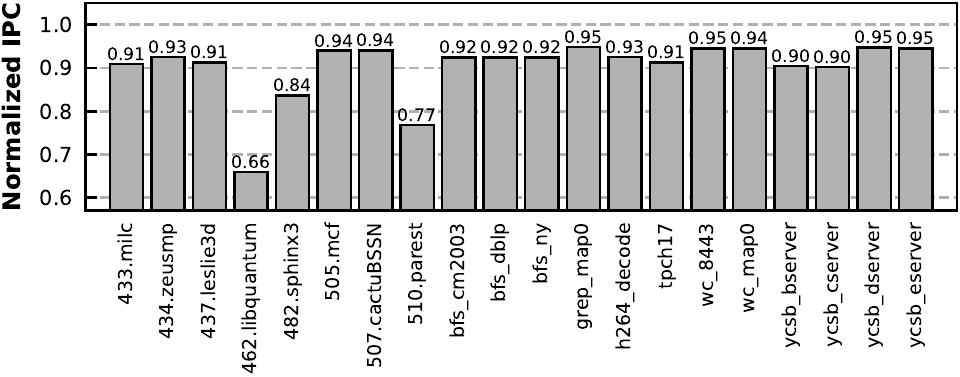}
    \caption{IPC of workloads when using the minimally-open-row policy, normalized to the baseline open-row policy.}
    \label{fig:max_slowdown}
\end{figure}

\begin{figure*}[t]
    \centering
    \includegraphics[width=0.99\linewidth]{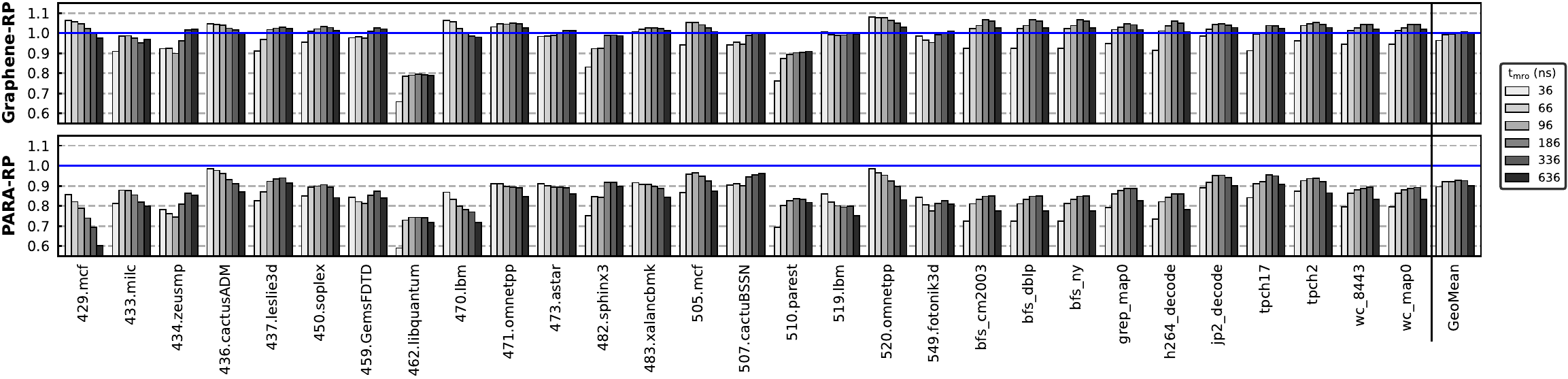}
    \caption{IPC of Graphene-RP and PARA-RP of single-core workloads (LLC-MPKI $>$ 5) with different $t_{mro}$ configurations, normalized to Graphene and PARA, respectively.}
    \label{fig:singlecore}
\end{figure*}

{
We observe that using the minimally-open-row policy can significantly reduce the performance of workloads with high row buffer locality. For example, the IPC of 462.libquantum reduces by $34.1\%$, {as} its row buffer misses per kilo instructions (RBMPKI) {increases} by $110\%$ from only 0.91 to 1.90. The performance of 510.parest reduces by $23.2\%$, {as} its RBMPKI {increases} by $62\%$. We conclude that using the minimally-open-row policy can significantly reduce system performance by reducing the row buffer hit rate.
}

{
Some existing row policy proposals adapt $t_{mro}$ based on row access patterns (e.g., keep the row open for longer when the row is predicted to be accessed soon in the future)~\cite{IntelAPM, Awasthi2011Prediction, Rokicki2022Method, park2003history, Kahn2004method, Sander2005dynamic, xu2009prediction}. Such row policies {\emph{cannot}} securely mitigate RowPress as $t_{mro}$ can be controlled by an attacker to be set to larger values than \DRAMTIMING{RAS}, as we show in~\secref{sec:real}. {We believe securely} mitigating RowPress requires co-designing existing RowHammer mitigations with a row policy that enforces $t_{mro}$, as \secref{sec:graphere_para} describes and evaluates.
}

\subsection{{Adapting Existing RowHammer Mitigations}}
\label{sec:extended_graphene_para}

\noindent
\textbf{Evaluation Methodology.}
We perform a sensitivity study of the performance overheads of Graphene-RP and PARA-RP over Graphene and PARA with the configurations shown in Table~\ref{tab:gp_params_full} \gf{using} Ramulator~\cite{kim2016ramulator,ramulatorgithub} {with} the same baseline system {configuration} in~\secref{sec:rowpolicy} {on} both single- and four-core multiprogrammed workloads. We evaluate both 1) homogeneous four-core workloads where we run four copies of {each} single-core {workload} on four cores, and 2) heterogeneous four-core workloads where we run different workloads on each core. To create the heterogeneous four-core workloads, we categorize the {memory-intensity of the single-core workloads {using two metrics:} last-level cache misses per kilo instructions, LLC-MPKI, \emph{and} row buffer misses per kilo instructions, RBMPKI. We group the single-core workloads} into high-memory-intensity (i.e., LLC-MPKI $\geq1$ \emph{and} RBMPKI $\geq$ 1), denoted as ``H'', and low-memory-intensity, (i.e., LLC-MPKI $<1$ \emph{and} RBMPKI $<$ 1), denoted as ``L''. We evaluate five different groups of heterogeneous workloads, denoted as HHHH, HHHL, HHLL, HLLL, and LLLL. For example, HHHH means all four workloads are from the ``H'' category, and HHLL means two workloads are from ``H'' and the other two are from ``L''. For each group, we evaluate eight different randomly picked workload mixes {for a total of 40 heterogeneous four-core workloads}. We use instruction per cycle (IPC) as the performance metric for single-core workloads, and weighted speedup~\cite{tullsen2001handling} for four-core workloads.

\begin{table}[h!]
\centering
\setlength{\tabcolsep}{0.42em}
{\fontsize{7.5}{9}\selectfont
\caption{{Graphene-RP and PARA-RP configurations {for different $t_{mro}$ values.}}}
\label{tab:gp_params_full}
\begin{tabular}{ccccccc}
\toprule[0.7pt]
$t_{mro}$ (ns)         & 36 (=\DRAMTIMING{RAS}) & 66  & 96  & 186   & 336    & 636    \\

\textbf{$T'_{RH}$}     & 1000 (=$T_{RH}$)  & 809    & 724   & 619   & 555    & 419    \\ 
\toprule[0.7pt]

\textbf{Graphene-RP $T$}      & 333  & 269 & 241 & 206  & 185    & 139    \\ 

\midrule

\textbf{PARA-RP $p$}          & 0.034 & 0.042 & 0.047 & 0.054 & 0.061 & 0.079\\

\bottomrule[0.7pt]

\end{tabular}}
\end{table}

\noindent
\textbf{Evaluation Results.}
{
\figref{fig:singlecore} shows the IPC of different Graphene-RP (top row) and PARA-RP (bottom row) configurations on the single-core workloads, normalized to Graphene and RP, respectively. For clarity, we only plot the workloads with more than five last-level-cache misses per kilo-instruction (LLC-MPKI $>$ 5). We show the average {(geometric mean)} normalized IPC across {\emph{all}} single-core workloads. We make the following observations.
}

{
First, Graphene-RP can mitigate RowPress with a {slightly} higher performance compared to Graphene. For single-core workloads, Graphene-RP slightly outperforms Graphene for all $t_{mro} \geq 66ns$, by up to $0.46\%$ on average when $t_{mro} = 336ns$. The reason for the small speedups of Graphene-RP over Graphene is that enforcing a $t_{mro}$ increases the performance of workloads with low row buffer hit rate. For example, the performance of 429.mcf (baseline RBMPKI $=68.6$) increases significantly (normalized IPC increases from 0.97 to 1.06) as $t_{mro}$ decreases from 636ns to 36ns. On the other hand, enforcing a $t_{mro}$ reduces the performance of workloads with high row buffer hit rate. For example, the performance of 462.libquantum (baseline RBMPKI of only $0.91$) decreases significantly (as low as 0.66 when $t_{mro}$ is 36ns) over Graphene for all of the $t_{mro}$ values we evaluate. 
}

{
Second, PARA-RP can mitigate RowPress at low \emph{additional} performance overhead for single-core workloads. For example, PARA-RP performs the best when $t_{mro} = 186ns$, with an average slowdown of only $7.3\%$. The reason for PARA-RP's consistent {slowdown} across different $t_{mro}$ values over PARA is that PARA does \emph{not} track aggressor rows deterministically (like Graphene). As a result, \emph{any} extra row activations (i.e., even if the activations are \emph{not} {concentrated} on a small number of rows) caused by the enforced $t_{mro}$ will increase the number of (false-positive) {preventive refreshes{\footnote{\prearxiv{A DRAM row is \emph{preventively} refreshed to prevent bitflips before its adjacent row is activated too many times (i.e., $T_{RH}$ times).}}}} issued by PARA. For example, when $t_{mro}$ is 636ns, the number of {preventive} refreshes issued by PARA-RP (427074) is $17.6\times$ more than Graphene-RP (23006), {even though both PARA-RP and Graphene-RP have} similar numbers of extra row activations (191447 for PARA-RP and 117229 for Graphene-RP) compared to the open-row baseline.
}

{
\figref{fig:multicore} shows the geometric means of the normalized weighted speedups of different Graphene-RP (top row) and PARA-RP (bottom row) configurations on homogeneous (left column) and heterogeneous (right column) four-core workloads. The error bars mark the lowest and highest normalized weighted speedups observed {within} a workload group. We make the following two observations.
}

\begin{figure}[h]
    \centering
    \includegraphics[width=\linewidth]{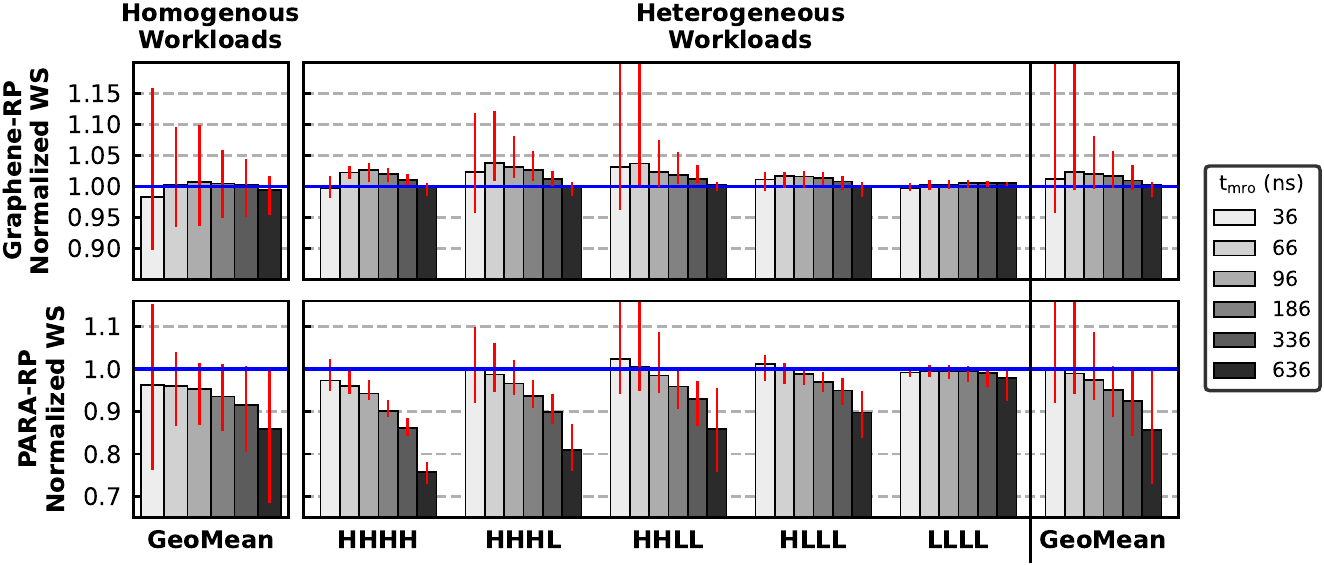}
    \caption{Geometric means of the normalized Weighted speedups of Graphene-RP and PARA-RP {for} homogeneous and heterogeneous four-core workloads with different $t_{mro}$ configurations.}
    \label{fig:multicore}
\end{figure}

{
First, both Graphene-RP and PARA-RP have small performance overhead compared to Graphene and PARA, respectively. For homogeneous workloads, Graphene-RP outperforms Graphene by $0.67\%$ when $t_{mro}$ is 96ns, and PARA-RP performs the best with only $3.8\%$ slowdown over PARA when $t_{mro}$ is 36ns. Across all heterogeneous workloads, Graphene-RP outperforms Graphene by $2.3\%$ when $t_{mro}$ is 66ns, and PARA-RP can perform PARA 
by $0.03\%$ when $t_{mro}$ is 36ns. We notice that when $t_{mro}$ is 36ns, both Graphene-RP and PARA-RP significantly {improve} the performance of certain workloads. For example, PARA-RP (Graphene-RP) has a speedup of $31.3\%$ (28.8\%) over PARA (Graphene) for a HHLL workload containing h264\_encode. This is because h264\_encode has a very high row buffer hit rate ($87.0\%$) in the baseline, {and thus gets} unfairly prioritized by the memory controller's FR-FCFS~\cite{rixner00, zuravleff1997controller} scheduling and open-row policy.\footnote{{Such (un)fairness and resulting performance issues are well-studied by prior works~\cite{mutlu2008parbs, mutlu2007stall, subramanian2016bliss, kim2010thread, ghose2019demystifying, moscibroda2007memory, kim2010atlas}.}} A low $t_{mro}$ value thus improves fairness between cores by allowing other workloads to progress and increases the weighted speedup.
}

{
Second, in contrast to single-core workloads, the performance overhead of PARA-RP always reduces as $t_{mro}$ increases beyond 36ns. The reason is that PARA's performance overhead does \agyh{\emph{not}} scale well with \agyh{reducing} $T'_{RH}$~\cite{kim2020revisiting, park2020graphene, yaglikci2021blockhammer}, {and thus the performance benefits} of longer row-open time and incrased row-buffer hit rate is outweighed by the performance {overheads} of the increased preventive refreshes.
}

{
We summarize our performance evaluation results of Graphene-RP and PARA-RP in Table~\ref{tab:evaluation_results_summary}. We conclude that existing RowHammer mitigations can be easily adapted to mitigate RowPress at low additional performance overhead. We expect future work to introduce and discuss new mitigation mechanisms in detail, as it has been happening analogously with RowHammer.
}

\begin{table}[h!]
\vspace{1em}
\centering
\scriptsize
\caption{Additional performance overhead of Graphene-RP and PARA-RP {over Graphene and PARA for single-core and multi-core workloads}.}
\label{tab:evaluation_results_summary}
\begin{tabular}{@{}ccccccc@{}}
\toprule
\textbf{$t_{mro}$ (ns)} & \textbf{36{($=$\DRAMTIMING{RAS})}} & \textbf{66} & \textbf{96} & \textbf{186} & \textbf{336} & \textbf{636} \\ \midrule
\multicolumn{7}{c}{\textbf{Average Graphene-RP Perf. Overhead Over Graphene}}   \\ \midrule
Single-core              & 3.7\%  & 0.8\%  & 0.5\%  & -0.4\% & -0.5\% & -0.05\% \\
Homogeneous Multi-core   & 1.7\%  & -0.3\% & -0.7\% & -0.5\% & -0.2\% & 0.6\%   \\
Heterogeneous Multi-core & -1.2\% & -2.3\% & -2.0\% & -1.7\% & -1.0\% & -0.2\%  \\ \midrule
\multicolumn{7}{c}{\textbf{Average PARA-RP Perf. Overhead Over PARA}}           \\ \midrule
Single-core              & 10.4\% & 8.0\%  & 7.9\%  & 7.3\%  & 7.4\%  & 9.9\%   \\
Homogeneous Multi-core   & 3.8\%  & 4.0\%  & 4.8\%  & 6.5\%  & 8.4\%  & 14.0\%  \\
Heterogeneous Multi-core & -0.0\% & 1.1\%  & 2.5\%  & 4.9\%  & 7.5\%  & 14.3\%  \\ \bottomrule
\end{tabular}
\end{table}

\pagebreak
\section{Repeatability of RowPress Bitflips}
\label{sec:repeatability}

We study the {\emph{repeatability}} of RowPress bitflips {across all five iterations of our experiments}. {We define repeatability as the number of occurrences of a bitflip across all five iterations (i.e., ranges from 1 to 5, the higher the number of occurrences, the higher the repeatability).} \figref{fig:stab_SS_50} {is a histogram of the distribution of the repeatability of RowPress bitflips from our experiments described in~\secref{sec:vulnerability_to_readdisturbance}. The y-axis shows the percentage of bitflips with different repeatability (from 1 to 5, x-axis).} We plot representative \DRAMTIMING{AggON} values in different rows of plots.

\begin{figure}[h]
    \centering
    \includegraphics[width=1.0\linewidth]{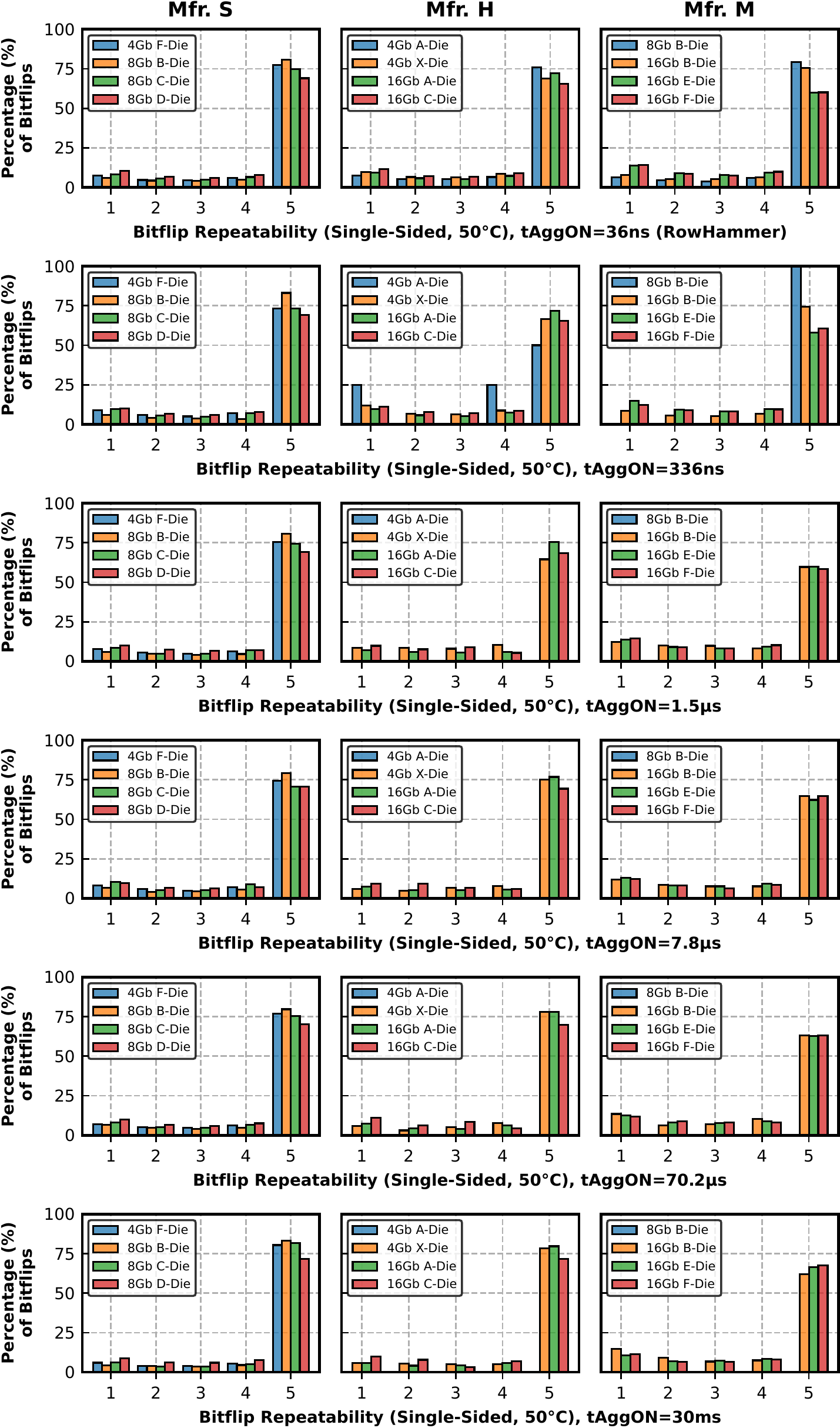}
    \caption{Repeatability of the single-sided RowPress (RowHammer) bitflips; $50^\circ C$}
    \label{fig:stab_SS_50}
\end{figure}

We observe that the majority of the RowPress bitflips are repeatable across all five iterations. For all \DRAMTIMING{AggON} values we test and for almost all die revisions from all three major DRAM manufacturers, at least 50\% of the bitflips occurs in all five iterations. {Even when \DRAMTIMING{AggON} is 30ms, the lowest percentage of bitflips that occur in all five iterations is still 61.9\% (observed from 16Gb B-Dies from Mfr. M).} We conclude that RowPress bitflips are repeatable, {similar to RowHammer bitflips~\cite{kim2014flipping}}.

\observation{RowPress bitflips are repeatable{, i.e., if they occur once in a cell, they are likely to occur again and again.}}

\figref{fig:stab_SS_80}, \figref{fig:stab_DS_50}, and \figref{fig:stab_DS_80} show the percentage of bitflips (y-axis) with different repeatability (x-axis) based on the single-sided access pattern at $80^\circ C$, double-sided access pattern at $50^\circ C$, and double-sided access pattern at $80^\circ C$, respectively. We plot representative \DRAMTIMING{AggON} values in different rows of the plots. 
{The lowest percentage of bitflips that occur in all five iterations is 56.8\% (observed from 16Gb C-Dies from Mfr. H) for the single-sided pattern at $80^\circ C$.} {For the double-sided pattern, the lowest percentage of bitflips that occur in all five iterations is 33.3\% at $50^\circ C$ (observed from 16Gb B-Dies from Mfr. M) and 47.2\% at $80^\circ C$ (observed from 16Gb E-Dies from Mfr. M).} 
We conclude that RowPress bitflips are {repeatable} with both single-sided and double-sided access patterns and at a higher temperature of $80^\circ C$.

\begin{figure}[H]

    \centering
    \includegraphics[width=1.0\linewidth]{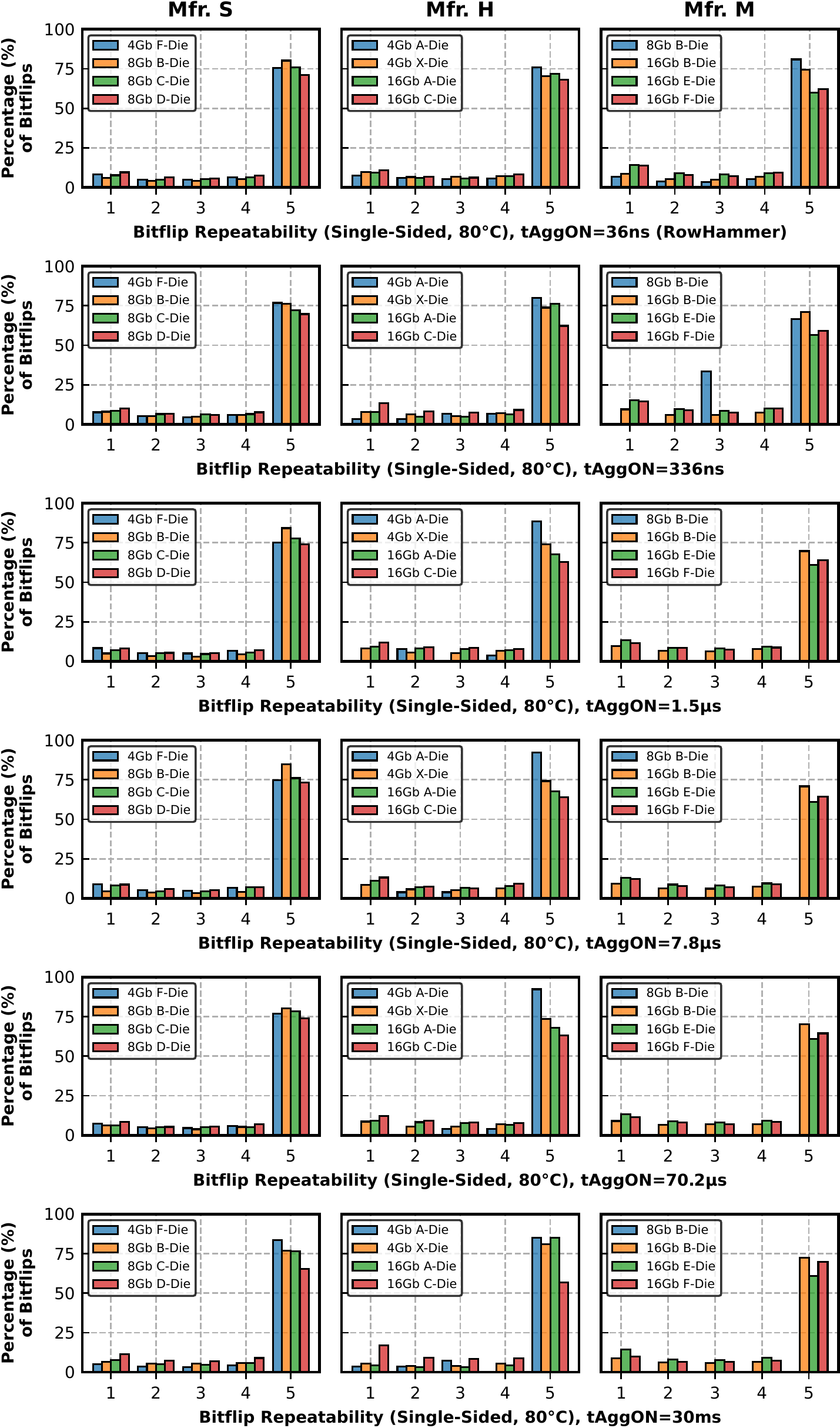}
    \caption{Repeatability of the single-sided RowPress (RowHammer) bitflips; $80^\circ C$}
    \label{fig:stab_SS_80}
\end{figure}

\begin{figure}[H]
    \centering
    \includegraphics[width=1.0\linewidth]{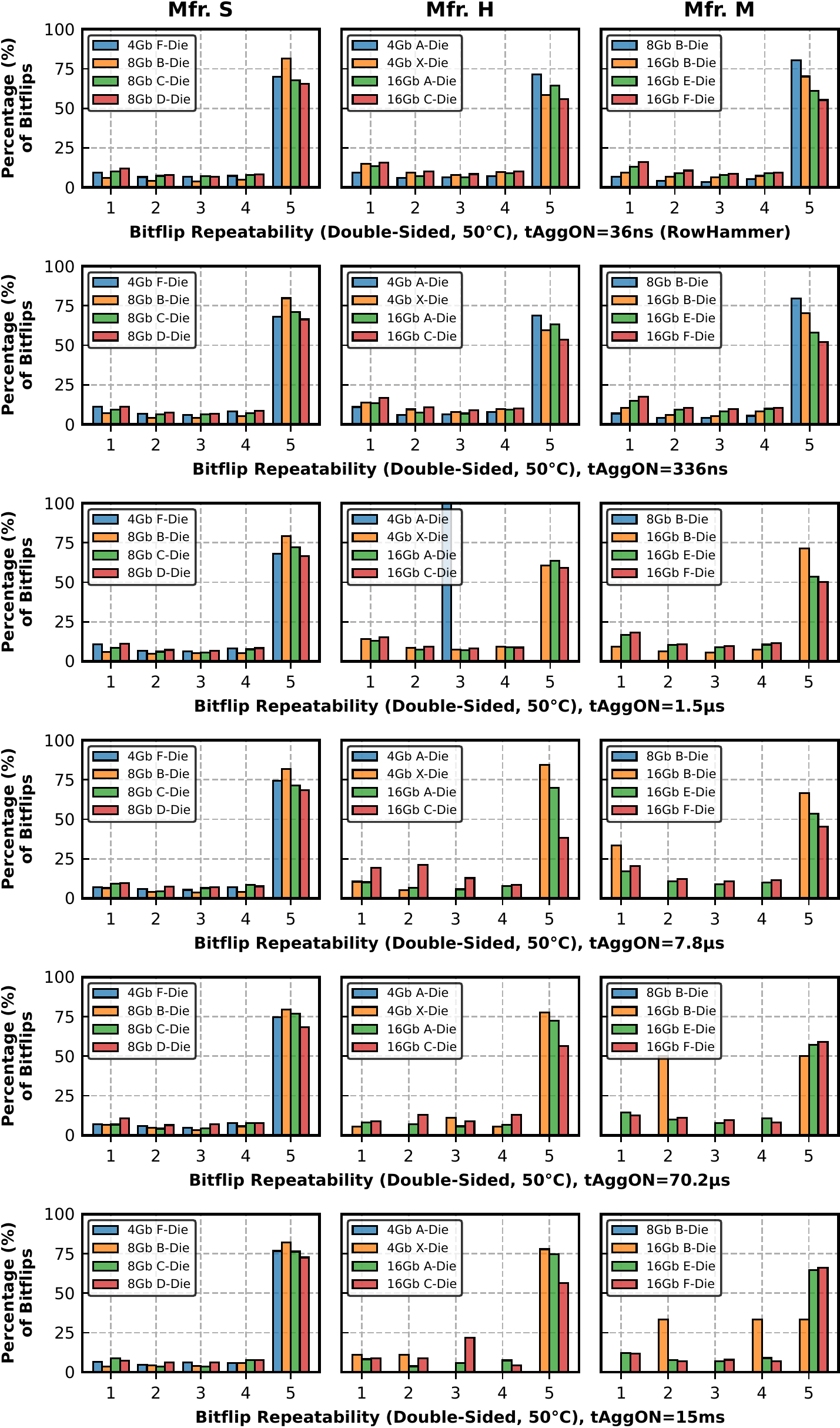}
    \caption{Repeatability of the double-sided RowPress (RowHammer) bitflips; $50^\circ C$}
    \label{fig:stab_DS_50}
\end{figure}

\begin{figure}[H]
    \centering
    \includegraphics[width=1.0\linewidth]{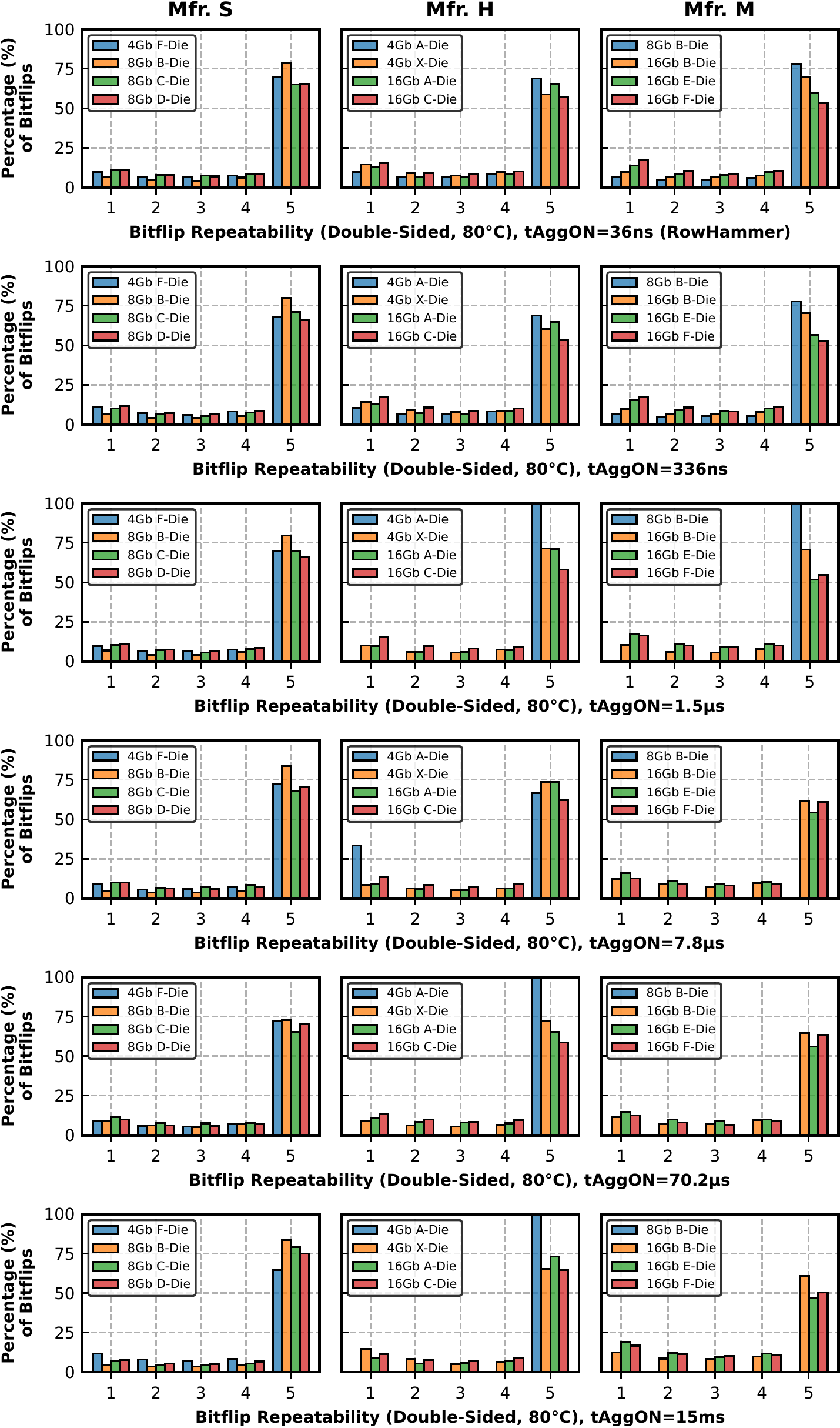}
    \caption{Repeatability of the double-sided RowPress (RowHammer) bitflips; $80^\circ C$}
    \label{fig:stab_DS_80}
\end{figure}
\nobalance
\pagebreak

\section{Extended Results on the Effect of Temperature on RowPress Bitflips}
\label{sec:65C}
We conduct further experiments to characterize RowPress bitflips at $65^\circ C$ to strengthen our observations that RowPress gets worse as temperature increases (Obsv. 9), and behaves differently compared to RowHammer as temperature and access pattern changes (Obsv. 13). 
\figref{fig:65_50} (\figref{fig:80_65}) shows the mean \gls{acmin} values we observe at $65^{\circ}C$ ($80^{\circ}C$) normalized to $50^{\circ}C$ ($65^{\circ}C$) as we sweep \DRAMTIMING{AggON} in {linear (y-axis) - log (x-axis)} scale, using the same {experimental} methodology as described in~\secref{sec:sen_temperature}. A data point below \gls{acmin} = 1 (highlighted with dashed red lines) means that for a given \DRAMTIMING{AggON}, it requires less aggressor row activations to induce at least one bitflip at a higher temperature.

\begin{figure}[H]
    \centering
    \includegraphics[width=1.0\linewidth]{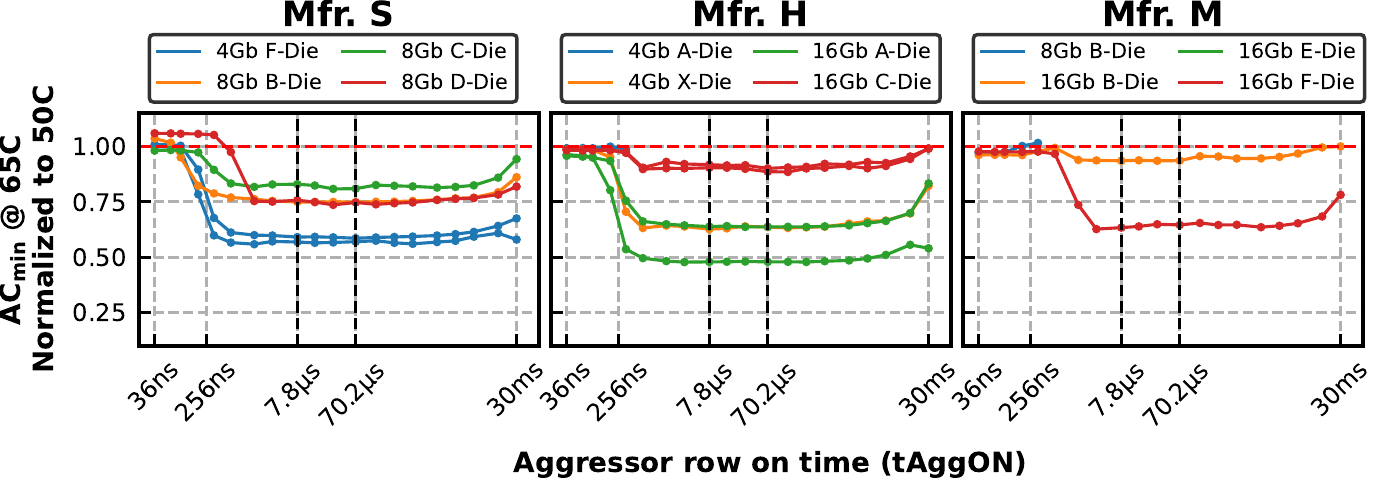}
    \caption{\gls{acmin} at $65^{\circ}C$ normalized to $50^{\circ}C$; single-sided RowPress.}
    \label{fig:65_50}
\end{figure}

\begin{figure}[H]
    \centering
    \includegraphics[width=1.0\linewidth]{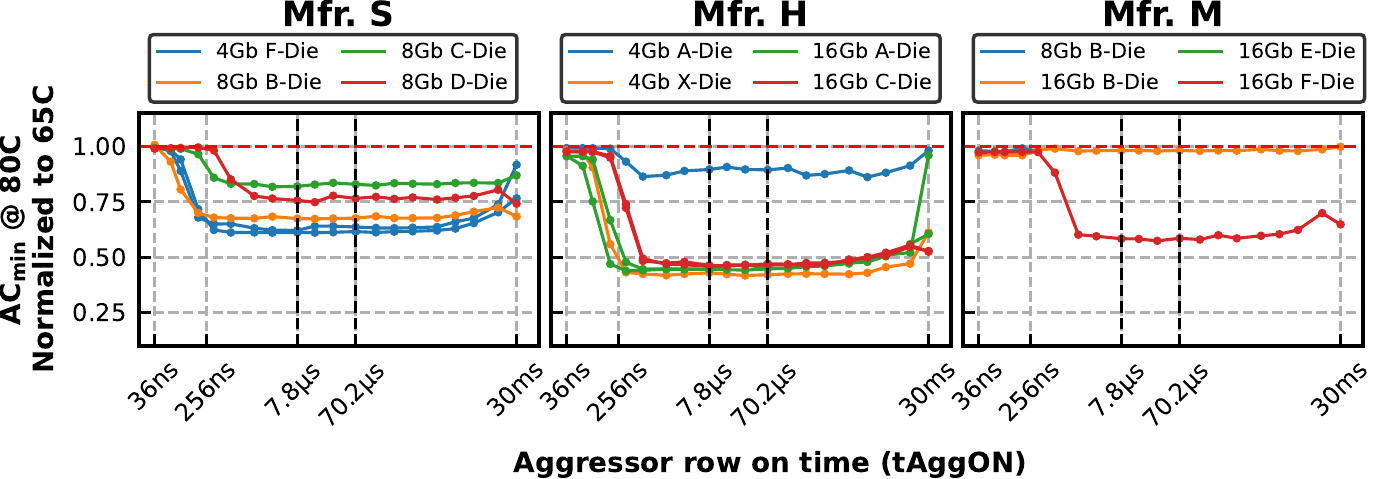}
    \caption{\gls{acmin} at $80^{\circ}C$ normalized to $65^{\circ}C$; single-sided RowPress.}
    \label{fig:80_65}
\end{figure}
We observe that for all die revisions vulnerable to RowPress, \gls{acmin} \emph{consistently} reduces for the same \DRAMTIMING{AggON} value as temperature increases from $50^{\circ}C$ to $65^{\circ}C$, {and then from $65^{\circ}C$ to $80^{\circ}C$} (i.e., Obsv. 9 still holds {when we consider three different temperatures, $50^{\circ}C$, $65^{\circ}C$, and $80^{\circ}C$}).

\figref{fig:ds_65} shows the difference between single- and double-sided \gls{acmin} (i.e., \gls{acmin}$(single)$ - \gls{acmin}$(double)$) at $50^{\circ}C$ (first row), $65^{\circ}C$ (second row) and $80^{\circ}C$ (third row), using the same {experimental} methodology as described in~\secref{sec:sen_acc_pattern}. A data point below $0$ means that the single-sided RowPress pattern needs fewer aggressor row activations in total to induce a bitflip compared to double-sided. We observe that, at $65^{\circ}C$, the single-sided RowPress pattern still needs fewer aggressor row activations in total to induce a bitflip compared to double-sided (i.e., Obsv. 13 still holds {when we consider three different temperatures, $50^{\circ}C$, $65^{\circ}C$, and $80^{\circ}C$}).

\begin{figure}[H]
    \centering
    \includegraphics[width=1.0\linewidth]{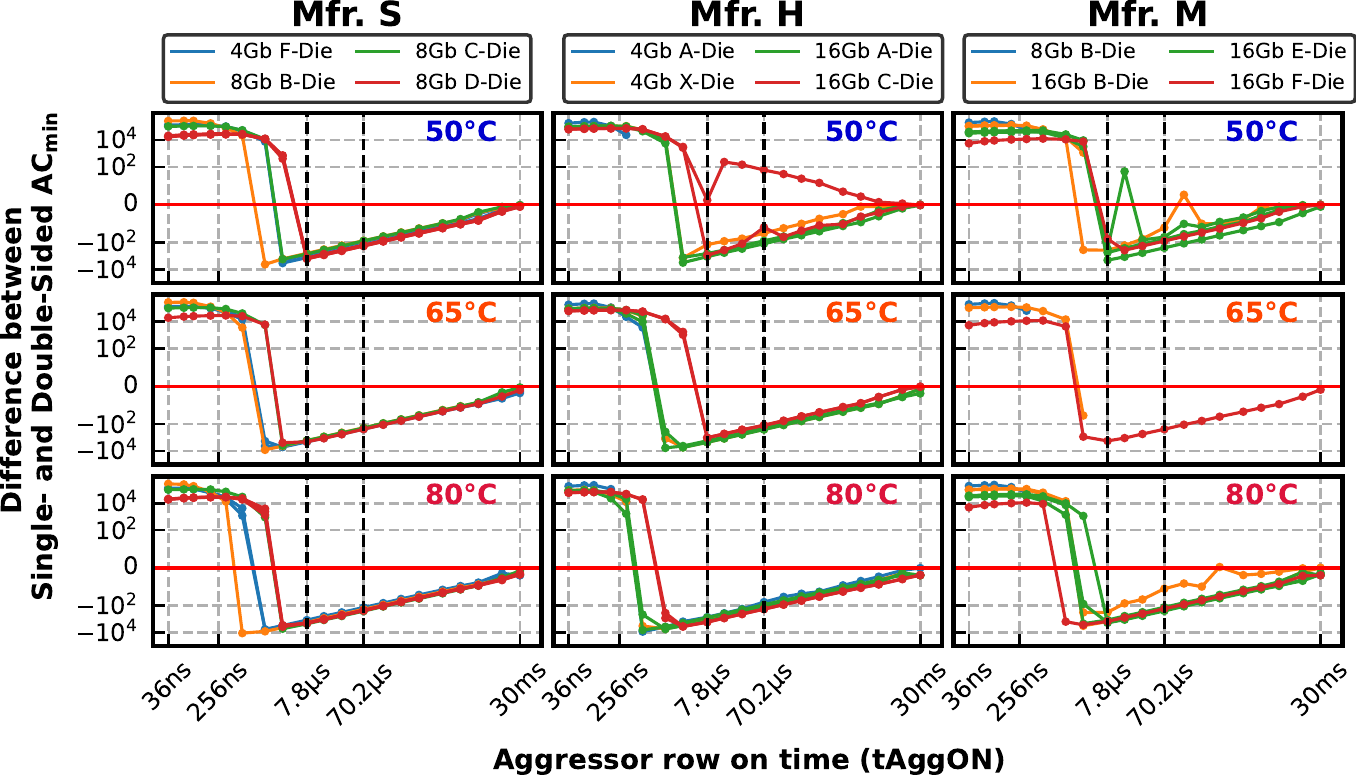}
    \caption{Single-sided \gls{acmin} minus double-sided \gls{acmin} at $50^{\circ}C$ (first row), $65^{\circ}C$ (second row) and $80^{\circ}C$ (third row).}
    \label{fig:ds_65}
\end{figure}

\nobalance
\pagebreak

\section{
Inducing Even More Bitflips on the Real System
}
\label{sec:realvariant}
Algorithm~\ref{alg:rowpress-program-variant} shows a variant of our real system RowPress test program (i.e., Algorithm~\ref{alg:rowpress-program} in~\secref{sec:real}) that changes the \emph{program order} of the accesses to the cache blocks and the \texttt{clflushopt} instructions to flush them from the cache. In the original Algorithm~\ref{alg:rowpress-program}, we flush the cache blocks {\emph{only after}} accessing \emph{all} cache blocks from \emph{both} aggressor rows (in program order, {lines 11-16} {in Algorithm~\ref{alg:rowpress-program}}). In Algorithm~\ref{alg:rowpress-program-variant}, we {\emph{immediately} flush each cache block \emph{after}} each cache block access (in program order, {lines 13-18} {in Algorithm~\ref{alg:rowpress-program-variant}}).

\renewcommand{\lstlistingname}{Algorithm} %
\begin{lstlisting}[caption={A variant of our RowPress test program that can induce many more bitflips than the test program in Algorithm 1.},captionpos=b,label={alg:rowpress-program-variant}, basicstyle=\scriptsize]
   // find two neighboring aggressor rows based on physical address mapping
   AGGRESSOR1, AGGRESSOR2 = find_aggressor_rows(|\textcolor{red}{{VICTIM}}|);
   // initialize the aggressor and the victim rows
   initialize(|\textcolor{red}{{VICTIM}}|, 0x55555555);
   initialize(AGGRESSOR1, AGGRESSOR2, 0xAAAAAAAA);
   // Synchronize with refresh
   for (iter = 0 ; iter < |\textcolor{red}{{NUM\_ITER}}| ; iter++):
     for (i = 0 ; i < |\textcolor{red}{{NUM\_AGGR\_ACTS}}| ; i++):
       // access multiple cache blocks in each aggressor row
       // to keep the aggressor row open longer
       // **************************************
       // *** MODIFIED PART START ***
       for (j = 0 ; j < |\textcolor{red}{{NUM\_READS}}| ; j++): 
            *AGGRESSOR1[j];
            clflushopt (AGGRESSOR1[j]); 
       for (j = 0 ; j < |\textcolor{red}{{NUM\_READS}}| ; j++):
            *AGGRESSOR2[j];
            clflushopt (AGGRESSOR2[j]);
       // *** MODIFIED PART END ***
       // **************************************
       mfence ();
     activate_dummy_rows();
   record_bitflips[|\textcolor{red}{{VICTIM}}|] = check_bitflips(|\textcolor{red}{{VICTIM}}|);
\end{lstlisting}

We run this variant of the test program (i.e., Algorithm 2) using the same methodology on the same system as described in~\secref{sec:real}. We plot the total number of bitflips (left) and the number of rows with bitflips (right) from both Algorithm 2 (purple bars) and Algorithm 1 (blue bars)\footnote{{The number of bitflips and the number of DRAM rows with bitflips from Algorithm 1 {depicted in~\figref{alg:rowpress-program-variant} differ slightly} from what we show {in~\figref{fig:real_rowpress_bitflips}} in~\secref{sec:real} because {these figures depict results from different runs of} our test program with Algorithm 1 \emph{on the real system}. Low-level events that are transparent to the program (e.g., the dynamic process scheduling decisions by the {operating system} and different synchronization points with the DRAM refresh commands) cause slight variations in the {experimental results across different runs of the same program.}}} for different {{numbers} of cache blocks read per aggressor row activation} ({\texttt{NUM\_READS};} x-axis) when {we activate each aggressor row four (top {plots}), three (middle {plots}), and two (bottom {plots}) times per iteration} in \figref{fig:real_comp_variant}.  We do not plot \texttt{NUM\_AGGR\_ACTS=1} because we do not observe any bitflips for all \texttt{NUM\_READS} we test. {The leftmost bar in {each graph} shows the number of \atb{\emph{conventional RowHammer-induced}} bitflips, \atb{where we read \emph{only a single} cache block {per aggressor row activation}, \atb{such that the aggressor row is kept open for a short time}}. {Remaining} bars {in each graph}  show results for RowPress-induced bitflips (with \joel{an} increasing number of cache block reads {from left to right}, such that the \atb{aggressor} row is kept open for an increasing amount of time).} We make the following {major observation} from \figref{fig:real_comp_variant}.

\begin{figure}[H]
    \centering
    \includegraphics[width=1.0\linewidth]{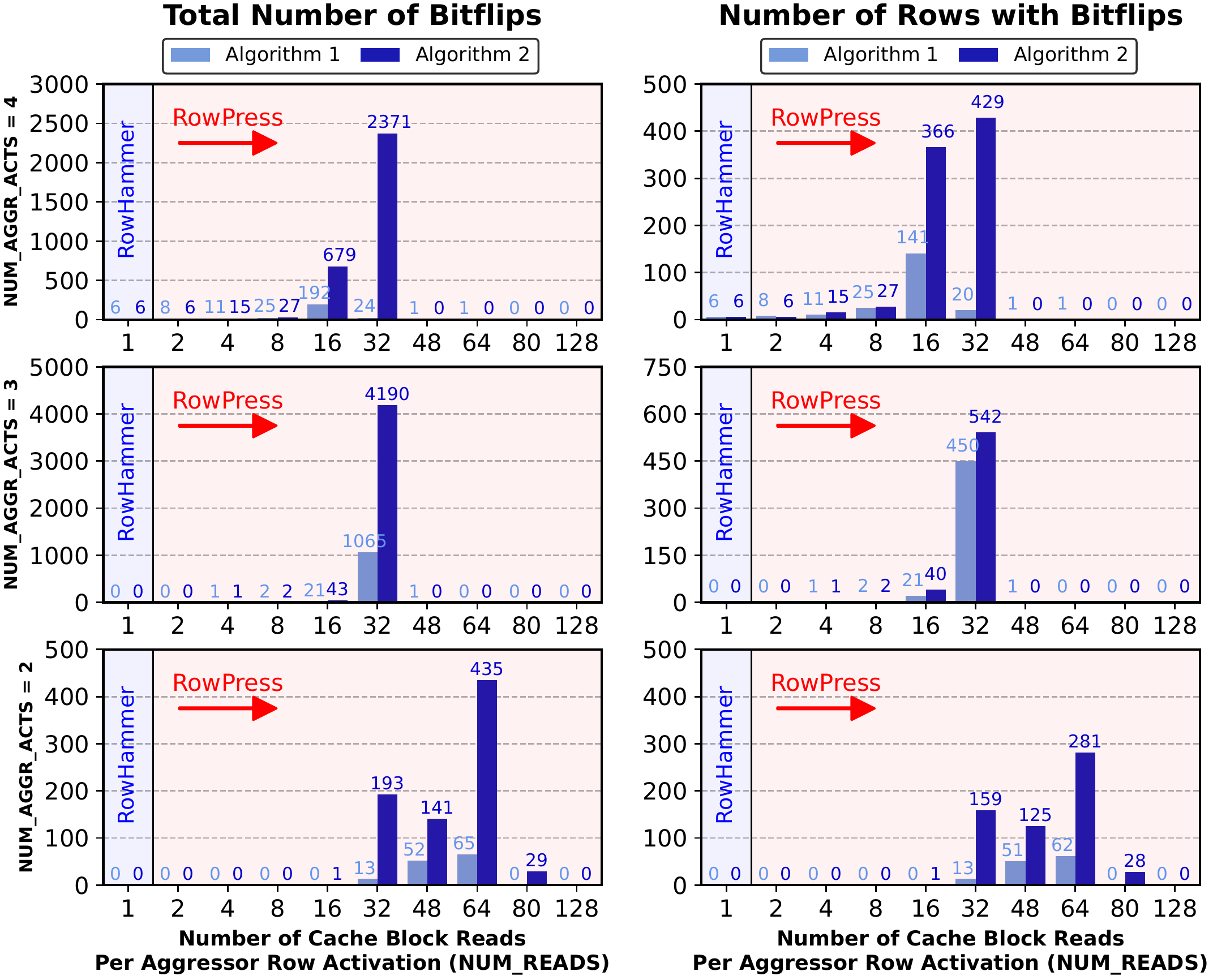}
    \caption{Number of RowHammer vs. RowPress bitflips (left) and number of rows with bitflips (right) we observe after running our {proof of concept test programs} with Algorithm 1 (blue bars) and Algorithm 2 (purple bars) {with four (top), three (middle), and two (bottom) activations per aggressor row per iteration}.}
    \label{fig:real_comp_variant}
\end{figure}

\observation{With Algorithm 2, the {proof of concept} test program induces {significantly more bitflips in many more DRAM rows} {in a real system}.}

With Algorithm 2, our test program induces significantly more bitflips in significantly more DRAM rows. For example, when \texttt{NUM\_AGGR\_ACTS=4} and \texttt{NUM\_READS=32}, with Algorithm 2, the test program induces 2371 bitflips in 429 DRAM rows, compared to {only} 24 bitflips in 20 rows with Algorithm 1, {amounting to an increase of 98.79$\times$ and 21.45$\times$, respectively.} When \texttt{NUM\_AGGR\_ACTS=3} and \texttt{NUM\_READS=32}, Algorithm 2 induces {4190 bitflips} in 542 DRAM rows, compared to 1065 bitflips in 450 DRAM rows with Algorithm 1, {amounting to an increase of 3.93$\times$ and 1.20$\times$, respectively.} {We hypothesize that the memory access pattern of Algorithm 2 causes the aggressor rows to be open longer than {that of Algorithm 1, leading to} many more bitflips in many more DRAM rows. Our results call for more investigation of how DRAM row open time is (and should be) handeled in modern memory controllers. To aid such research, we open source all our proof of concept programs (including Algorithm 2) in our Github repository at~\cite{rowpress-artifact-github}.}

\end{document}